\documentclass[10pt-default]{article}
\usepackage{amsmath}
\usepackage{amssymb}
\usepackage{achemso}

\input{tcilatex}

\begin{document}

{\huge \ }

{\huge New Strings for Old Veneziano }

\ \ \ \ \ \ \ \ \ \ \ \ \ \ \ \ \ \ \ \ \ \ \ \ {\huge Amplitudes }

\ \ \ \ \ \ {\huge II. Group-Theoretic Treatment}

$\ \ \ \ \ \ \ \ \ \ \ \ \ \ \ \ \ \ \ \ \ \ \ \ \ \ \ \ \ \ \ \ \ $

\ \ \ \ \ \ \ \ \ \ \ \ \ \ \ \ \ \ \ \ \ \ \ \ \ \ \ A.L. Kholodenko

\textit{375 H.L.Hunter Laboratories, Clemson University, Clemson, }

\textit{SC} 29634-0973, USA

\bigskip \textbf{Abstract}

In this part of our four parts work we use theory of polynomial invariants
of finite pseudo-reflection groups in order to reconstruct both the
Veneziano and Veneziano-like (tachyon-free) amplitudes and the generating
function reproducing these amplitudes. We demonstrate that such generating
function and amplitudes associated with it can be recovered with help of 
\textit{finite dimensional} \textit{exactly solvable} N=2 supersymmetric
quantum mechanical model known earlier from works of Witten, Stone and
others. Using the Lefschetz isomorphism theorem we replace traditional
supersymmetric calculations by \ the group-theoretic thus solving the
Veneziano model exactly using standard methods of representation theory.
Mathematical correctness of our arguments relies on important theorems by
Shepard and Todd, Serre and Solomon proven respectively in the early fifties
and sixties and documented in the monograph by Bourbaki. Based on these
theorems, we explain why the developed formalism leaves all known results of
conformal field theories unchanged. We also explain why \ these theorems
impose stringent requirements connecting analytical properties of scattering
amplitudes with symmetries of space-time in which such amplitudes act.

\ 

MSC: 81T30; 13A50; 20F55: 20G45

\ 

\textit{Subj Class}.: String theory, polynomial invariants of finite

pseudo-reflection groups

\ 

\textit{Keywords}: Veneziano and Veneziano-like amplitudes,

Weyl-Coxeter reflection and pseudo reflection groups and their invariants,

combinatorics of \ complex Grassmannians, supersymmetric quantum

mechanics, Lefschetz isomorphism theorem, Riemann's zeta function zeros,

Bergman metric, complex hyperbolic geometry

\bigskip

\pagebreak

\bigskip

\section{Introduction}

\subsection{Motivation}

In our earlier work, Ref.[1], which will be called Part I, while discussing
analytical properties of the Veneziano and Veneziano-like amplitudes we
noticed that the Veneziano condition for the 4-particle amplitude is given
by 
\begin{equation}
\alpha (s)+\alpha (t)+\alpha (u)=-1,  \tag{1.1}
\end{equation}%
where $\alpha (s)$, $\alpha (t),\alpha (u)$ $\in \mathbf{Z}$ . This result
can be rewritten in more general and mathematically suggestive form. To this
purpose, following Ref.[2], we would like to consider additional homogenous
equation of the type 
\begin{equation}
\alpha (s)m+\alpha (t)n+\alpha (u)l+k\cdot 1=0  \tag{1.2}
\end{equation}%
with $m,n,l,k$ being some integers. By adding this equation to Eq.(1.1) we
obtain, 
\begin{equation}
\alpha (s)\tilde{m}+\alpha (t)\tilde{n}+\alpha (u)\tilde{l}=\tilde{k}. 
\tag{1.3a}
\end{equation}%
so that we formally obtain,%
\begin{equation}
n_{1}+n_{2}+n_{3}=\hat{N},  \tag{1.3b}
\end{equation}%
where all entries \textit{by design} are nonnegative integers. For the
multiparticle case this equation should be replaced by%
\begin{equation}
n_{0}+\cdot \cdot \cdot +n_{k}=N  \tag{1.4}
\end{equation}%
so that combinatorially the task lies in finding all nonnegative integer
combinations of $n_{0},...,n_{k}$ producing Eq.(1.4). It should be noted
that such a task makes sense as long as $N$ is assigned. But the actual
value of $N$ is \textit{not} \textit{fixed} and, hence, can be chosen quite
arbitrarily.

\textbf{Remark 1.1}. In view of the results of Part I, one can argue that
the value of $N$ should coincide with the exponent of the Fermat
(hyper)surface. This observation is superficial, however, in view of
Eq.(3.29) of Part I. Indeed, in this Introduction we are talking about the
mathematical statements \textit{before} the bracket operation \TEXTsymbol{<}%
...\TEXTsymbol{>} defined in Part I is applied. This means that we shall be
working mainly with precursors of the period integrals in the projective
space discussed in some detail in Section 3 of Part I. Evidently, this makes
sense only if, in contrast with traditional string-theoretic treatments, we
interpret the Veneziano amplitudes as periods of the Fermat (hyper) surfaces.

\textbf{Remark 1.2}. Eq.(1.1) is a simple statement about the energy
-momentum conservation. Although the numerical entries in this equation can
be changed as we just have explained to make them more suitable for
theoretical treatments, the actual physical values can be re obtained
subsequently by the appropriate coordinate shift. Such a procedure is not
applicable to amplitudes in conformal field theories (CFT) where the
periodic (antiperiodic, etc.) boundary conditions cause energy and momenta
to become a \textit{quasi -energy} and a \textit{quasi momenta} as is well
known from the solid state physics. This fact was noticed already in Part I
where Eq.(3.22) used in CFT replaces the standard Veneziano condition, e.g.
Eq.(1.2).

This arbitrariness of choosing $N$ represents a kind of gauge freedom in
physics terminology. As in other gauge theories, we can fix the gauge by
using some physical considerations. These include, for example, an
observation made in Part I that the 4 particle amplitude is zero if any two
entries into Eq.(1.1) (or, which is the same, into Eq.(1.3b)) are the same.
This \ fact prompts us to arrange the entries in Eq.(1.3b) in accordance
with their magnitude, i.e. $n_{1}\geq n_{2}\geq n_{3}.$ More generally, in
view of Eq.(1.4), we can write: $n_{0}\geq n_{1}\geq \cdot \cdot \cdot \geq
n_{k}\geq 1\footnote{%
The last inequality: $n_{k}\geq 1,$ is chosen only for the sake of
comparison with the existing literature conventions, e.g. see Ref.[3\textbf{]%
}.}$. Provided that Eq.(1.4) holds, we shall call such a sequence a \textit{%
partition }and shall denote it as\textit{\ }$\mathit{n\equiv }%
(n_{0},...,n_{k})$. If $n$ is partition of $N$ , then we shall write $%
n\vdash N$ . It is well known [3,4] that there is one- to -one
correspondence between the Young diagrams and partitions. We would like to
use this fact in order to design a new partition function capable of
reproducing the Veneziano (and Veneziano-like) amplitudes. Clearly, such a
partition function should also make physical sense. This is the primary goal
of our paper. In this Introduction\ we would like to provide some convincing
qualitative arguments that such a goal can indeed be achieved. The rest of
the paper provides more rigorous mathematical results supporting our claim%
\footnote{%
We would like to warn our readers that, actually, there are several \
interrelated formulations of this partition function. Ref.[6] and Part III
provide some examples of such formulations. Part IV provides additional
requirements aimed to connect our formulations with the experimental data.
\par
{}}.

We begin with observation (taken from our earlier study of the
Witten-Kontsevich model, Ref.[7]) that there is one- to- one correspondence
between the Young tableaux and directed random walks. Let us recall details
of this correspondence now. To this purpose we need to consider a square
lattice and to place on it the Young diagram associated with some specific
partition which belongs to $n$. To do so, let us choose some $\tilde{n}%
\times \tilde{m}$ rectangle\footnote{%
The parameters $\tilde{n}$ and $\tilde{m}$ will be specified shortly below.}
so that the Young diagram occupies the left part of this rectangle. We
choose the upper left vertex of the rectangle as the origin of the $xy$
coordinate system whose $y$ axis (south direction) is directed downwards and 
$x$ axis is directed eastwards. Then, the south-east boundary of the Young
diagram can be interpreted as directed (that is without self intersections)
random walk which begins at $(0,-\tilde{m})$ and ends at $(\tilde{n},0).$%
Clearly, such a walk completely determines the diagram. The walk can be
described by a sequence of 0's and 1's,\ say, $0$ for the $x-$ step move and
1 for the $y-$ step move. The totality $\mathcal{N}$ of Young diagrams which
can be placed in the rectangle is in one-to-one correspondence with the
number of arrangements of 0's and 1's whose total number is $\tilde{m}+%
\tilde{n}$. The logarithm of the number $\mathcal{N}$ of possible
combinations of 0's and 1's \ is just the entropy associated with the Fermi
statistic (or, equivalently, the entropy of mixing for the binary mixture)
used in physics literature. The number $\mathcal{N}$ is given by $\mathcal{N}%
=(m+n)!/m!n!\footnote{%
We have suppressed the tildas for $n$ and $m$ in this expression since these
parameters are going to be redefined below anyway.}$. It can be represented
in two equivalent ways%
\begin{eqnarray}
(m+n)!/m!n! &=&\frac{(n+1)(n+2)\cdot \cdot \cdot (n+m)}{m!}\equiv \left( 
\begin{array}{c}
n+m \\ 
m%
\end{array}%
\right)  \notag \\
&=&\frac{(m+1)(m+2)\cdot \cdot \cdot (n+m)}{n!}\equiv \left( 
\begin{array}{c}
m+n \\ 
n%
\end{array}%
\right) .  \TCItag{1.5}
\end{eqnarray}%
In Part I, Eq-s (1.21)-(1.23) \ explain how $\mathcal{N}$ is entering the
Veneziano amplitude. Additional physical significance of this number in
connection with the Veneziano amplitude and its partition function is
developed in Ref.[6] and in Part III. \ For such a development it is
absolutely essential that the number $\mathcal{N}$ is integer \textit{for
all nonnegative m's and n's\footnote{%
That this should be the case can be seen by noticing that using symmetry
considerations we can always write $(m+n)!=m!n!C(m,n).$ The constant $C(m,n)$
represents all permutations between sets $m$ and $n$. It should be a
positive integer since the l.h.s.in the above equation \textit{is} a
positive integer.} and can be presented in two ways}. In Part I we noticed
that $\mathcal{N}$ can be interpreted as the total number of \ points with
integer coordinates enclosed by the dilated (with dilation coefficient $n$) $%
m$- dimensional simplex $\mathit{n}\Delta _{m}$ whose vertices are located
at the nodes of \textbf{Z}$^{m}.$ This observation is crucial for
development of our formalism, especially in Part III.

Let now $p(N;k,m)$ be the number of partitions of $N$ into $\leq k$ \
nonnegative parts, each not larger than $m$. Consider the generating
function of the following type: 
\begin{equation}
\mathcal{F}(k,m\mid q)=\dsum\limits_{N=0}^{S}p(N;k,m)q^{N}.  \tag{1.6}
\end{equation}%
where the upper limit $S$\ will be determined shortly below. It is shown in
Refs.[3-5,8] that $\mathcal{F}(k,m\mid q)=\left[ 
\begin{array}{c}
k+m \\ 
m%
\end{array}%
\right] _{q}\equiv \left[ 
\begin{array}{c}
k+m \\ 
k%
\end{array}%
\right] _{q}$ where, for instance,$\left[ 
\begin{array}{c}
k+m \\ 
m%
\end{array}%
\right] _{q=1}=\left( 
\begin{array}{c}
k+m \\ 
m%
\end{array}%
\right) \footnote{%
On page 15 of the book by Stanley, Ref.[4], one can find that the number of
solutions $N(n,k)$ in \textit{positive} integers to $y_{1}+...+y_{k}=n+k$ is
given by $\left( 
\begin{array}{c}
n+k-1 \\ 
k-1%
\end{array}%
\right) $ while the number of solutions in \textit{nonnegative} integers to $%
x_{1}+...+x_{k}=n$ is $\left( 
\begin{array}{c}
n+k \\ 
k%
\end{array}%
\right) .$Careful reading of Page 15 indicates however that the last number
refers to solution in nonnegative integers of the equation $%
x_{0}+...+x_{k}=n $. We have used this fact in Part I, e.g. see Eq.(1.21).}.$
It should be clear from this result that the expression $\left[ 
\begin{array}{c}
k+m \\ 
m%
\end{array}%
\right] _{q}$ is a $q-$analog of the binomial coefficient $\left( 
\begin{array}{c}
k+m \\ 
m%
\end{array}%
\right) .$ In literature [3-5,8] this $q-$ analog is known as the \textit{%
Gaussian} coefficient. Explicitly,%
\begin{equation}
\left[ 
\begin{array}{c}
k \\ 
m%
\end{array}%
\right] _{q}=\frac{(q^{k}-1)(q^{k-1}-1)\cdot \cdot \cdot (q^{k-m+1}-1)}{%
(q^{m}-1)(q^{m-1}-1)\cdot \cdot \cdot (q-1)}.  \tag{1.7}
\end{equation}%
From this definition, it should be intuitively clear that the sum defining
the generating function $\mathcal{F}(k,m\mid q)$ in Eq.(1.6) should have
only \textit{finite} number of terms. Eq.(1.7) allows easy determination of
the upper limit $S$ in the sum, Eq.(1.6). It is given by $km$. This is just
the area of the $k\times m$ rectangle. Evidently, in view of the definition
of $p(N;k,m)$, the number $m=N-k$. Using this fact, Eq.(1.6) can be
rewritten as: $\mathcal{F}(N,N-k\mid q)=\left[ 
\begin{array}{c}
N \\ 
k%
\end{array}%
\right] _{q}.$ This expression happens to be the Poincare$^{\prime }$
polynomial for the complex Grassmannian $Gr(m,k)$. This can be found on page
292 of the famous book by Bott and Tu, Ref.[9]\footnote{%
To make a comparison it is sufficient to replace parameters $t^{2}$ and $n$
in \ Bott and Tu book by $q$ and $N.$}. From this point of view the
numerical coefficients, i.e. $p(N;k,m),$ in the $q$ expansion of Eq.(1.6)
should be interpreted as the Betti numbers of this Grassmannian. They can be
determined recursively using the following property of the Gaussian
coefficients [4, page 26]%
\begin{equation}
\left[ 
\begin{array}{c}
n+1 \\ 
k+1%
\end{array}%
\right] _{q}=\left[ 
\begin{array}{c}
n \\ 
k+1%
\end{array}%
\right] _{q}+q^{n-k}\left[ 
\begin{array}{c}
k \\ 
m%
\end{array}%
\right] _{q}  \tag{1.8}
\end{equation}%
and taking into account that $\left[ 
\begin{array}{c}
n \\ 
0%
\end{array}%
\right] _{q}=1.$ To demonstrate that $\mathcal{F}(N,N-k\mid q)$ is indeed
the Poincare$^{\prime }$ polynomial for the Grassmannian we would like to
use some results from the number theory. For readers unfamiliar with number
theory a concise summary of relevant results can be found, for instance, in
the Appendix A of our earlier work, Ref.[10]. Given this, let $q$ be some
prime and consider the finite field \textbf{F}$_{q}$ of $q$ elements.
Consider next the field extension. This is effectively accomplished \ by
constructing an $N$ dimensional vector space via prescription: 
\begin{equation}
\mathbf{F}_{q}^{N}=\{\alpha _{1},...,\alpha _{N}\}:\alpha _{i}\in \mathbf{F}%
_{q}.  \tag{1.9}
\end{equation}%
Any number which belongs to this new(extended) number field is expandable in
terms of the basis "vectors" just specified. It can be shown [2,8], that the
number of $k-$dimensional subspaces of the vector space $\mathbf{F}_{q}^{N}$
is given exactly by $\mathcal{F}(N,N-k\mid q).$ The arguments leading to
such a conclusion can be found already in the classical paper by Andre Weil,
Ref.[11], written in 1949. Incidentally, in his paper he studies the number
of solutions in the field \textbf{F}$_{q}$ for the Fermat hypersurface $%
\mathcal{F}$ 
\begin{equation}
a_{0}z_{0}^{\hat{N}}+\cdot \cdot \cdot +a_{n+1}z_{n+1}^{\hat{N}}=0 
\tag{1.10}
\end{equation}%
living in the complex projective space \textbf{CP}$^{n+1}$. Such a
hypersurface was discussed in Part I (e.g. see Eq.(3.6)) in connection with
our calculations of the Veneziano (and Veneziano-like) amplitudes. For the
sake of space, with exception of \ Subsection 8.3, in this work we avoid the
number- theoretic aspects related to the Veneziano amplitudes and their
partition functions. We shall explain shortly below the rationale behind
such an exception.

In the meantime, going back to our discussion, we notice that due to
relation $m=N-k$ it is more advantageous \ for us to use parameters $m$ and $%
k$ than $N$ and $k$. With this in mind we obtain, 
\begin{eqnarray}
\mathcal{F}(k,m &\mid &q)=\left[ 
\begin{array}{c}
k+m \\ 
k%
\end{array}%
\right] _{q}=\frac{(q^{k+m}-1)(q^{k+m-1-1}-1)\cdot \cdot \cdot (q^{m+1}-1)}{%
(q^{k}-1)(q^{k-1}-1)\cdot \cdot \cdot (q-1)}  \notag \\
&=&\dprod\limits_{i=1}^{k}\frac{1-q^{m+i}}{1-q^{i}}  \TCItag{1.11}
\end{eqnarray}%
This result is re obtained in the main text using different \ mathematical
arguments. It is of \textit{central importance} for this work since it is
obtainable from the supersymmetric partition function capable of reproducing
the Veneziano and Veneziano-like amplitudes. In the limit : $q\rightarrow 1$
Eq.(1.11) reduces to the number $\mathcal{N}$ as required. To make
connections with results already known in physics we need to rescale $%
q^{\prime }s$ in Eq.(1.11), e.g. let $q=t^{\frac{1}{i}}.$ Substitution of
such an expression back into Eq.(1.11) and taking the limit $t\rightarrow 1$
again produces $\mathcal{N}$ in view of Eq.(1.5). This time, however, we can
accomplish much more. By noticing that in Eq.(1.4) the actual value of $N$
by design is not fixed thus far and taking into account that $m=N-k$ we can
fix $N$ by fixing $m$. Specifically, we would like to choose $m=1\cdot
2\cdot 3\cdot \cdot \cdot k$ and with such an $m$ \ to consider a particular
term in the product Eq.(1.11), e.g.%
\begin{equation}
S(i)=\frac{1-t^{1+\frac{m}{i}}}{1-t}.  \tag{1.12}
\end{equation}%
In view of our "gauge fixing" the ratio $m/i$ is a positive integer by
design. This means that we are having a sum for the geometric progression.
Indeed, if we rescale $t$ again : $t\rightarrow t^{2},$ then we obtain%
\begin{equation}
S(i)=1+t^{2}+\cdot \cdot \cdot +t^{2\hat{m}}  \tag{1.13}
\end{equation}%
with $\hat{m}=\frac{m}{i}.$ Written in such a form the sum above is just the
Poincare$^{\prime }$ polynomial for the complex projective space \textbf{CP}$%
^{\hat{m}}.$This can be seen by comparing pages 177 and 269 of the book by
Bott and Tu, Ref.[9]. Hence, at least for some $m$'s, \textit{the Poincare}$%
^{\prime }$\textit{\ polynomial for the Grassmannian in just the product of} 
\textit{the Poincare}$^{\prime }$\textit{\ polynomials for the complex
projective spaces of known dimensionalities}. For $m$ just chosen in the
limit : $t\rightarrow 1,$ we re obtain back the number $\mathcal{N}$ as
required. This physically motivating process of gauge fixing we have just
described can be replaced by more rigorous mathematical arguments. The
recursion relation, Eq.(1.8), introduced earlier indicates that this is
possible. The \ mathematical details leading to just described factorization
can be found, for instance, in the lecture notes by Schwartz, Ref.[12, Ch-r
3]. Nevertheless, in Section 7, in view of their simplicity and intuitive
appeal (as compared with arguments by Schwartz), we use different chain of
arguments to arrive at the same conclusions. The topological significance of
the Poincare$^{\prime }$ polynomial decomposition into the product of
Poincare$^{\prime }$ polynomials is discussed in general terms in Section 4
and is used in Sections 7and 8 to recover the relevant physics. The relevant
physics emerges by noticing that the partition function $Z(J)$ for the
particle with spin $J$ is given by [13] 
\begin{eqnarray}
Z(J) &=&tr(e^{-\beta H(\sigma )})=e^{cJ}+e^{c(J-1)}+\cdot \cdot \cdot
+e^{-cJ}  \notag \\
&=&e^{cJ}(1+e^{-c}+e^{-2c}+\cdot \cdot \cdot +e^{-2cJ})  \TCItag{1.14}
\end{eqnarray}%
where $c$ is known constant. Evidently, up to a constant, $Z(J)\simeq S(i).$
But the result Eq.(1.14) is the Weyl character formula! This fact is \ to be
discussed at length in Part III. \ The observation just made brings the
classical group theory into our arguments. More importantly, because the
partition function for the particle with spin $J$ can be written in the
language of N=2 supersymmetric quantum mechanical model\footnote{%
We hope that no confusion is made about the meaning of N in the present case.%
}as demonstrated by Stone [13] and others [14], the connections between the
supersymmetry and the classical group theory are evident. We develop these
connections further in this work. Part III (see also Ref.[6]) contains many
additional results.

In view of arguments presented above, the Poincare$^{\prime }$ polynomial
for the Grassmannian can be interpreted as a partition function for a kind
of a spin chain made of spins of various magnitudes\footnote{%
In such a context it can be vaguely considered as a variation on the theme
of the Polyakov rigid string (Grassmann $\sigma )$ model, Ref.[15], pages
283-287, except that now it is \textit{exactly solvable}. A somewhat
different interpretation of the rigid string model was developed in our
earlier work, Ref.[16].} caused by gauge fixing just described. In fact, the
spin analogy is actually unnecessary since the formalism developed in
Ref.[14] is valid for any finite dimensional homogenous space. It remains to
demonstrate that the \textit{finite dimensional} supersymmetric model just
sketched can be used for reproduction of the Veneziano and Veneziano-like
amplitudes. Such a task is accomplished in the rest of this work. The major
reason for finite dimensionality is given by \ important theorems by
Solomon, Shepard and Todd, Lefschetz and Serre discussed in the main text.
In addition, the important theorem by Serre discussed in Section 9 not only
provides needed support to our qualitative conclusions about finite
dimensionality but also explains the connection between analytical
properties of the Veneziano (and Veneziano-like) amplitude and the
properties of space-time in which such amplitude "lives".

\bigskip\ 

\subsection{Organization of the rest of this paper}

The rest of this paper provides needed mathematical justifications
supporting intuitive ideas just discussed. These justifications come mainly
from the theory of invariants of finite pseudo-reflection groups. For the
sake of uninterrupted reading, major ingredients of this theory are provided
in the text along with important facts (in Appendix) about the Weyl-Coxeter
reflection groups and their generalization to pseudo-reflection groups by
Shepard and Todd and others. Sections 2 through 6 contain all needed
information needed for recovery of Eq.(1.11). In view of the recursion
relation, Eq.(1.8), it can be interpreted as the Weyl character formula
(this will be proven in Part III). In view of this observation, in Section 7
we accomplish several tasks. First, we provide needed mathematical
justification for Eq.(1.12) thus connecting our results with those known
earlier for spins and spin chains. Second, we investigate if this connection
is the only option available or if there are other options. We find these
other options as well. They allow us to bring into picture the formalism of
exactly integrable systems and, in particular, to connect the obtained
results with the tau function of the Kadomtsev-Petviashvili hierarchy.
Through such a connection it is possible, in principle, to establish links
with the existing string-theoretic formalism. Obtained results supply us
with still other options however. This point of bifurcation from traditional
formalism is studied further in Section 8. In it we discuss new exact
solution of the Veneziano model. The obtained result \ happens to have
additional uses which we also discuss in some detail. In particular, earlier
searches for quantum mechanical systems whose spectrum reproduces zeros of
the Riemann zeta function had resulted in the likely candidate : H=xp. In
part I, Eq.(1.12), does represents the Veneziano amplitude as product of
zeta functions. In Section 8 we argue that this is not a curiosity: there is
a deep reason for such a representation. Thus, we explain how the
Hamiltonian H=xp is related to the string-theoretic results we have
obtained. After this, in Section 9 we discuss from various angles the
important theorem by Serre. This theorem explains why there is not much
freedom left to improve (replace, change) the Veneziano amplitude. This
result is further strengthened by our observation that the function
generating all Veneziano amplitudes is obtainable as a deformation retract
for the Bergman kernel. Such a kernel has been used recently in connection
with complex- hyperbolic geometry. The metric obtainable with such a kernel
is an analog of the Lobachevskii metric in the real hyperbolic space. In
accord with the ball model for the real hyperbolic space, there is analogous
ball model for the complex hyperbolic space. The isometries of the boundary
of such a ball model are described by the Heisenberg group. Since the real
hyperbolic space is just a part of the complex- hyperbolic and since the
real-hyperbolic is connected with the Minkowski space-time, we obtain the
unusually tight connections between the Veneziano amplitudes and the
properties of space -time in which these amplitudes act.

\section{Selected exercises from Bourbaki (begining)}

In this section we begin our explanation of how group-theoretic methods can
be used for reconstruction of both the Veneziano amplitudes and their
generating/partition function. To accomplish this task, we need \ to work
out some problems listed at the end of Chapter 5, paragraph 5 (problem set
\# 3) of the monograph by Bourbaki, Ref.[17]. Fortunately, answers to these
problems to a large extent (but not completely!) can be extracted from the
paper by Solomon [18]. In view of their crucial mathematical and physical
importance, we reproduce some of his results in this section and will
complete our treatment in Sections 3- 6\textbf{.}

Let $K$ be the field of characteristic zero (e.g.\textbf{C}) and $V$ be the
vector space of finite dimension $l$\textit{\ }over it. Let $G$ be a
subgroup of $GL_{l}(V)$ acting on $V$. Let $q$ be the cardinality of $G%
\footnote{%
I.e. the \ number of elements of $G$.}$. Introduce \ now the symmetric $S(V)$
and the exterior $E(V)$ algebra of $V$ in order to construct invariants of
the group $G$ made of $S(V)$ and $E(V)\footnote{%
The fact that we actually need to use the pseudo-reflection groups (e.g. see
Appendix, part d)) will be explained mathematically only later, in Section
9. Hence, the formalism we are describing \ in this and following sections
has actually a wider use.}$. This task requires several steps. First, the
multiplication of polynomials leads to the notion of a graded ring $R$%
\footnote{%
Surely, once the definition of such ring is given, there is no need to use
polynomials. But in the present case this analogy is useful.}. For example,
if the polynomial\ $P_{i}(x)$ of degree $i$ belongs to the polynomial ring $%
\mathbf{F}[x],$ then the product $P_{i}(x)P_{j}(x)\in P_{i+j}(x)\in \mathbf{F%
}[x]$.

\textbf{Definition 2.1.} A graded ring $R$ is a ring with decomposition $%
R=\oplus _{j=\mathbf{Z}}R_{j}$ compatible with addition and multiplication.

Next, for the vector space $V$ if $x=x_{1}\otimes \cdot \cdot \cdot \otimes $
$x_{s}\in V^{\otimes _{s}}$and $y=y_{1}\otimes \cdot \cdot \cdot \otimes $ $%
y_{t}\in V^{\oplus _{t}}$, then the product $x\otimes y\in V^{\oplus
_{s+t}}. $ A multitude of such type of tensor products forms the
noncommutative associative algebra $T(V)$. Finally, the \textit{symmetric}
algebra $S(V)$ is defined by $S(V)=T(V)/I$, where the ideal $I$ is made of $%
x\otimes y-y\otimes x$ (with both $x$ and $y$ $\in V).$ In practical terms $%
S(V)$ is made of \ symmetric polynomials \textbf{F}[$t_{1},...,t_{l}]$ with $%
t_{1},...,t_{l}$ being in one-to-one correspondence with the basis elements
of $V$ (that is each of $t_{i}^{\prime }s$ is entering into $S(V)$ with
power one). The exterior algebra $E(V)$ can be now defined analogously. For
this we need to map the vector space $V$ into the Grassmann algebra of $V$.
In particular, if $x\in V$, then its image in the Grassmann algebra $\tilde{x%
}$ possess a familiar property $:$ $\tilde{x}^{2}=0.$ The graded two- sided
ideal $I$ can be defined now as 
\begin{equation}
I=\{\tilde{x}^{2}=0\mid \text{ }x\rightarrow \tilde{x};x\in V\text{ }\} 
\tag{2.1}
\end{equation}%
so that $E(V)=T(V)/I$. To complicate matters a little bit, we would like to
consider a map $d:$ $x\rightarrow dx$ for $x\in V$ and $dx$ belonging to the
Grassmann algebra. If $t_{1},...,t_{l}$ is the basis of $V$,then $%
dt_{i_{1}}\wedge \cdot \cdot \cdot \wedge dt_{i_{k}}$ is the basis of $%
E_{k}(V)$ with $0\leq k\leq l$ and, accordingly, the graded algebra $E(V)$
admits the following decomposition: $E(V)$ $=\oplus _{k=0}^{l}E_{k}(V).$
Next, we need to construct the invariants of a (pseudo-reflection) group $G$
made out of $S(V)$ and $E(V)$ and, more importantly, out of the tensor
product $S(V)\otimes E(V)$. Toward this goal we need to look if the action
of the map $d:V\rightarrow E(V)$ extends to a differential map 
\begin{equation}
d:S(V)\otimes E(V)\rightarrow S(V)\otimes E(V).  \tag{2.2}
\end{equation}%
Clearly, $\forall x\in E(V)$ \ we have $d(x)=0.$ Therefore, $\forall x,y\in
S(V)\otimes E(V)$ we can write $d(x,y)=d(x)y+xd(y)$. By combining these two
results together we obtain, 
\begin{equation}
d:S_{i}(V)\otimes E_{j}(V)\rightarrow S_{i-1}(V)\otimes E_{j+1}(V), 
\tag{2.3}
\end{equation}%
i.e. the differentiation is compatible with grading. Now we are ready to
formulate the theorem by Solomon [18] which is of central importance for our
work. It is formulated in the form stated in Bourbaki, Ref.[17].

\bigskip

\textbf{Theorem} \textbf{2.2.} (Solomon [18] ) \textit{Let }$P_{1},...,P_{k%
\text{ }}$\textit{\ be algebraically independent polynomial forms made of
symmetric combinations of }$t_{1},...,t_{k}$\textit{\ generating the ring }$%
S(V)^{G}$\textit{\ of invariants of }$G$\textit{. Then, every invariant
differential p-form }$\omega ^{(p)}$\textit{\ may be written uniquely as a
sum} 
\begin{equation}
\omega ^{(p)}=\sum\limits_{i_{1}<...<i_{p}}c_{i_{1}...i_{p}}dP_{i_{1}}\cdot
\cdot \cdot dP_{i_{p}}\text{ \ ; }1\leq p\leq k,  \tag{2.4}
\end{equation}%
\textit{with }$c_{i_{1}...i_{p}}\in S(V)^{G}.$\textit{Moreover, actually,
the differential forms }$\Omega ^{(p)}=dP_{i_{1}}\wedge \cdot \cdot \cdot
\wedge dP_{i_{p}}$\textit{\ with }$1\leq p\leq k$\textit{\ generate the
entire algebra of }$G-$\textit{\ invariants of }$S(V)\otimes E(V).$

\bigskip

\textbf{Corollary\textit{\ }2.3.} Let \textit{\ }$t_{1},...,t_{k}$ be the
basis of $V$. Furthermore, let $S(V)=$ \textbf{F}[$t_{1},...,t_{k}]$ be its
algebra of symmetric polynomials and $S(V)^{G}=$\textbf{F}[$P_{1},...,P_{l}]$
its\textbf{\ }\textit{finite} algebra of $G-$invariants\footnote{%
More details about $S(V)^{G}$ are given later in Section 9. The fact that
the number of polynomial forms $P_{i}$ is equal to the rank \emph{l }of $G$
is not a trivial fact. The proof can be found in Ref.[19, p.128].
Incidentally, this proof implies immediately the result, Eq.(2.6), given
below. It is essential that to arrive at this result requires for the
algebra of $G$-invariants to be finite.}. Then, since $dP_{i}=\sum\limits_{j}%
\frac{\partial P_{i}}{\partial t_{j}}dt_{j}$ , we have 
\begin{equation}
dP_{1}\wedge \cdot \cdot \cdot \wedge dP_{k}=J(dt_{1}\wedge \cdot \cdot
\cdot \wedge dt_{k}),  \tag{2.5}
\end{equation}%
where, up to a constant factor $c\in K$, the Jacobian $J$ \ is given by $%
J=c\Omega $ with 
\begin{equation}
\Omega =\prod\limits_{i=1}^{\nu }L_{i}^{c_{i}-1}.  \tag{2.6}
\end{equation}%
In this equation $L_{i}$ is the $linear$ form defining $i-th$ reflecting
hyperplane $H_{i}$ (it is assumed that the set of $H_{1},...,H_{\nu }$ of
reflecting hyperplanes is associated with $G$) , i.e. $H_{i}=\{\alpha \in
V\mid L_{i}(\alpha )=0\}$ as defined in the Appendix. In the same Appendix,
part c), one \ finds out that the set of all elements of $G$ fixing $H_{i}$
pointwise forms a cyclic subgroup of order $c_{i}$ generated by
pseudoreflections.

\ 

\ The result, Eq.(2.6), as well as the proportionality, $J=c\Omega ,$ can be
found in the paper by Stanley [20\textbf{]}. It can be also found in much
earlier paper by Solomon [18] where it is attributed to Steinberg and
Shephard and Todd. Stanley's paper contains some details missing in earlier
papers however.

\textbf{Remark 2.4.} The results given by Eq.s(2.5),(2.6) play a key role in
the theory of hyperplane arrangements to be briefly discussed in Section 9.

Using Theorem 2.2. by Solomon, Ginzburg proved the following

\ 

\textbf{Theorem 2.5.}(Ginzburg, Ref.[21, page 358]) \ \textit{Let }$\omega
_{x}(\xi _{1},\xi _{2})$\textit{\ be a symplectic (Kirillov-Kostant) two-
form \ (to be defined in Part III), let }$\Omega ^{N}=\omega _{x}^{N}$%
\textit{\ be its }$N$\textit{-th exterior power -the volume form, with }$N$%
\textit{\ being the number of positive roots of the associated Weyl-Coxeter
reflection group, then} 
\begin{equation*}
\ast \left( \Omega ^{N}\right) =const\cdot dP_{1}\wedge \cdot \cdot \cdot
\wedge dP_{k},
\end{equation*}%
\textit{where the star }$\ast $\textit{\ denotes the standard }$Hodge-type$ 
\textit{star operator.}

\ 

\textbf{Corollary 2.6. }As it is argued in Part III, every nonsingular
algebraic variety in projective space is symplectic. The symplectic
structure gives raise to the complex K\"{a}hler structure which, in turn, is
of the Hodge-type for the Kirillov -type symplectic manifolds. Alternative
arguments leading to the same conclusion are presented in Section 9.2.2.

\ 

In the famous paper, Ref.[22, page 10, Eq.(4.1)] Atiyah and Bott \ argued
that $\omega ^{(p)}$ can be used for construction of the basis of the
equivariant cohomology ring. Their results will be discussed in some detail
in Part III. We refer our readers to the monograph [23] by Guillemin and
Sternberg where the concepts of equivariant cohomology are pedagogically
explained along with many other helpful mathematical facts of immediate
relevance to our work.

The results just presented are essential for reconstruction of both the
multiparticle Veneziano and Veneziano-like amplitudes. They also provide the
needed mathematical background for adequate physical interpretation of these
amplitudes. The next section illustrates these claims. \ \ \ \ \ \ \ 

\section{Veneziano amplitudes and Solomon's theorem}

Let $V$ be the complex affine space of dimension $\mathit{l}$\emph{\ }and
let $L_{i}(v),v\in V$ be the linear form defining the $i-$th hyperplane H$%
_{i},$ i.e. 
\begin{equation}
\text{H}_{i}=\{v\in V\mid L_{i}(v)=0\}\text{ , }i=1,\text{..., }\mathit{l}%
\emph{.}  \tag{3.1}
\end{equation}%
Results of Ref.s[24,25] and those in Appendix allow us to connect the set of
hyperplanes, Eq.(3.1), with the complete fan (see also Section 9) and, using
this fan, to associate it with it the polyhedron $\mathcal{P}$ . Taking
these facts into account, let us consider now an integral $I$ of the type 
\begin{equation}
I=\int\limits_{\mathcal{P}}c_{i_{1}...i_{p}}dP_{i_{1}}\wedge \cdot \cdot
\cdot \wedge dP_{i_{p}}\text{; }1\leq p\leq l.  \tag{3.2}
\end{equation}%
Such an integral is connected with the period integrals of the type 
\begin{equation}
\Pi (\lambda )=\oint\limits_{\Gamma }\frac{P\left( z_{1},...,z_{n}\right) }{%
Q\left( z_{1},...,z_{n}\right) }dz_{1}\wedge dz_{2}\cdot \cdot \cdot \wedge
dz_{n}  \tag{3.3}
\end{equation}%
discussed in Part I. To avoid duplications, we are only presenting results
of immediate relevance. In particular, in Part I we demonstrated that for
the Fermat variety whose affine form is written as 
\begin{equation}
\text{ }\mathcal{F}_{aff}(N):\text{\ \ }Y_{1}^{N}+\cdot \cdot \cdot
+Y_{n+1}^{N}=1,\text{ }Y_{i}=x_{i}/x_{0}\equiv z_{i},  \tag{3.4}
\end{equation}%
the period integral $\Pi (\lambda )$ is reduced (after calculation of the
Leray residue) to the Veneziano(or Dirichlet)- like integral \ $I$ \footnote{%
E.g. see Eq.(3.24) of Part I.} given explicitly by 
\begin{equation}
I\dot{=}\int\limits_{\Delta }t_{1}^{\frac{<c_{1}>}{N}-1}\cdot \cdot \cdot
t_{n+1}^{\frac{<c_{n+1}>}{N}-1}dt_{1}\wedge \cdot \cdot \cdot \wedge dt_{n}.
\tag{3.5}
\end{equation}%
In view of Eq.s(2.5),(2.6), it is of the type given by Eq.(3.2). In the
present case the polyhedron $\mathcal{P}$ is the $n+1$ simplex $\Delta .$
Not surprisingly, it is the deformation retract for the Fermat variety (as
it is for \textbf{CP}$^{n})$ since the Fermat variety is embedable into 
\textbf{CP}$^{n}$ [24]\footnote{%
We shall discuss this issue also in Section 9.}. In accord with Part I, the
symbol \.{=} denotes the statement: \textquotedblright with accuracy up to
some constant (a phase factor)\textquotedblright .The phase factors are
important. We have discussed them at length in Part I without much theory
behind them. Such a theory is well described, for example, in the monograph
by Fulton, Ref.[24], and will be used and further discussed in Part III. In
Section 9 we shall use some facts from this theory in order to prove that
developments in this work do require use of the complex pseudo-reflection
groups.

In connection with Part III (see also Ref.[6]) \ and in view of the Theorems
2.2 and 2.5 we would like now to re derive result, Eq.(3.5), making emphasis
on symplectic aspects of the Veneziano amplitudes. To this purpose, we would
like to consider\ an auxiliary problem of calculation of the volume of $k$%
-dimensional simplex $\Delta _{k}$. It is given by the integral of the type 
\begin{equation}
\text{vol}(\Delta _{k})=\int_{x_{i}\geq 0}dx_{1}\cdot \cdot \cdot
dx_{k+1}\delta (1-x_{1}-\cdot \cdot \cdot -x_{k+1}).  \tag{3.6}
\end{equation}%
Using results from symplectic geometry [26], it is straightforward to show
that the above integral (up to unimportant constant) is just the
microcanonical partition function for the system of $k+1$ harmonic
oscillators whose total energy is equal to $1$. To calculate such a
partition function it is sufficient to take into account the integral
representation of the delta function. Then, the standard manipulations with
integrals produce the following anticipated result: 
\begin{equation}
\text{vol}(\Delta _{k})=\frac{1}{2\pi }\oint \frac{dy\exp (iy)}{(iy)^{k+1}}=%
\frac{1}{k!}.  \tag{3.7a}
\end{equation}%
Clearly, for the dilated volume we would obtain instead: vol$(n\Delta _{k})=%
\dfrac{n^{k}}{k!}$, where $n$ is the dilatation coefficient. This result was
discussed already in Part I, e.g. see Eq.(1.24). This calculation allows us
to obtain as well the volume of $k$-dimensional hypercube (or, perhaps more
generally, the convex polytope $\mathcal{P}$) as 
\begin{equation}
n^{k}=k!vol(n\Delta _{k}).  \tag{3.7b}
\end{equation}%
This result was obtained in famous paper by Atiyah [27] inspired by earlier
result by Koushnirenko [28]. It is discussed at length both in Ref.[6] and
Part III, in connection with alternative symplectic formulation of the
partition function reproducing the Veneziano (and Veneziano -like)
amplitudes. In the meantime, the observations just made allow us to re
derive the Veneziano amplitude in a much simpler way. To do so, we extend
our analysis of Eq.(3.6) having in mind both Theorems 2.2. and 2.5. This
leads us to consideration of the integral of the type 
\begin{equation}
I=\int_{x_{i}\geq 0}dx_{1}^{<c_{1}>}\wedge \cdot \cdot \cdot \wedge
dx_{k+1}^{<c_{k+1>}}\delta (1-x_{1}^{N}-\cdot \cdot \cdot -x_{k+1}^{N}) 
\tag{3.8}
\end{equation}%
written in accord with notations of Part I. The presence of $\delta $
function reminds us about the procedure (discussed in Part I) of taking the
Leray-type residue in the period integral, Eq.(3.3). We would like to
demonstrate now that the integral, Eq.(3.8) can be calculated much easier as
compared to calculations described in detail in Part I. \ To simplify
notations, let $<c_{i}>=p_{i}$ and, furthermore, let $n_{i}=\frac{N}{p_{i}}. 
$\ \ By analogy with (3.7a) we obtain (up to a constant as before), 
\begin{equation}
I\dot{=}\frac{1}{2\pi }\oint dy[\prod\limits_{i=1}^{k+1}\left( \frac{1}{%
iyn_{i}}\right) ^{\dfrac{1}{n_{i}}}\Gamma (\frac{1}{n_{i}})]\exp (iy)\dot{=}%
\frac{\Gamma (\frac{p_{1}}{N})\cdot \cdot \cdot \Gamma (\frac{p_{k+1}}{N})}{%
\Gamma (\sum\limits_{i}\frac{p_{i}}{N})},  \tag{3.9}
\end{equation}%
in agreement with Eq.(3.27) of Part I as required.

\section{\protect\bigskip Selected excersises from Bourbaki (continuation)}

The results of previous section indicate that the Veneziano and
Veneziano-like amplitudes can be reconstructed from the algebra of
invariants $(S(V)\otimes E(V))^{G}$ of the group $G$ not yet specified. The
question naturally arises: Can we use the same invariance principle in order
to reconstruct the meaningful physical model reproducing the Veneziano and
Veneziano-like amplitudes? We provide positive answer to the above posed
question in this section and in Sections 6-9.

To begin, we need to discus properties of the ring $S(V)^{G}$ of symmetric
invariants composed of algebraically independent polynomial forms $%
P_{1},...,P_{l}$ made of symmetric combinations of \ $t_{1},...t_{l}$ raised
to some powers $d_{i}$ ( $i=1,...,l$ ) different for different reflection
groups [29]. The ring of invariants is graded and it admits a decomposition
(which actually is always finite) : $S(V)^{G}$ =$\bigoplus\nolimits_{j=0}^{%
\infty }S_{j}(V)^{G}$. Provided that $\dim _{K}V_{j}^{G}$ is the dimension
of the graded invariant subspace $S_{j}(V)^{G}$ defined over the field $K,$
the following definition can be given.

\textbf{Definition 4.1}. The Poincare$^{\prime }$ polynomial $P(S(V)^{G},t)$
is defined by 
\begin{equation}
P(S(V)^{G},t)=\sum\limits_{i=0}^{\infty }(\dim _{K}V_{j}^{G})t^{i}. 
\tag{4.1}
\end{equation}%
The Poincare$^{\prime }$ polynomial possesses the splitting property \ [29]
(the most useful for applications in K-theory [12]). This can be described
as follows. If the total vector space $M$ is made as a product $V\otimes
_{K}V^{\prime }$ of vector spaces $V$ and $V^{\prime }$, then the Poincare$%
^{\prime }$ polynomial of such a product is given by 
\begin{equation}
P(V\otimes _{K}V^{\prime },t)=P(V,t)P(V^{\prime },t).  \tag{4.2}
\end{equation}%
This splitting property is of topological nature [9\textbf{]} since it
reflects the decomposition property of the total topological space into
pieces and for this reason is extremely useful in actual calculations. In
particular, let us consider the polynomial ring F$[x]$ made of monomials of
degree $d$ which \ are the building blocks of the graded vector space $V$ as
discussed in Section 2. The Poincare$^{\prime }$ polynomial for such a space
is given by 
\begin{equation}
P(V,t)=1+t^{d}+t^{2d}+....=\frac{1}{1-t^{d}}.  \tag{4.3}
\end{equation}%
Consider now the multivariable polynomial ring F[$x_{1},...,x_{n}]$ made of
monomials of respective degrees $d_{i}$ . Then, using the splitting
property, we obtain at once 
\begin{equation*}
V^{T}=\text{F[}x_{1}]\otimes _{K}\text{F[}x_{2}]\otimes _{K}\text{F[}%
x_{3}]\otimes _{K}\cdot \cdot \cdot \otimes _{K}\text{F[}x_{n}]
\end{equation*}%
and, of course, 
\begin{equation}
P(V^{T},t)=\frac{1}{1-t^{d_{1}}}\cdot \cdot \cdot \frac{1}{1-t^{d_{n}}}. 
\tag{4.4}
\end{equation}%
In particular, if all $d_{i}$ in Eq.(4.4) are equal to one, which is
characteristic for $S(V)$ defined in Section 2, e.g. read Ref.[29, page
171], then we reobtain back Eq.(1.22) of Part I.

\textbf{Remark 4.2}. Since the Laplace transform of Eq.(1.22) of Part I
produces the (nonsymmerized) Veneziano amplitude, the connection between the
theory of invariants of finite (pseudo) reflection groups and the physical
model reproducing such an amplitude becomes apparent already at this point.

This remark allows us to make several additional steps to make our
presentation mathematically self contained and focused on \ physics. To this
purpose, let $G\subset GL(V)$ be one of such reflection groups. Suppose that
its cardinality $\left\vert G\right\vert =\prod\limits_{i}d_{i}$. Next, we
introduce the averaging operator $Av:V\rightarrow V$ via 
\begin{equation}
Av(x)=\frac{1}{\left\vert G\right\vert }\sum\limits_{\varphi \in G}\varphi
\circ x.  \tag{4.5}
\end{equation}%
By definition, $x$ is the group invariant, $x$ $\in V^{G}$ , if $Av(x)=x.$
In particular, 
\begin{equation}
\dim _{K}V_{j}^{G}=\frac{1}{\left\vert G\right\vert }\sum\limits_{\varphi
\in G}tr(\varphi _{j}).  \tag{4.6}
\end{equation}%
This result \ can be explained as follows. Suppose $x$ $\in V^{G}$, then $Av$%
($Av(x)=x)\rightarrow Av^{2}(x)=Av(x)=x.$Thus, the $Av$ operator is
indepotent. Such indepotent operator has evidently only 2 eigenvalues :1 and
0. Using this fact in Eq.(4.5) produces Eq.(4.6).

By combining Eq.(4.6) result with Eq.(4.1) we obtain, 
\begin{eqnarray}
P(S(V)^{G},t) &=&\sum\limits_{i=0}^{\infty }(\dim
_{K}V_{j}^{G})t^{i}=\sum\limits_{i=0}^{\infty }\frac{1}{\left\vert
G\right\vert }\sum\limits_{\varphi \in G}tr(\varphi _{i})t^{i}  \notag \\
&=&\frac{1}{\left\vert G\right\vert }\sum\limits_{\varphi \in
G}[\sum\limits_{i=0}^{\infty }tr(\varphi _{i})t^{i}]=\frac{1}{\left\vert
G\right\vert }\sum\limits_{\varphi \in G}\frac{1}{\det (1-\varphi t)}. 
\TCItag{4.7}
\end{eqnarray}%
The obtained result is known as the Molien theorem [29]. It is based on the
following nontrivial identity 
\begin{equation}
\sum\limits_{i=0}^{\infty }tr(\varphi _{i})t^{i}=\frac{1}{\det (1-\varphi t)}
\tag{4.8}
\end{equation}%
valid for the upper triangular matrices, i.e. for matrices which belong to
the Borel subgroup $B$ of $G\footnote{%
In Part III we shall provide important details on the subgroup $B$.}$. \ For
such matrices 
\begin{equation}
tr(\varphi _{i})=\sum\limits_{j_{1}+j_{2}+...+j_{n}=i}\lambda
_{1}^{j_{1}}\cdot \cdot \cdot \lambda _{n}^{j_{n}}  \tag{4.9}
\end{equation}%
where the Borel-type matrix $\varphi $ of dimension $n$ has $\lambda
_{1},...,\lambda _{n}$ on its diagonal. Substitution of Eq.(4.9) into
Eq.(4.7) produces 
\begin{eqnarray}
\sum\limits_{i=0}^{\infty }tr(\varphi _{i})t^{i}
&=&\sum\limits_{i=0}^{\infty }\left[ \sum\limits_{j_{1}+j_{2}+...+j_{n}=i}%
\lambda _{1}^{j_{10}}\cdot \cdot \cdot \lambda _{n}^{j_{n}}\right] t^{i} 
\notag \\
&=&\left[ \sum\limits_{j_{1}=0}^{\infty }\lambda _{1}^{j_{1}}t^{j_{1}}\right]
\cdot \cdot \cdot \left[ \sum\limits_{j_{n}=0}^{\infty }\lambda
_{n}^{j_{n}}t^{j_{n}}\right]  \notag \\
&=&\frac{1}{1-\lambda _{1}t}\cdot \cdot \cdot \frac{1}{1-\lambda _{n}t}=%
\frac{1}{\det (1-\varphi t)}.  \TCItag{4.10}
\end{eqnarray}%
We have gone through all details in order to demonstrate the \textit{bosonic
nature} of the obtained result: by replacing $t$ with exp($-\varepsilon )$
with $0\leq \varepsilon \leq \infty $ and associating numbers $j_{i}$ with
the \ Bose statistic occupation numbers we have obtained the partition
function for the set of $n$ independent harmonic oscillators (up to zero
point energy). This result will be re obtained in Part III using different
arguments.

In view of Eq.(4.7), thus obtained result should be additionally group
averaged. In particular, by combining \ Eq.s.(4.4) and (4.7) we obtain, 
\begin{equation}
P(S(V)^{G},t)=\frac{1}{\left\vert G\right\vert }\sum\limits_{\varphi \in G}%
\frac{1}{\det (1-\varphi t)}=\prod\limits_{i=1}^{n}\frac{1}{1-t^{d_{i}}}. 
\tag{4.11}
\end{equation}%
This result is valid for any (pseudo) reflection group $G\subset GL(V)$
whose cardinality $\left\vert G\right\vert =\prod\limits_{i}d_{i}$ in accord
with the conditions of the Theorem 2.2. by Solomon. Following Humphreys
[30], it is useful to re interpret Eq.(4.11) using the following physically
motivated arguments. We consider an action of the averaging operator,
Eq.(4.5), on the monomials 
\begin{equation*}
x=z_{1}^{j_{1}}\cdot \cdot \cdot z_{n}^{j_{n}},\text{ where \ }%
j_{1}+j_{2}+...+j_{n}=i.
\end{equation*}%
These are the eigenvectors for $\varphi _{i}$ with the corresponding
eigenvalues $\lambda _{1}^{j_{1}}\cdot \cdot \cdot \lambda _{n}^{j_{n}}.$
The weighted sum of these eigenvalues is the trace of the linear operator $%
Av(x)$ acting on monomials from $S_{i}(V)^{G}$. But, according to Eq.(4.6),
this is just the dimension of space $S_{i}(V)^{G}.$ This dimension has the
following physical meaning. If for the moment we assume (and later, in
Section 9, we prove) that all eigenvalues in Eq.(4.9) are $i-th$ roots of
unity, then, by combining Eq.s (9.15)-(9.18) of Section\textbf{\ }9 and the
results of Appendix, part c), we arrive at the Veneziano condition 
\begin{equation}
\sum\limits_{k}m_{k}j_{k}=i  \tag{4.12a}
\end{equation}%
again. Since in this equation $m_{i}=d_{i}-1\func{mod}i,$ it is equivalent
to 
\begin{equation}
\sum\limits_{k}d_{k}j_{k}=0\func{mod}i.  \tag{4.12b}
\end{equation}%
In particular, if $d_{k}j_{k}=\omega i$ \ (for $k=1,...,n$ and $\omega $ $%
\in \mathbf{Z})$ then, using this equation along with Eq.s(4.8),(4.9), we
arrive at the following result: 
\begin{eqnarray}
\sum\limits_{i=0}^{\infty }tr(\varphi _{i})t^{i}
&=&\sum\limits_{i=0}^{\infty
}[\sum\limits_{j_{1}d_{1}+j_{2}d_{2}+...+j_{n}d_{n}=i}\lambda
_{1}^{j_{1}d_{1}}\cdot \cdot \cdot \lambda _{n}^{j_{n}d_{n}}]t^{i}  \notag \\
&=&\sum\limits_{j_{1}=0}^{\infty
}t^{j_{1}d_{1}}\sum\limits_{j_{2}=0}^{\infty }t^{j_{2}d_{2}}\cdot \cdot
\cdot \sum\limits_{j_{n}=0}^{\infty }t^{j_{n}d_{n}}  \notag \\
&=&\prod\limits_{i=1}^{n}\frac{1}{1-t^{d_{i}}},  \TCItag{4.13}
\end{eqnarray}%
to be compared with earlier obtained Eq.(4.11). Again, we have gone through
all these details in order to demonstrate \textit{the bosonic nature} of the
obtained result, Eq.(4.11). Clearly, the result, Eq.(4.11), can be
reproduced using the path integrals for $n$ independent bose-like particles,
e.g harmonic oscillators. Such a conclusion is going to be strengthened in
Part III devoted to the symplectic interpretation of the obtained results.
However, the obtained results are incomplete since thus far we were dealing
only with $S(V)^{G}$ -type of invariants made of monomials raised to $d_{i}$
-th powers. Theorem 2.2. requires us to construct the Poincare$^{\prime }$
polynomials for invariants of $\left( S(V)\otimes E(V)\right) ^{G}$type. To
design such polynomials we need to discuss several additional topics. These
are presented in the next two sections.

\section{Additional facts from \ the theory of pseudo-reflection groups}

In Appendix, part c), we have listed some basic facts about
pseudo-reflection groups. At this point we would like to extend this
information. To this purpose we would like to use some results from the
classical paper by Shepard and Todd [31] (S-T).We shall use these results
along with those from the monograph by McMullen [32] containing the up to
date developments related to the S-T work.

Adopting S-T notations, let $N\geq 1,n\geq 2$, and let $p$ be a divisor of $%
N $, i.e. $N=pq$. In addition, let $\xi $ be a primitive $N$-th root of
unity. Then, $\ $the unitary group$\ \ G(N,p,n)$ is defined as the group of
all monomial transformations in \textbf{C}$^{n}$ of the form (e.g. see
Section 9\textbf{\ }below) 
\begin{equation}
x_{i}^{\prime }=\xi ^{\nu _{i}}x_{\sigma (i)},\text{ }i=1,...,n,  \tag{5.1}
\end{equation}%
where $\sigma (1),...,\sigma (n)$ is permutation $\sigma $ of $(1,,...,n)$,
i.e. $\sigma \in S_{n},$ and 
\begin{equation}
\sum\nolimits_{i}\nu _{i}=0(\func{mod}N).  \tag{5.2a}
\end{equation}%
In the case if $N=pq$ the above condition should be changed to 
\begin{equation}
\sum\nolimits_{i}\nu _{i}=0(\func{mod}p).  \tag{5.2b}
\end{equation}%
The group $G(N,p,n)$ has order (or cardinality) $\left\vert G\right\vert =$ $%
qN^{n}n!$. The order of the group $G(N,p,n)$ can be determined by
considering the set of 2-fold reflections given by 
\begin{equation}
x_{i}^{\prime }=\xi ^{\nu }x_{j},\text{ \ }x_{j}^{\prime }=\xi ^{-\nu }x_{i},%
\text{ \ }x_{k}^{\prime }=x_{k}\text{ },k\neq i,j.  \tag{5.3}
\end{equation}%
Such set generates the normal subgroup of order $N^{n}n!$. The other
reflections, if any, are of the form 
\begin{equation}
x_{j}^{\prime }=\xi ^{\tfrac{\nu N}{r}}x_{j},\text{ }x_{i}^{\prime }=x_{i}%
\text{ for }i\neq j,  \tag{5.4}
\end{equation}%
where $j=1,...,n$, $(\nu ,N/r)=1$ and $r\mid q$ if $q>1$. The following
theorem can be found in McMullen's book [32, page 292].

\ 

\textbf{Theorem} \textbf{5}.\textbf{1}. \textit{If }$n\geq 2$\textit{\ ,
then up to conjugacy within the group of all unitary transformations, the
only \textbf{finite} irreducible unitary reflection groups in }\textbf{C}$%
^{n}$\textit{\ which are imprimitive are the groups }$G(m,p,n)$\textit{\
with }$\mathit{m}\geq 2,$ $p\mid m$\textit{\ and }$(m,p,n)\neq (2,2,2).$

\ 

\textbf{Definition 5.2. } A group $G$ of unitary transformations of \textbf{C%
}$^{n}$ is called \textit{imprimitive} if \textbf{C}$^{n}$ is the direct sum 
\textbf{C}$^{n}=E_{1}\oplus \cdot \cdot \cdot \oplus E_{k}$ of non-trivial
proper linear subspaces $E_{1},...,E_{k}$ such that the family$%
\{E_{1},...,E_{k}\}$ is invariant under $G$.

For the purposes of this work, it is sufficient to consider only the case $%
p=1$. The group $G(N,1,n)$, traditionally denoted as $\gamma _{n}^{N}$ , is
the group of symmetries of the complex $n-$ cube. Actually, it is the same
as that for the real $n$-cube [29,31] which is inflated by the factor of $N.$
In the standard notations [30] the cubic symmetry is denoted as $B_{n+1}$
while in the S-T notations it corresponds to just mentioned group $G(N,1,n)$%
\ whose exponents $d_{i}=Ni$ with $i=1,...n$ [29,31]. In the typical case of 
\textit{real} space with cubic symmetry we have $N=2$ (e.g. see Eq.(A.6) of
Appendix ) so that these exponents $d_{i}$ coincide with those for $B_{n+1}$
in accord with the exponents for this group listed in the book by Humphreys
[30, page 59], as required.\ 

In order to utilize all these observations efficiently, we need to re obtain
the same results from another perspective. To this purpose, following S-T,
we introduce an auxiliary function $g_{r}(m,p,n)$ describing the number of \
reflexion operations made with help of $G(m,p,n)$ which leave fixed every
point of the subspace of dimensionality $n-r,r=0,1,...,n\footnote{%
This construction reminds us about the Grassmannians considered in the
Introduction. We shall take the full advantage of this observation
momentarily.}.$ By definition, $g_{0}(m,p,n)=1.$ Moreover, let 
\begin{equation}
G(m,p,n;t)\equiv \sum\limits_{r=0}^{n}g_{r}(m,p,n)t^{r}.  \tag{5.5}
\end{equation}%
On one hand, the above equation serves as the definition of the generating
function $G(m,p,n;t)$, on another, in view of Eq.(1.8) of the Introduction,
we can reinterpret the r.h.s. as the Weyl character formula. The full proof
of this fact is given in Part III. Shepard and Todd calculate $G(m,p,n;t)$
explicitly. Their derivation is less physically adaptable however than that
obtained by Solomon, Ref. [18]. Hence, we would like to discuss Solomon's
results now.

\section{Selected exercises from Burbaki (end)}

Our main objective at \ this point is to obtain the explicit form of $%
G(m,p,n;t)$ defined in Eq.(5.5) and to explain its physical meaning. To this
purpose, let us recall that according to the Theorem 2.2. the differential
form $\omega ^{(p)}$, Eq.(2.4), \ belongs to the set of $G$-invariants of
the product $S(V)\otimes E(V)$. The splitting property, Eq.(4.2), of the
Poincare$^{\prime }$ polynomials requires some minor changes for the present
case. In particular, if by analogy with $S(V)^{G\text{ }}$decomposition we
would write ($S(V)\otimes
E(V))^{G}=\bigoplus\limits_{i,j}S_{i}(V)^{G}\otimes E_{j}(V)^{G}$, then the
associated Poincare$^{\prime }$ polynomial can be defined by 
\begin{equation}
P((S(V)\otimes E(V))^{G};\text{ }x,\text{ }y)=\sum\limits_{i,j\geq 0}(\dim
_{K}S_{i}^{G}\otimes E_{j}^{G})x^{i}y^{j}.  \tag{6.1}
\end{equation}%
Following Solomon [18], by analogy with Eq.(4.6) we introduce 
\begin{equation}
\dim _{K}(S_{i}^{G}\otimes E_{j}^{G})=\frac{1}{\left\vert G\right\vert }%
\sum\limits_{\varphi \in G}tr(\varphi _{i})tr(\varphi _{j}).  \tag{6.2}
\end{equation}%
In order to use this result we need to take into account that 
\begin{equation}
\sum\limits_{j=0}^{n}tr(\varphi _{j})y^{j}=\det (1+\varphi y)  \tag{6.3}
\end{equation}%
to be contrasted with Eq.(4.8). To prove that this is indeed the case it is
sufficient to recall that for fermions the occupation numbers $j_{i}$ are
just $0$ and $1.$ Hence, in view of Eq.(4.13), but accounting for the
fermionic nature of the occupation numbers in the present case, we obtain, 
\begin{eqnarray}
\sum\limits_{j=0}^{n}tr(\varphi _{j})y^{j}
&=&\sum\limits_{j=0}^{n}[\sum\limits_{j_{1}+j_{2}+...+j_{n}=j}\lambda
_{1}^{j_{1}}\cdot \cdot \cdot \lambda _{n}^{j_{n}}]y^{j}  \notag \\
&=&\prod\limits_{i=1}^{n}\sum\limits_{j_{i}=0}^{1}\lambda
_{i}^{j_{i}}y^{j_{i}}=\prod\limits_{i=1}^{n}(1+\lambda _{i}y).  \TCItag{6.4}
\end{eqnarray}%
As in the bosonic case, the result, Eq.(6.4), can be obtained using the
fermionic path integrals for $n$ independent particles (say, fermionic
oscillators) obeying the Fermi-type statistics. Using Eq.(6.2) in (6.1) and
taking into account the rest of the results obtained in Section 4, the
following expression for the Poincare$^{\prime }$ polynomial \ is obtained 
\begin{equation}
P((S(V)\otimes E(V))^{G};x,y)=\frac{1}{\left\vert G\right\vert }%
\sum\limits_{\varphi \in G}\frac{\det (1+\varphi y)}{\det (1-\varphi x)}%
=\prod\limits_{i=1}^{n}\frac{(1+yx^{d_{i}-1})}{(1-x^{d_{i}})}  \tag{6.5}
\end{equation}%
in accord with Bourbaki [17]. To check its correctness we can: a) put $y=0$
thus obtaining back Eq.s(4.4)and (4.11) or, b) put $y=-x$ thus obtaining
identity $1=1$ between the second and the third terms above.

\textbf{Remark 6.1}. Since the result, Eq.(6.5), is just the ratio of
determinants, its supersymmetric nature should be clear to everybody
familiar with the path integrals.

Eq.(6.5), can now be used for several tasks. First, for completeness of
presentation, we would like to recover the major S-T result: 
\begin{equation}
G(m,p,n;t)=\prod\limits_{i=1}^{n}(m_{i}t+1)  \tag{6.6}
\end{equation}%
extensively used in theory of hyperplane arrangements [33.34] to be
discussed further in Section\textbf{\ }9. Taking into account notations
introduced in Eqs.(4.12), the above equation produces (for $t=1)$ the
following result: $G(m,p,n;t=1)=\left\vert G\right\vert
=\prod\limits_{i=1}^{n}d_{i}$ ,\ which is in accord with Section 4, as
required. Moreover, in view of \ Eq.(5.5), it allows us to recover $%
g_{r}(m,p,n)$. After this is done, we need to discuss its physical meaning.

To recover the S-T results let us rewrite Eq.(6.5) as follows 
\begin{equation}
\sum\limits_{\varphi \in G}\frac{\det (1+\varphi y)}{\det (1-\varphi x)}%
=\left\vert G\right\vert \prod\limits_{i=1}^{n}\frac{(1+yx^{d_{i}-1})}{%
(1-x^{d_{i}})}  \tag{6.7}
\end{equation}%
and let us treat the right (R) and the left (L) hand sides separately.
Following Bourbaki\textbf{\ [}17\textbf{],} we put $y=-1+t(1-x)$.
Substitution of this result to R produces at once 
\begin{equation}
R\mid
_{x=1}=\prod\limits_{i=1}^{n}(d_{i}-1+t)=\prod\limits_{i=1}^{n}(m_{i}+t) 
\tag{6.8}
\end{equation}%
To do the same for L requires us to keep in mind that $\det AB=\det A\det B$
and, hence, $\det AA^{-1}=1$ leads to $\det A^{-1}=1/\det A$. Therefore,
after few steps we arrive at 
\begin{equation}
L\mid _{x=1}=\sum\limits_{r=1}^{n}h_{r}t^{r}.  \tag{6.9}
\end{equation}%
Equating L with R, replacing $t$ by $1/T$ and relabeling $1/T$ again by $t$
\ and $h_{l}$ by $\tilde{h}_{l}$ = $g_{r}(m,p,n)$ we obtain the S-T result,
Eq.(5.5). To obtain physically useful result we have to take into account
that for the cubic symmetry we had quoted already in the previous section
the result : $d_{i}=iN.$ Therefore, let $y=-x^{Nq+1}$ in Eq.(6.5), then we
obtain, 
\begin{equation}
P((S(V)\otimes E(V))^{G};z)=\prod\limits_{i=1}^{n}\frac{1-z^{q+i}}{1-z^{i}}.
\tag{6.10}
\end{equation}%
\textbf{Remark 5.2}. The result almost identical to our Eq.(6.10) was
obtained some time ago in the paper by Lerche et al [35, page 444,
Eq.(4.4)]. To obtain their result, it is sufficient to replace $z^{q+i}$ by
\ $z^{q-i}.$ Clearly, such a substitution is \textit{not permissible} in our
case. Nevertheless, some ideas discussed in the paper by Lerche et all
happen to be helpful in obtaining the correct partition function for the
Veneziano (and Veneziano-like) amplitudes. This will be discussed in detail
in the following two sections.

Taking into account the cubic symmetry in Eq.(6.10) in the limit $z=x^{N}=1$
we obtain, 
\begin{equation}
P((S(V)\otimes E(V))^{G};\text{ }z=1)=\frac{(q+1)(q+2)\cdot \cdot \cdot (q+n)%
}{n!},  \tag{6.11}
\end{equation}%
Both, Eq.s(6.10) and (6.11) have been obtained in the Introduction in a much
simpler way so that there is no need to repeat the arguments presented there.

From the Introduction we know already that $P((S(V)\otimes E(V))^{G};z)$
given by Eq.(6.10) is the Poincare$^{\prime }$ polynomial for the complex
Grassmann manifold. The question arises: is the method of obtaining such a
polynomial specific only for Grassmannians? The answer is clearly "No"! This
had been demonstrated mathematically rigorously by Hiller, Ref.[36], based
on earlier fundamental results by Bernstein, Gelfand and Gelfand, Ref.[37]
(BGG). Incidentally, Hiller \textit{does} obtain our main result, Eq.(6.10),
using BGG formalism, e.g. see Ref.[36, page 155 (top)]. His derivation is
entirely different and is considerably more complicated than ours\footnote{%
We encorage our readers to make such a comparison.}. The invariant algebra $%
(S(V)\otimes E(V))^{G}$ considered by Solomon, Ref.[18], is called "the
topological algebra" in Ref.[38]. This reference explains in detail the
universal nature of such an algebra which makes it absolutely indispensable
in the theory of fiber bundles, K-theory, theory of characteristic classes
and equivariant cohomology. \ We discuss this topic further in Section 9.2.2.

\section{Designing Veneziano partition function using algebraic geometry}

\subsection{General considerations}

The paper by Lerche et al, Ref.[35], provides plausible arguments implying
that (up to some unimportant constant) \textit{any} one -variable Poincare$%
^{\prime }$polynomial can be actually interpreted as some kind of the Weil
character formula. In Part III we shall reach the same conclusions using
entirely different arguments. These are presented in conjunction with the
symplectic development of our formalism. To actually use our result,
Eq.(6.10) (or, which is the same, Eq.(1.11)) we do need to take into account
the connection with the Weyl character formula just mentioned. The recursion
relation, Eq.(1.8), provides already an indication that, indeed, the product
given in the r.h.s of Eq.(1.11) is in fact a polynomial in $q$ whose highest
degree is $km.$ Such a polynomial can be already interpreted as the Weyl
character formula. In principle, because of the noticed connections with the
Weyl character formula, one can use the supersymmeric formalism for
reproduction of this formula. Such formalism, developed for any homogenous
space (including that for the Grassmanniann) in Refs[13,14], can be used for
reproduction of the result Eq.(6.10). In view of the results to be presented
in the next section, this is not the most illuminating way however to arrive
at our final destination. Such a derivation will not take into account the
"gauge" freedom and the "gauge fixing" discussed in the Introduction. In
view of this discussion, it is more advantageous to take advantage of the
fact that the Grassmannian can be mapped into the product of complex
projective spaces of prescribed dimensionalities. At the level of
classifying spaces such a possibility is indicated on page 303 in the book
by Bott and Tu, Ref.\textbf{[}9\textbf{],} and proven in the book by
Husemoller, Ref.[39, page 297, Proposition 3.1]. There is however another,
more direct, way to obtain the desired result without recourse to the
classifying space. In the Introduction we indicated that such a possibility
does indeed exist: to this purpose it is sufficient to "fix the gauge" by
choosing $m=1\cdot 2\cdot \cdot \cdot k$ in Eq.(1.11). Then, the Poincare$%
^{\prime }$ polynomial for the complex Grassmannian becomes manifestly
decomposable into the product of Poincare$^{\prime }$ polynomials for the
complex projective spaces of prescribed dimensionalities. Although such a
decomposition is plausible it is a bit restricted. Below we would like to
discuss another less restricted way to arrive at the desired result. By
doing so we shall accomplish several tasks. First, we can then formally use
the results by Stone, Ref.[13], for the partition function for a particle
with spin in the magnetic field. Second, and more important for us, by
embedding the complex Grassmannian into the complex projective space of
prescribed dimensionality we shall obtain important results to be used in
the rest of this paper.

To embed the Grassmannian into the complex projective space requires several
steps. We would like to describe them now.

\subsection{The Pl\"{u}cker embedding}

\bigskip

For reader's convenience we would like to summarize the idea behind such an
embedding. In particular, let $V$ be an $n$dimensional vector space and let $%
E(V)$ be its exterior algebra, as described in Section 2, so that $E(V)$ $%
=\oplus _{k=1}^{l}E_{k}(V)$ where $1\leq k\leq l.$ While the space $E_{l}(V)$
is one dimensional, the dimensionality of the subspace $E_{k}(V),$ is known
to be $\mathcal{N}$=$\dim E_{k}(V)=\left( 
\begin{array}{c}
\text{\textit{l}} \\ 
k%
\end{array}%
\right) .$ This should be contrasted with \ the dimensionality of the usual
vector subspace which is just $k$. The number just produced reminds us about
the number of subspaces of the \ (complex) Grassmannian $G(k,l)$ we have
discussed in the Introduction. This means that the totality of $k-$%
dimensional subspaces of $l-$dimensional vector space $V$ can be identified
with $\dim E_{k}(V).$ Let now $\hat{E}_{k}(V)$ be a particular member of the
exterior algebra so that if $\mathbf{w}_{1},...,\mathbf{w}_{k}$ represents a
set of linearly independent vectors defining the chosen $k-$dimensional
subspace, then $\hat{E}_{k}(V)=\mathbf{w}_{1}\wedge ...\wedge \mathbf{w}_{k}$%
. Clearly, if $\mathbf{w}_{i}\in \mathbf{C}_{i\text{ }}$, then the product $%
\mathbf{w}_{1}\wedge ...\wedge \mathbf{w}_{k}$ represents some point in 
\textbf{C}$^{\mathcal{N}}$ and, at the same time, it represents some
particular subspace of the vector space V and, as such, can be used to
describe the Grassmannian. Moreover, if $\mathbf{e}_{1},...,\mathbf{e}_{l}$
is an ordered basis for $V$, then there are some $k\times l$ matrices 
\textbf{M}=$\left( a_{i,j}\right) $ such that 
\begin{equation}
\mathbf{w}_{j}=\dsum\limits_{i=1}^{l}a_{i,j}\mathbf{e}_{i}  \tag{7.1}
\end{equation}%
and, accordingly, 
\begin{equation}
\mathbf{w}_{1}\wedge ...\wedge \mathbf{w}_{k}=\dsum\limits_{\left(
i_{1},...,i_{k}\right) }m_{i_{1},...,i_{k}}\mathbf{e}_{i_{1}}\wedge
...\wedge \mathbf{e}_{i_{k}}  \tag{7.2}
\end{equation}%
where $m_{i_{1},...,i_{k}}$ is made out of $k\times k$ minors of the matrix 
\textbf{M} formed by columns of \textbf{M} with indices $i_{1},...,i_{k}.$
Since both $\mathbf{w}_{1}\wedge ...\wedge \mathbf{w}_{k}$ and $\mathbf{e}%
_{i_{1}}\wedge ...\wedge \mathbf{e}_{i_{k}}$ are some points in \textbf{C}$^{%
\mathcal{N}},$ Eq.(7.2) can be interpreted as an equivalence relation and
thus provides a projective embedding of the Grassmannian into \textbf{CP}$^{%
\mathcal{N}-1}.$ The r.h.s. of Eq.(7.2) represents a kind of a basis
expansion of the tensor $\mathbf{w}_{1}\wedge ...\wedge \mathbf{w}_{k}$ so
that $m_{i_{1},...,i_{k}}$ are some (actually, Pl\"{u}cker) coordinates with
respect to the standard basis. Suppose we have another tensor $\mathbf{w}%
_{1}^{\prime }\wedge ...\wedge \mathbf{w}_{k}^{\prime }$ and interested in
relating it with $\mathbf{w}_{1}\wedge ...\wedge \mathbf{w}_{k}$. This can
be achieved if there is another $k\times k$ matrix $\mathbf{A}$ such that $%
\mathbf{M}=\mathbf{AM}^{\prime }$ and, accordingly,%
\begin{equation}
m_{i_{1},...,i_{k}}=\left[ \det \mathbf{A}\right] m_{i_{1},...,i_{k}}^{%
\prime }  \tag{7.3}
\end{equation}%
Finally, following Fulton, Ref.[40, page 108], the Sylvester theorem
(discovered in 1851) should be used to arrive at Pl\"{u}cker relations.
These relations provide guarantee that such an embedding of the Grassmannian
is permissible. To describe the Sylvester theorem we should notice that,
actually, by symmetry the matrix \textbf{A} is of the same kind as the $%
k\times k$ minor of \textbf{M}. That is it should be a $k\times k$ matrix.
Because of this, the Sylvester theorem can be stated as follows

\textbf{Theorem 7.1}.(Sylvester) \textit{Let} \textbf{M} \textit{and }%
\textbf{N}\textit{\ be any} $k\times k$ $matrices$ $and$ $let1\leq \lambda
\leq k,$ $then$ 
\begin{equation}
\left[ \det \mathbf{M}\right] \left[ \det \mathbf{N}\right] =\sum \det 
\mathbf{M}^{\prime }\det \mathbf{N}^{\prime }  \tag{7.4}
\end{equation}%
\textit{where the sum is taken over all pairs of matrices} $\mathbf{M}%
^{\prime }$ \textit{and} $\mathbf{N}^{\prime }$ \textit{obtained from 
\textbf{M} and \textbf{N} by interchanging a fixed set of }$\lambda $\textit{%
\ columns of }\textbf{N}\textit{\ with any }$\mathit{\lambda }$\textit{\
columns of} \textbf{M}, \textit{preserving the ordering of the columns.}

\textit{\bigskip }

This concludes our description of the Pl\"{u}cker embedding. For the goals
we would like to accomplish, such an embedding is not sufficient. Hence, now
we would like to discuss the Segre embedding.

\subsection{The Segre embedding}

\bigskip

The idea of this kind of embedding is rather simple and has its origins in a
simple problem which can be formulated as follows. If the complex space 
\textbf{C}$^{2}$ can be thought as a Cartesian product \textbf{C}$^{1}\times 
\mathbf{C}^{1},$ is it possible to construct, say, \textbf{CP}$^{2},$ as 
\textbf{CP}$^{1}\times \mathbf{CP}^{1}?$ Our experience with the Pl\"{u}cker
embedding suggests that this may be possible if we look at the operation "$%
\times "$ group-theoretically. Specifically, let \{$z_{0},...,z_{n}\}$
represent a point in \textbf{CP}$^{n}$ while \{$z_{0}^{\prime
},...,z_{m}^{\prime }\}$represent a point in \textbf{CP}$^{m}\footnote{%
One should keep in mind that points in the projective space are the
equivalence classes. The notations in the text refer to some set of
coverings of \textbf{CP}$^{n}$ by \textbf{C}$^{n+1\prime }s$ such that at
least one of z$_{i}^{\prime }s$ is strictly nonzero, as usual.},$then the
Segre embedding $s_{n,m}$ :\textbf{CP}$^{n}\times $\textbf{CP}$%
^{m}\rightarrow $\textbf{CP}$^{N},N=(n+1)(m+1)-1,$ is described explicitly
as 
\begin{equation}
s_{n,m}:(\{z_{0},...,z_{n}\},\{z_{0}^{\prime },...,z_{m}^{\prime
}\})\rightarrow (\{\cdot \cdot \cdot ,z_{i}z_{j}^{^{\prime }},\cdot \cdot
\cdot \}).  \tag{7.5}
\end{equation}%
Let\ $R_{ij}$ $=z_{i}z_{j}^{^{\prime }},$then the analogs of Pl\"{u}cker
relations in the present case are the conditions%
\begin{equation}
R_{ij}R_{kl}=R_{il}R_{kj}.  \tag{7.6}
\end{equation}

Finally, we need to describe the Veronese embedding.

\subsection{ The Veronese embedding}

\bigskip\ 

It can be described as follows. Let \{$z_{0},...,z_{k}\}$ be a point in 
\textbf{CP}$^{k}$ and let $\mathit{v}_{\text{\textsf{n}}}$ be the map $%
\mathit{v}_{\text{\textsf{n}}}:$ \textbf{CP}$^{k}\rightarrow $\textbf{CP}$^{%
\mathcal{N}}$, where $\mathcal{N}=\left( 
\begin{array}{c}
\text{\textsf{n}}+k \\ 
k%
\end{array}%
\right) -1\footnote{%
In the literature on algebraic geometry, e.g. see Ref.[41], \ one finds an
alternative way of writing $\mathcal{N}$, e.g.$\mathcal{N}=\left( 
\begin{array}{c}
\text{\textsf{n}}+k \\ 
\text{\textsf{n}}%
\end{array}%
\right) -1$ but, in view of Eq.(1.5), both are the same numbers.},$
explicitly described as 
\begin{equation}
\mathit{v}_{\text{\textsf{n}}}:\{z_{0},...,z_{k}\}\rightarrow \{x_{0},...,x_{%
\mathcal{N}}\},  \tag{7.7}
\end{equation}%
with x$_{i}=x_{0}^{i_{0}}x_{1}^{i_{1}}\cdot \cdot \cdot x_{k}^{i_{k}}$ $%
\equiv X^{\mathbf{I}}$ and $i_{0}+\cdot \cdot \cdot +i_{k}=$\textsf{n}$,$%
then this is the Veronese map provided that for any quadruple of
multiindices $\mathbf{I},\mathbf{J},\mathbf{K}$ and $\mathbf{L}$ the
following relation%
\begin{equation}
X^{\mathbf{I}}X^{\mathbf{J}}=X^{\mathbf{K}}X^{\mathbf{L}}  \tag{7.8}
\end{equation}%
holds.

\subsection{From analysis to synthesis}

\bigskip

Being armed with these descriptions of the respective embeddings we are
ready now to use all three of them. We begin with observation that the
dimensionalities of projective spaces in the case of Pl\"{u}cker and
Veronese embeddings can be made the same due to the same type of
combinatorics in both cases. This happens when we require: (Veronese) 
\textsf{n}$+k=l($Pl\"{u}c$\ker ).$ Evidently, relations of the type given by
Eq.(7.8) will be also satisfied by the Pl\"{u}cker relations (since in both
cases we are dealing with the same multiindex sets). Hence, we can identify
point by point both projective spaces. It should be clear that this can be
done only with some restriction on ordering of indices but this is
sufficient for our physical purposes. We shall discuss this topic further in
the next subsection. Next, we can think about making a projective space of
dimensionality $\mathcal{N}$ out of projective spaces of smaller
dimensionality using the Segre embedding. This fact is important physically
since it is connected with the fusion rules for scattering (e.g. Veneziano)
amplitudes. Suppose, we would like to compose a larger space out of complex
projective spaces of \ dimensionalities $i_{0},...,i_{k}$. Then, the Segre
embedding can be described schematically as the follows : 
\begin{eqnarray}
V_{i_{0}}\times \cdot \cdot \cdot \times V_{i_{k}} &\rightarrow
&V_{i_{0}}\otimes \cdot \cdot \cdot \otimes V_{i_{k}}\text{ \ causing} 
\notag \\
\mathbf{CP}(V_{i_{0}})\times \cdot \cdot \cdot \times \mathbf{CP}(V_{i_{k}})
&\hookrightarrow &\mathbf{CP(}V_{i_{0}}\otimes \cdot \cdot \cdot \otimes
V_{i_{k}})  \TCItag{7.9}
\end{eqnarray}%
with dimensionality of the final complex projective space being equal to $%
\mathcal{N}=(i_{0}+1)(i_{1}+1)\cdot \cdot \cdot (i_{k}+1)-1.$ To compare
this dimensionality with that for, say, the Veronese-type space we have to
require $(i_{0}+1)(i_{1}+1)\cdot \cdot \cdot (i_{k}+1)=\left( 
\begin{array}{c}
\text{\textsf{n}}+k \\ 
k%
\end{array}%
\right) $. In complete agreement with arguments made in the Introduction, if
we choose \textsf{n}$=1\cdot 2\cdot 3\cdot \cdot \cdot k$ and then identify $%
i_{l}$ with \textsf{n}$/l$ , provided that $0\ <l\leq k,$ we indeed obtain
the required decomposition. This result allows us to think about the
partition function for the Veneziano amplitudes in terms of the spin model
which is the some kind of reduction of the rigid string model \ proposed
some time ago by Polyakov [15]. Unlike his model, our (spin chain ) model is
exactly solvable. The noticed connection is important in view of the
potential physical applications: in Ref.[42], based on our earlier developed
variant [16] of the Polyakov rigid string model, interesting applications to
QCD were considered. Since the rigid string model proposed by Polyakov is
also of Grassmann-type (albeit, apparently, for different reasons) it is of
interest to study its reduction to the exactly solvable spin chain models.

\subsection{Some physical applications}

\subsubsection{\protect\bigskip\ The Poincare$^{\prime }$ polynomial and the
partition function}

\bigskip

Although the arguments presented above are standard, they cannot be used for
our immediate tasks. This is so because of the fact that in our case we have
to consider the differential forms. Specifically, let us consider a
differential form of the type $df_{0}\wedge df_{1}\wedge \cdot \cdot \cdot
\wedge df_{k\text{ }}$ with entries $f_{i}$ , $f_{2},etc.$ considered as
independent varibles. In which case we are dealing with just \textit{one}
differential form. Let now $%
f_{0}=z_{0}^{n_{0}},f_{2}=z_{2}^{n_{2}},...,f_{k}=z_{k}^{n_{k}}$provided
that $n_{0}+n_{1}+\cdot \cdot \cdot +n_{k}=$\textsf{n}$.$ Clearly, under
such conditions we shall obtain $\left( 
\begin{array}{c}
\text{\textsf{n}}+k \\ 
k%
\end{array}%
\right) $ different differential forms which can be looked upon as Pl\"{u}%
cker embedding of the space $(z_{0},...,z_{k})$ into the space of
differential forms of the type $dz_{0}^{n_{0}}\wedge dz_{1}^{n_{1}}\wedge
\cdot \cdot \cdot \wedge dz_{k\text{ }}^{n_{k}}\simeq
z_{0}^{n_{o}}z_{1}^{n_{1}}\cdot \cdot \cdot z_{k}^{n_{k}}\frac{dz_{0}}{z_{0}}
$ $\wedge \cdot \cdot \cdot \wedge \frac{dz_{k}}{z_{k}}.$ Coordinates $%
z_{0}^{n_{o}-1}z_{1}^{n_{1}-1}\cdot \cdot \cdot z_{k}^{n_{k}-1}$ (taken in a
prescribed order) can be viewed as Pl\"{u}cker coordinates so that,
apparently, we have essentially obtained the Pl\"{u}cker embedding. This is
not quite the case yet. To obtain the desired result several additional
steps are needed. For instance we can look at projective transformations of
the type $z_{0}\rightarrow lz_{0}$ , etc. Upon such a replacement the
combination $\frac{dz_{0}}{z_{0}}$ will stay the same while the combination $%
z_{0}^{n_{o}}z_{1}^{n_{1}}\cdot \cdot \cdot z_{k}^{n_{k}}$ will formally
change into $l^{n}$ $z_{0}^{n_{o}}z_{1}^{n_{1}}\cdot \cdot \cdot
z_{k}^{n_{k}}$ and is not invariant with respect to such transformations. To
correct the problem we have to divide $z_{0}^{n_{o}}z_{1}^{n_{1}}\cdot \cdot
\cdot z_{k}^{n_{k}}$ by the combination which scales the same way. In view
of results Part I, this will be the Fermat variety $\mathcal{F}$(\textbf{z})
=$z_{0}^{n}+\cdot \cdot \cdot +z_{k}^{n}.$ The Poincare$^{\prime }$
polynomial obtained in Eq.(6.11) in the limit $z\rightarrow 1$ counts the
number of such distinct invatiant forms. This topic will be discussed
further in Section 9 where we define the projective toric varieties and in
Part III. The rescaling \ just described subdivides the complex projective
space into equivalence classes. This procedure is essentially eqivalent to
the Pl\"{u}cker embedding. At the same time, in view of invariance of the
combination $\frac{dz_{0}}{z_{0}}$ $\wedge \cdot \cdot \cdot \wedge \frac{%
dz_{k}}{z_{k}}$ with respect to scale transformation, one can think about
the equivalence classes only between the monomials of the type $%
z_{0}^{n_{o}}z_{1}^{n_{1}}\cdot \cdot \cdot z_{k}^{n_{k}}$ and, from this
point of view, one obtains the Veronese embedding. \ Since combinatorially
in both cases we have been working with the same objects, not surprisingly,
the number of equvalence classes in both cases came out the same.
Physically, however, it is more adavntageous to use the Poincare$^{\prime }$
polynomial for the Veronese embedding since, as discussed in the
Introduction, the Poincare$^{\prime }$ polynomial for the complex projective
space of dimensionality $\mathcal{N}$ (up to numerical prefactor) coincides
with the partition function for a particle with spin $\mathcal{N}+1$ placed
into constant magnetic field. Such a partitiuon function was obtained by
Stone using N=2 finite dimensional supersymmetric model. In our case we are
interested not only in the partition function counting the number of entries
(summands) in the total Veneziano amplitude but also in the possibility of
reobtaining these amplitudes with help of this partition function. If this
can be achieved, we might consider the Veneziano model as exactly solved.
Since we know already that such a partition function up to a constant
coincides with the Weyl character formula, we must look for the
group-theoretic aspects of results we have just obtained. This is done in
the next subsection and the section which follows.

\subsubsection{Connections with KP hierarchy}

For the sake of space we shall assume familiarity of our readers with the
theory of symmetric functions. Excellent exposition can be found in
Ref.[43], while the basic facts can be found in \ Ref.[44]. From these
sources, it is known that the Schur functions \textbf{s}$_{\lambda }$(%
\textbf{x}) play the central role in this theory in view of their mutual
orthogonality with respect to carefully chosen scalar product $<$ $,$ $>,$
i.e. $<\mathbf{s}_{\lambda },\mathbf{s}_{\mu }>=\delta _{\lambda ,\mu }$ 
\footnote{%
We have suppressed the arguments, e.g. $\mathbf{x=\{}x_{1},x_{2},...x_{m}\},$
in this product for brevity.}. To make a connection with previous
discussion, let us consider the following generating function%
\begin{eqnarray}
f(\mathbf{z}) &=&\frac{1}{N!}(z_{0}+...+z_{k})^{N}=\dsum\limits_{_{\substack{
(n_{0},n_{1,...,}n_{k}\}  \\ N=n_{0}+n_{1}+\cdot \cdot \cdot +n_{k}}}}\frac{1%
}{n_{0}!n_{1}!\cdot \cdot \cdot n_{k}!}z_{0}^{n_{0}}\cdot \cdot \cdot
z_{k}^{n_{k}}  \notag \\
&\equiv &\dsum\limits_{\lambda \vdash N}c_{\lambda }\mathbf{z}^{\lambda }, 
\TCItag{7.10}
\end{eqnarray}%
where notations introduced in Section 1 were used.\footnote{%
We have suppressed an auxiliary variable, say \textbf{t}, which is normally
used in generating functions. Clearly, it can be restored whenever it is
needed.} Such an expansion can be considered as some kind of a basis
expansion in which the basis vectors belong to the set $\mathbf{z}^{\lambda
} $. From previous subsection we know that such an expansion makes sense
since the monomials $\mathbf{z}^{\lambda }$ represent well defined
equivalence classes in complex projective space. In general, however, such
monomials are not orthogonal with respect to the scalar product just
introduced. Evidently, each of these monomials can be re expanded with help
of the Schur polynomials, i.e.%
\begin{equation}
\mathbf{z}^{\lambda }=\dsum\limits_{\mu \vdash N}\tilde{c}_{\mu ,\lambda }%
\mathbf{s}_{\mu }.  \tag{7.11}
\end{equation}%
But from the book by Miwa et al, Ref.[45, page 90], we find out that under
such circumstances $\mathbf{z}^{\lambda }$ represents the tau function of
the KP hierarchy. Using this observation, perhaps superimposed with our
earlier treatment of the Witten-Konsevich model, Ref.[7], one can develop,
in principle, some quantum mechanical model whose partition function will
coincide with the Poincare$^{\prime }$ polynomial discussed earlier. There
is much faster way however to arrive at the final destination. It is
described in the next section. In the meantime, we would like to provide
some qualitative arguments in favour of such an alternative approach. To
this purpose, using Eq.(7.11) in (7.10) we can expand $f(\mathbf{z})$ in
terms of the orthogonal basis. Suppose now that there is an operator such
that $\mathbf{s}_{\mu }$ is its eigenfunction. Hence, $f(\mathbf{z})$ is
also an eigenfunction of such an operator. Since the Hilbert space is finite
dimensional in the present case, we may look, using the analogy with\ the
commutator algebra for angular momentum, for some kind of rasing and
lowering operators. If they indeed exist, then, as for the angular momentum
(or spin), there will be the upper and the lower vacuum states. The
dimension of the Hilbert state in this case can be determined
straightforwardly from the partition function whose Hamiltonian H$%
=B_{z}\cdot S_{z}$ \ where \textbf{B} is some external "magnetic" field
whose direction, as usual, is chosen to be along the "z -axis". The
partition function Z is obtained now in a usual way as 
\begin{equation}
Z=tr(\exp (-\beta \text{H}))  \tag{7.12}
\end{equation}%
and coincides with the Weyl character formula. In the limit B$_{z}\mathbf{%
\rightarrow }0$ one obtains the dimensionality of the Hilbert space as
expected. The same result can be obtained differently. For this purpose it
is sufficient to choose the Hamiltonian as $H=\mathbf{S}^{2}-const$ where
the $const$ is determined by some assigned fixed eigenvalue for the square
of the total spin (or angular momentum). \ Under such circumstances taking
trace in Eq.(7.12) using eigenfunctions of H will also produce the
dimensionality of the Hilbert space. In the next section we shall implement
the first procedure explicitly. To do so, we need to describe a model for
the complex projective space. It is known that there is number of such
models [46]. So, we have to select one of them which is the most convenient
for us. It will be used also in Part III.

\subsubsection{Description of particular model describing the projective
space}

To describe the model, we notice that each complex line in \textbf{C}$^{n+1}$
passing through the origin can be characterized by the unit vector $\omega
_{\nu }^{0}=\dfrac{\omega _{\nu }}{\left\vert \omega _{\nu }\right\vert }%
,\nu =0,...,n$, so that \ parametrically it can be represented as $z_{\nu
}=\omega _{\nu }^{0}\xi $ with $\xi $ being some complex parameter. By
definition, the projective space \textbf{CP}$^{n}$ is made of equivalence
classes of points $\mathbf{z}$ $\in $ \textbf{C}$^{n+1}\backslash \{0\}$
such that $\mathbf{z}^{\prime }$ =$\lambda \mathbf{z}$ with $\lambda $ being
some nonzero complex number. In the present case such a definition
essentially implies $\lambda =\xi .$ Consider now a unit sphere of real
dimension 2n+1\ living in \textbf{C}$^{n+1}$ and centered at the arbitrarily
chosen origin. It is characterized by the equation \ $\dsum\limits_{\nu
=0}^{n}z_{\nu }\bar{z}_{\nu }=1.$ The projective space \textbf{CP}$^{n}$ can
be realized as the set of \ points originating from the intersection of such
a sphere with the complex line just described. This results in an equation%
\begin{equation}
\left\vert \xi \right\vert ^{2}\dsum\limits_{\nu =0}^{n}\left\vert \omega
_{\nu }^{0}\right\vert ^{2}=1  \tag{7.13}
\end{equation}%
from which it follows that $\left\vert \xi \right\vert ^{2}=1$ and, hence, $%
\xi =e^{i\varphi }$. Thus, the points of \textbf{C}$^{n+1}$ can be
parametrized by $\omega _{\nu }^{0}$ $e^{i\varphi }$ with $\omega _{\nu
}^{0} $ being some nonnegative numbers subject to the constraint $%
\dsum\limits_{\nu =0}^{n}\left( \omega _{\nu }^{0}\right) ^{2}=1.$ For the
purposes of our discussion sometimes it will be convenient to redefine $%
\left( \omega _{\nu }^{0}\right) ^{2}$ as $t_{\nu \text{ }}$ so that in
terms of such variables the constraint \ describes a simplex instead of a
sphere.\footnote{%
The simplex was used already in Part I. It will be further discussed in
Section 9 and in Part III.} Two points of \textbf{C}$^{n+1}$ differing by
their phase factors belong to the same equivalence class so that the whole 
\textbf{C}$^{n+1}$ is divided into equivalence classes which are labeled by $%
\omega _{\nu }^{0\prime }s$. Because $\omega _{\nu }^{0\prime }s$ are
subject to the constraint, the dimensionality of thus formed projective
space \textbf{CP}$^{n}$ is $n$. Evidently, earlier discussed quotient $%
z_{0}^{n_{o}}z_{1}^{n_{1}}\cdot \cdot \cdot z_{k}^{n_{k}}/$ $\mathcal{F}$($%
\mathbf{z}$) will remain invariant with respect to such parametrization.
This fact was crucial for reconstruction of the Veneziano amplitudes from
periods of the Fermat varieties. Further implications of this invariance
will be discussed in Section 9 and in Part III.

\textbf{Remark 7.2}. From the Appendix, part d), it follows that the action
of elements of pseudo-reflection groups on some prescribed positive definite
Hermitian form leaves this form invariant. This fact can be taken as
defining property of the pseudo-reflection groups. The model of projective
space just described is compatible with actions of elements of the
pseudo-reflection groups. One can say as well that such groups can exist
only in certain spaces thus reflecting properties of these spaces. This fact
will be analyzed further in Section 9 (also in connection with Remark 7.4).
In fact, all results of previous sections are just consequences of this
observation.

\textbf{Remark 7.3}. From the previous remark it follows that the action of
group elements is taking place component wise. This fact will be used in the
next section.

\textbf{Remark 7.4. }The quadratic form\textbf{\ }$\dsum\limits_{\nu
=0}^{n}z_{\nu }\bar{z}_{\nu }=1$ can be extended to $\ \dsum\limits_{\nu
=0}^{n}z_{\nu }\bar{z}_{\nu }=z_{n+1}\bar{z}_{n+1}$. For $z_{n+1}\bar{z}%
_{n+1}\neq 0$ this form can be reduced back to the original. However, if we
keep it with this extra term, it becomes an invariant for groups of
isometries of the complex hyperbolic space. Such a space was analyzed
thoroughly by Goldman, Ref.[47]. Its profound physical significance will be
discussed in Section 9.

.

\section{Exact solution of the Veneziano model}

\subsection{From Witten to Lefschetz}

We begin with the following observations. First, given that Veneziano (and
Veneziano-like) amplitudes are periods of the Fermat (hyper)surfaces, the
associated with such periods invariant differential forms are living in the
complex projective space. The theorem by Kodaira asserts that a compact
complex manifold $X$ is projective algebraic if it is a Hodge manifold. For
the sake of space, we refer our readers to the monograph by Wells, Ref.[48],
for more information. We shall use this reference extensively in what
follows.

Second, the Theorem 2.17 of this reference states that under conditions of
Kodaira's theorem the manifold $X$ can be embedded into complex Grassmannian
so that the differential forms will be also living in the Grassmannian as
discussed earlier.

Third, for a complex Hermitian manifold $X$ let $\mathcal{E}^{p+q}(X)$
denote the set of complex -valued differential forms (sections) of the type $%
(p,q),p+q=r,$ living on $X$. The Hodge decomposition insures that $\mathcal{E%
}^{r}(X)$=$\sum\nolimits_{p+q=r}\mathcal{E}^{p+q}(X).$ The Dolbeault
operators $\partial $ and $\bar{\partial}$ act on $\mathcal{E}^{p+q}(X)$ as
follows \ $\partial :\mathcal{E}^{p+q}(X)\rightarrow \mathcal{E}^{p+1,q}(X)$
and $\bar{\partial}:\mathcal{E}^{p+q}(X)\rightarrow \mathcal{E}^{p,q+1}(X)$.
The exterior derivative operator is defined as $d=\partial +\bar{\partial}$.
Let now $\varphi _{p}$,$\psi _{p}\in \mathcal{E}^{p}(X)$ where $\mathcal{E}%
^{p}(X)$ belongs to the elliptic complex \ $\mathcal{E}^{\ast }\mathcal{(}X)%
\mathcal{=}\dbigoplus\nolimits_{p=0}^{r}\mathcal{E}^{p}(X)$ of differential
forms on $X$ forming a complex vector space of dimension $r+1$\footnote{%
Actually, one should consider instead a more complicated object $\mathcal{E}%
^{\ast }\mathcal{(}X,E)$ of differential forms with coefficients in $E$
where $E$ is the hermitian vector bundle over $X$. This would lead us to the
discussion involving the sheaf theory, \v{C}hech cohomology, etc. These are
beautifully explained in Ref.[46]. Since the final results which we obtain
are not going to be affected, we are not going to complicate matters by
these intricacies.}. By analogy with traditional quantum mechanics one can
define (using Dirac's notations) the inner product 
\begin{equation}
<\varphi _{p}\mid \psi _{p}>=\int\limits_{M}\varphi _{q}\wedge \ast \bar{\psi%
}_{p}  \tag{8.1}
\end{equation}%
where the bar means the complex conjugation and the star $\ast $ means the
Hodge conjugation as usual. The period integrals, e.g. those for the
Veneziano-like amplitudes, are expressible through such inner products [48].
Fortunately, such a product possesses properties typical for the finite
dimensional quantum mechanical Hilbert spaces. In particular, 
\begin{equation}
<\varphi _{p}\mid \psi _{q}>=C\delta _{p,q}\text{ and }<\varphi _{p}\mid
\varphi _{p}>>0,  \tag{8.2}
\end{equation}%
where $C$ is some known positive constant.

Fourth, with respect to such defined scalar product it is possible to define
all conjugate operators, e.g. $d^{\ast }$, etc. and, most importantly, the
Laplacians 
\begin{align}
\Delta & =dd^{\ast }+d^{\ast }d,  \notag \\
\square & =\partial \partial ^{\ast }+\partial ^{\ast }\partial ,  \tag{8.3}
\\
\bar{\square}& =\bar{\partial}\bar{\partial}^{\ast }+\bar{\partial}^{\ast }%
\bar{\partial}.  \notag
\end{align}%
All this was known to mathematicians before Witten's work [49\textbf{].} The
unexpected twist occurred when Witten suggested to extend the notion of the
exterior derivative $d$. Within the de Rham picture (valid for both real and
complex manifolds) let $M$ be a compact Riemannian manifold and $K$ be the
Killing vector field which is just one of the generators of isometry of $M.$
Then Witten suggested to replace the exterior derivative operator $d$ by the
extended operator 
\begin{equation}
d_{s}=d+si(K).  \tag{8.4}
\end{equation}%
Here $s$ is real nonzero parameter conveniently chosen. Witten argued that
one can construct the modified Laplacian by replacing conventional $\Delta $
given in Eq.(8.3) by $\Delta _{s}=d_{s}d_{s}^{\ast }+d_{s}^{\ast }d_{s}$.
This is possible if and only if $d_{s}^{2}=d_{s}^{\ast 2}$ $=0$ or, since $%
d_{s}^{2}=s\mathcal{L}(K)$, where $\mathcal{L}(K)$ the Lie derivative along
the field $K$, acting on the corresponding differential form, vanishes. The
details are beautifully explained in the much earlier paper by Frankel [50]
to be discussed in Part III. Atiyah and Bott, Ref.[22],\ observed that
replacement of the operator $d$ by $d_{s}$ causes replacement of the de Rham
cohomology by the equivariant cohomology. This topic is mentioned in Ref.[6]
\ and will be discussed in more detail in Part III in connection with
designing of the symplectic model reproducing the Veneziano amplitudes. In
this work we shall use more traditional methods however based on Eq.s.(8.3).

Looking at these equations and following Ref.s[26,51,52] we define the
(Dirac) operator $\acute{\partial}=\bar{\partial}+\bar{\partial}^{\ast }$
and its adjoint with respect to scalar product, Eq.(8.2). Then, use of the
above references \ allows us to determine the dimension $Q$ of the quantum
Hilbert space for which the scalar product, Eq.(8.2), was defined. It is
given by 
\begin{equation}
Q=\ker \acute{\partial}-co\ker \acute{\partial}^{\ast }=Q^{+}-Q^{-}. 
\tag{8.5}
\end{equation}%
We would like to arrive at the same result differently using earlier
introduced partition function, Eq.(7.12). To this purpose we notice that
according to Theorem 4.7. in the book by Wells [48] we have $\Delta
=2\square =2\bar{\square}$ with respect to the K\"{a}hler metric on $X$.
Next, according to the Corollary 4.11. of the same reference $\Delta $
commutes with $d,d^{\ast },\partial ,\partial ^{\ast },\bar{\partial}$ and $%
\bar{\partial}^{\ast }.$ From these facts it follows immediately that if we,
in accord with Witten, choose $\Delta $ as our Hamiltonian, then the
supercharges can be selected as $Q^{+}=d+d^{\ast }$ and $Q^{-}=i\left(
d-d^{\ast }\right) .$ Evidently, this is not the only choice as Witten also
indicates. If the Hamiltonian \ H is acting in \textit{finite} dimensional
Hilbert space one may require axiomatically that : a) there is a vacuum
state (or states) $\mid \alpha >$ such that H$\mid \alpha >=0$ (i.e. this
state is the harmonic differential form) and $Q^{+}\mid \alpha >=Q^{-}\mid
\alpha >=0$. This implies, of course, that [H,$Q^{+}]=[$H,$Q^{-}]=0.$
Finally, \ once again, following Witten, we require that $\left(
Q^{+}\right) ^{2}=\left( Q^{-}\right) ^{2}=$H. Then, the equivariant
extension, Eq.(8.4), leads to $\left( Q_{s}^{+}\right) ^{2}=$ H+$2is\mathcal{%
L}$($K$). Fortunately, we can avoid this extension by noticing that the
above \ supersymmetry algebra can be extended. This can be done with help of
the Lefschetz isomorphism theorem whose exact formulation is given as
Theorem 3.12. in Wells, Ref.[48]. We shall only use some parts of this
theorem in our work. In particular, using notations of Ref.[48], we
introduce the operator $L$ commuting with $\Delta $ and its adjoint $L^{\ast
}\equiv \Lambda $.\ It can be shown [48, page 159], that $L^{\ast }=w\ast
L\ast $ where, as before, $\ast $ denotes the Hodge star operator and the
operator $w$ can be formally defined through the relation $\ast \ast =w$
[48, page 156]. From these definitions it should be clear that $L^{\ast }$
also commutes with $\Delta $ on the space of harmonic differential forms (in
accord with page 195 of [48]).

As part of preparation for proving of the Lefschetz isomorphism theorem, it
can be shown [48], that 
\begin{equation}
\lbrack \Lambda ,L]=B\text{ and }[B,\Lambda ]=2\Lambda \text{, }[B,L]=-2L. 
\tag{8.6}
\end{equation}%
This commutator algebra (up to a constant) coincides with the $sl_{2}(%
\mathbf{C})$ Lie algebra given in the canonical form, e.g. see Ref.[53, page
37], as follows

\bigskip 
\begin{equation}
\lbrack h,e]=2e\text{ , }[h,f]=-2f\text{ , \ }[e,f]=h\ .  \tag{8.7}
\end{equation}%
Comparison between the above two expressions leads to the isomorphism of Lie
algebras, i.e. the operators $h,$ $f$ and $e$ act on the vector space $\{v\}$
to be described below while the operators $\Lambda ,L$ and $B$ obeying the
same commutation relations act on the space of differential forms.

\textbf{Remark 8.1}. For such an isomorphism to exist the elliptic complex
should be finite dimensional. This requirement of finite dimensionality
comes from important result by Serre to be described in the next subsection

\subsection{From Lefschetz to Veneziano via Serre and Ginzburg}

\bigskip

Now we would like to recall, e.g. Ref.[53], page 25, that all semisimple Lie
algebras are made of copies of $sl_{2}(\mathbf{C}).$ Assuming our readers
familiarity with the Lie algebras and, in particular, with semisimple Lie
algebras, we would like now to adopt the Lefschetz correspondence to our
needs. In particular, let $z_{1},...,z_{n}$ denote a basis of the vector
space $V$ in \textbf{C}$^{n}$. In terms of this basis consider a polynomial $%
f(\mathbf{z})$ given by 
\begin{equation}
f(\mathbf{z})=f(z_{1},...,z_{n})=\sum\limits_{\mathbf{i}}\mathbf{\lambda }_{%
\mathbf{i}}\mathbf{z}^{\mathbf{i}}\equiv \sum\limits_{\mathbf{i}}\lambda
_{i_{1}....i_{n}}z_{1}^{i_{1}}\cdot \cdot \cdot z_{n}^{i_{n}},\text{ (}%
\lambda _{\mathbf{i}},\text{ }z_{m}^{i_{m}}\in \mathbf{C,}1\leq m\leq n). 
\tag{8.8}
\end{equation}%
Let $\mathbf{z}$\textbf{\ }be treated as a column vector(or,better, as a set
of column vectors, e.g. see Section 7.1). Then, by definition (e.g. see
Ref.[54\textbf{]}) compatible with earlier made Remark 7.3.) we obtain, $%
M\circ f(\mathbf{z})=f(M\mathbf{z})\equiv f(Mz_{1},...,Mz_{n}$ ), where $M$ $%
\subset G.$ Here $G$ belongs to some matrix group. In particular, let $%
M\subset sl_{2}(\mathbf{C}),$then \ following Dixmier [55, Ch-r 8], we
introduce operators $h=\sum\nolimits_{\alpha =1}^{n}a_{\alpha }h_{\alpha }$, 
$e=\sum\nolimits_{\alpha =1}^{n}b_{\alpha }e_{\alpha }$, $%
f=\sum\nolimits_{\alpha =1}^{n}c_{\alpha }f_{\alpha }.$ Provided that the
constants are subject to the constraint: $b_{\alpha }c_{\alpha }=a_{\alpha }$
, the commutation relations between the operators $h$, $e$ and $f$ are 
\textit{exactly the same} as respectively for $B$, $\Lambda $ and $L$. To
avoid unnecessary complications, we choose $a_{\alpha }=b_{\alpha
}=c_{\alpha }=1$.

Next, following Serre [56, Ch-r 4], we need to introduce the \textit{%
primitive} vector (or element). This is the vector $v$ such that $hv$=$%
\lambda v$ but $ev=0.$ The number $\lambda $ is the weight of the module $%
V^{\lambda }=\{v\in V\mid hv$=$\lambda v\}.$ If the vector space is \textit{%
finite dimensional}, then $V=\sum\nolimits_{\lambda }V^{\lambda }$.
Moreover, only if $V^{\lambda }$ is finite dimensional it is straightforward
to prove that the primitive element \textit{does} exist in accord with
Remark 8.1. The proof is based on observation that if $x$ is the eigenvector
of $h$ with weight $\lambda ,$ then $ex$ is also an eigenvector of $h$ with
eigenvalue $\lambda -2,$ etc. Moreover, from the book by Kac [57, Chr.3], it
follows that if $\lambda $ is the weight of $V,$ then $\lambda -<\lambda
,\alpha _{i}^{\vee }>\alpha _{i}$ is also the weight with the same
multiplicity. Since according to Eq.(A.2) of Appendix $<\lambda ,\alpha
_{i}^{\vee }>\in \mathbf{Z}$\textbf{, }Kac introduces another module: $%
U=\sum\nolimits_{k\in \mathbf{Z}}$ $V^{\lambda +k\alpha _{i}}$. Such a
module is finite for finite reflection groups and is infinite for the affine
reflection groups. We would like to argue that for our purposes, in view of
the Theorem 2.2. by Solomon it is sufficient to use only finite reflection
(or pseudo-reflection) groups.

\textbf{Remark 8.2.} It should be remembered at this point that in Solomon's
theorem the requirement of finiteness of the (pseudo)reflection group is
stated explicitly.

\textbf{Remark 8.3}. From the book by Kac, it should be clear that the
infinite dimensional version of the module $U$ straightforwardly leads to
all known string-theoretic results. Development of connections with KP
hierarchy discussed in Section 7.2. also ultimately leads to the
conventional string-theoretic formulations. In the case of CFT this is
essential and will be explained further in Section 9 and in Part IV, but for
calculation of the Veneziano-like amplitudes this is \textit{not} essential.
By accepting \ the traditional option we\ \ loose connections with the
Lefschetz isomorphism theorem ( relying heavily on the existence of
primitive elements) and with the Hodge theory in its traditional form. The
infinite dimensional extensions of the Hodge-de Rham theory involving loop
groups, etc. relevant for the CFT can be found in Ref.[58,59]. Fortunately,
they are not needed for the purposes of this work. Hence, below we work only
with finite dimensional spaces.

In particular, let now $v$ be a primitive element of weight $\lambda .$
Then, following Serre, we let $v_{n}=\frac{1}{n!}e^{n}v$ for $n\geq 0$ and $%
v_{-1}=0,$ so that 
\begin{align}
hv_{n}& =(\lambda -2n)v_{n},  \tag{8.9} \\
ev_{n}& =(n+1)v_{n+1},  \notag \\
fv_{n}& =(\lambda -n+1)v_{n-1}.  \notag
\end{align}%
Clearly, the operators $e$ and $f$ are the creation and the annihilation
operators according to existing in physics terminology while the vector $v$
can be interpreted as the vacuum state vector. The question arises: how this
vector is related to earlier introduced vector $\mid \alpha >?$ Before
providing the answer to this question we need, following Serre, to settle
the related issue. In particular, we can either: a) assume that for all $%
n\geq 0$ the first of Eq.s(8.9) has solutions and all vectors $v,v_{1},v_{2}$
, ...., are linearly independent or b) beginning from some $m+1\geq 0,$ all
vectors $v_{n\text{ }}$are zero, i.e. $v_{m}\neq 0$ but $v_{m+1}=0.$ The
first option leads to the infinite dimensional representations associated
with affine Kac-Moody algebras just mentioned. The second option leads to
the finite dimensional representations and to the requirement $\lambda =m$
with $m$ being an integer. We shall adjust this integer to our needs shortly
below. In the meantime, following Serre, this observation can be exploited
further thus leading us to the crucial physical identifications. Serre
observes that with respect to $n=0$ Eq.s (8.9) possess a (\textquotedblright
super\textquotedblright )symmetry. That is the linear mappings 
\begin{equation}
e^{m}:V^{m}\rightarrow V^{-m}\text{ and \ }f^{m}:V^{-m}\rightarrow V^{m} 
\tag{8.10}
\end{equation}%
are isomorphisms and the dimensionality of $V^{m}$ and $V^{-m}$ are the
same. Serre provides an operator (the analog of Witten's $F$ operator) $%
\theta =\exp (f)\exp (e)\exp (-f)$ such that $\theta \cdot f=-e\cdot \theta $%
, $\theta \cdot e=-\theta \cdot f$ and $\theta \cdot h=-h\cdot \theta .$ In
view of such an operator, it is convenient to redefine the $h$ operator : $%
h\rightarrow \hat{h}=h-\lambda $. Then, for such redefined operator the
vacuum state is just $v$. Since both $L$ and $L^{\ast }=\Lambda $ commute
with the supersymmetric Hamiltonian H and, because of the group isomorphism,
we conclude that the vacuum state $\mid \alpha >$ for H corresponds to the
primitive state vector $v$, moreover, $-m\leq n\leq m$ in Eq.s(8.9)$.$

Now we are ready to apply yet another isomorphism following Ginzburg,
Ref.[21, pages 205-206],\footnote{%
Unfortunately, the original source contains very minor mistakes (misprints).
These are easily correctable. The corrected results are given in the text.}.
To this purpose we make the following identification 
\begin{equation}
e_{i}\rightarrow t_{i+1}\frac{\partial }{\partial t_{i}}\text{ , }%
f_{i}\rightarrow t_{i}\frac{\partial }{\partial t_{i+1}}\text{ , }%
h_{i}\rightarrow t_{i}\frac{\partial }{\partial t_{i}}+2\left( t_{i+1}\frac{%
\partial }{\partial t_{i+1}}-t_{i}\frac{\partial }{\partial t_{i}}\right) , 
\tag{8.11}
\end{equation}%
$i=0,...,m$. Such operators are acting on the vector space made of monomials
of the type 
\begin{equation}
v_{n}\rightarrow \mathcal{F}_{N}=\frac{1}{n_{0}!n_{2}!\cdot \cdot \cdot
n_{k}!}t_{0}^{n_{0}}\cdot \cdot \cdot t_{k}^{n_{k}},  \tag{8.12}
\end{equation}%
where $n_{0}+...+n_{k}=N$.This result is in accord with. Eq.(7.10).
Moreover, now we have analogs of Eq.s (8.9).These are given by 
\begin{align}
h_{i}\ast \mathcal{F}_{n}(i)& =2(n_{i+1}-n_{i})\mathcal{F}_{n}(i),  \notag \\
e_{i}\ast \mathcal{F}_{n}(i)& =2n_{i}\mathcal{F}_{n}(i+1),  \tag{8.13} \\
\text{ }f_{i}\ast \mathcal{F}_{n}(i)& =2n_{m+1}\mathcal{F}_{n}(i-1),  \notag
\end{align}%
where $\mathcal{F}_{n}(i)$ is a part of the wave function relevant to action
of operators $e_{i}$, $f_{i}$, $h_{i}$. Clearly, at this point one should
make the following identifications: $m(i)-2n(i)=2\left( n_{i+1}-n_{i}\right) 
$ , $2n_{i}=n(i)+1$ and $m(i)-n(i)+1=2n_{i+1}$ in order to be consistent
with Eq.s(8.9). Next, we define the total Hamiltonian: $h=$ $%
\sum\nolimits_{i=0}^{k}h_{i}$ \footnote{%
In accord with general rules of construction of the Lie algebras out of
copies of $sl_{2}(\mathbf{C})$ thus designed Hamiltonian represents the
standard action of $sl_{k}(\mathbf{C})$ on the vector space made out of
monomials, Eq.(8.12).}\ and redefine individual Hamiltonians as described
above. This causes $m(i)$ to be effectively zero in the above equations. The
operators $\dfrac{\partial }{\partial t_{i}}$ act on the total set of
monomials $1,t,\frac{1}{2!}t^{2},...,\frac{1}{N!}t^{N}$. Such monomials are
forming the basis of vector space analogous to $\nu _{n}$.This leads to
identification of $m(i)$ with $N$ $\ \forall i$. Incidentally, $N$ is the
total energy according to Veneziano condition, Eq.(1.4). Based on these
remarks let us consider an action of such redefined total Hamiltonian on the
individual wave function $\mathcal{\hat{F}}_{N}$ of the type given by
Eq.(8.12). Using Eq.s (8.13) we obtain,%
\begin{equation}
h\mathcal{\hat{F}}_{N}=2\left( n_{k}-n_{0}\right) \mathcal{\hat{F}}_{N}. 
\tag{8.14}
\end{equation}%
Since $0\leq n_{i}\leq N$ \ $\forall i$ we obtain: $0\leq \left\vert
n_{k}-n_{0}\right\vert $ $\leq N$. In view of this Eq.(8.14) becomes very
much analogous to the equation for the z-component of the angular momentum
(or spin) of magnitude $N$. Unlike the case of angular momentum, here there
is an additional degeneracy which we would like to describe now. To this
purpose we need to describe in some detail the wave functions entering
Eq.(8.14).

Firstly, we notice that the operators introduced in Eq.(8.11) do not change
the total power of monomials of the type given in Eq.(8.12). This is in
accord with the requirement that for such monomials the constraint $%
n_{0}+...+n_{k}=N$ should always hold. Next, for completeness of our
presentation we would like to restore the subtracted term in the total
Hamiltonian. Then, its action on the monomial $\frac{1}{N!}t_{0}^{N}$
produces an eigenvalue $-N$. Furthermore, consider now another monomial $%
\frac{1}{\left( N-1\right) !}t_{0}^{N-1}t_{1}.$ The action of a Hamiltonian
on such a monomial will produce an \ eigenvalue $-N+2$ as required [56]. But
the same eigenvalue will be produced also by the monomials of the type $%
\frac{1}{\left( N-1\right) !}t_{0}^{N-1}t_{i}$ , $i=2,3,...,k$. Hence we
have obtained a degeneracy. The next generation of wave functions can be
constructed as follows $\frac{1}{\left( N-2\right) !}t_{0}^{N-2}t_{i}$ $%
t_{j} $, $\frac{1}{\left( N-3\right) !}t_{0}^{N-3}\frac{1}{2!}t_{i}^{2}$ $%
t_{j}$, etc. This process will end when we shall reach the situation when we
would have $t_{0}^{0}\cdot \cdot \cdot $. Such wave function will have an
eigenvalue zero. Next, we can obtain another series of wave functions which
begins with $\frac{1}{N!}t_{k}^{N}$. To construct this series we have to
switch signs in the corresponding equations according to rules implied by
Eq.(8.10). These two series exhaust all the combinations satisfying $%
n_{0}+...+n_{k}=N.$ Clearly, the number of such combinations $\mathcal{N=}%
\left( 
\begin{array}{c}
N+k \\ 
k%
\end{array}%
\right) $ in accord with Section 1. Since all wave functions given by
Eq.(8.12) possess the same energy $N$, the partition function, Eq.(7.12), in
the limit $\beta \rightarrow 0$ reproduces $\mathcal{N}$ as required. Since
thus constructed wave functions are in one-to one relation with the
corresponding Veneziano amplitudes, obtained results provide a complete
solution of the Veneziano model.

\subsection{Connections with chaotic dynamical systems and problem of zeros
of the Riemann zeta function}

Some of our readers may ask at this point the following question: All this
is fine but what kind of physics the Hamiltonian introduced in Eq.(8.11)
represents? We would like to address this important issue now. First, even
if this Hamiltonian would be a formality, because of the Lefschets
isomorphism theorem we can always go back to the traditional supersymmetric
formulation of the problem. Such a formulation, although physically useful,
leaves certain aspects of the problem undetected. This is especially true in
the present case since by using the supersymmetric formulation the fact that
4-particle Veneziano amplitude can be equivalently presented as the product
of Riemann zeta functions (e.g. see Eq.(1.12) of Part I) seems only as a
curiosity. This curiosity happens to be intrinsically related to
Hamiltonians of the type given by Eq.(8.11). The simplest Hamiltonian of
this type is $H=xp.$ It was recently considered by Berry and Keating,
Ref.[60], and, more comprehensively, in Ref.[61]. These authors notice that
at the classical level the system described by such Hamiltonian has a
hyperbolic point at the origin of the $(x,p)$ phase space plane. The
trajectories $x(t)=x(0)exp(t)$ and $p(t)=p(0)exp(-t)$ are uniformly unstable
with stretching in $x$ and contraction in $p$. The motion has the desired
lack of time reversal symmetry so that the orbit cannot be retracted.%
\footnote{%
In Ref.[62] it is demonstrated that the Hamiltonian $H=xp$ is canonically
equivalent to the Hamiltonian for the "inverted" harmonic oscillator : $H=%
\frac{1}{2}(P^{2}-Q^{2})$, obtained upon the symplectic rotation of the
type: $x=\frac{P+Q}{\sqrt{2}}$ , $p=\frac{P-Q}{\sqrt{2}}.$} Quantization of
classically chaotic systems is currently in the focus of attention in
physics literature[63] and in the case of the system just described can be
directly connected with zeros of the Riemann zeta function and, hence, with
the Riemann hypothesis about these zeros [60]. Okubo [64] noticed the
Lorentz invariance of two dimensional Hamiltonians of the type given by
Eq.(8.11) and made a conjecture about some intrinsic connections between the
Lorentz invariance and the Riemann hypothesis. The Lorentz invariance of N=2
supersymmetric quantum mechanics was noticed already in the seminal paper by
Witten, Ref.[49, page 662]. No connections with the Riemann hypothesis were
made however in his paper. We shall discuss further the issues related to
the Lorentz invariance in Section 9, mainly elaborating on the Remark 7.4.
made earlier.

In mathematics literature the ergodic properties of the dynamical systems
associated with semisimple Lie groups and algebras have received
considerable attention recently, e.g. see monograph by Feres, Ref.\textbf{[}%
65]. In our opinion, the dynamical issues in the present case intrinsically
are of the number-theoretic nature. The number-theoretic aspects of the
Veneziano amplitudes discussed in our earlier publication, Ref.[10\textbf{]}%
, provide a natural link between dynamics, Rieamnn's zeta function and, more
general, L-functions. For the sake of space, we refer our readers to just
mentioned literature containing, in addition, a large number of relevant
references of major importance.

\section{The theorem by Serre and its physical significance}

\bigskip

\subsection{\protect\bigskip Statement of the theorem and physically
motivated proof}

In the previous sections we repeatedly mentioned the theorem by Serre. In
this section \ we would like to state the theorem explicitly, to explain
using physical arguments its proof, and to discuss some physical
consequences of this theorem not mentioned thus far.

Before stating the theorem, we need to recall that any linear algebraic
group $G$ is isomorphic to a closed subgroup of $GL_{n}(V,$K$)$ \ acting on
a vector space $V$ of dimension $n\geq 1$ by matrices $M$ whose entries
belong \ to any closed number field K such as \textbf{C} or $p$-adic [66].
With such an observation, we are ready to formulate the theorem, e.g see
Bourbaki, Ref.[17], Chapter 5, paragraph 5 (problem set \#8).

\ 

\textbf{Theorem 9.1}. (Serre [67]) \textit{Let }$V$\textit{\ be a vector
space of dimension }$n$\textit{\ over the field }K\textit{\ and let }$%
S(V)^{G}$\textit{\ \ be a graded ring of invariants of the group }$G$\textit{%
\ acting on symmetric algebra }$S$\textit{(}$V)$\textit{\ (defined in
Section 2). Then }$S(V)^{G}$\textit{\ is a \textbf{polynomial} algebra if
and only if }$G$\textit{\ is \textbf{finite} group generated by
pseudo-reflections.}

\ 

\textbf{Remark 9.2}. It is important that the theorem by Serre involves only
finite pseudo-reflection groups. This requirement is consistent with \
earlier stated Theorem 5.1.by McMullen. Below we shall mention the
conditions under which it should be\ amended in order to reproduce the
results of CFT.

Although the proof of this theorem can be found in many places, e.g. see
Refs.[68,69] or the original paper by Serre, Ref.[67], we would like to
provide arguments \ leading to a physically motivated proof.

To begin, let us recall that in Section 4 we defined $x\in V^{G}$ as a group
invariant (for some group $G$) if $Av(x)=x$. The averaging (over group $G$)
operator $Av$ was defined in Eq.(4.5). Following Stanley, Ref.[54], we would
like to provide few additional details. For instance, taking into account
Eq.(8.8) and discussion which follows this equation the polynomial ring $A[%
\mathbf{z}]$, $\mathbf{z}\in C^{n},$ contains a subring $S(V)^{G}$ of
invariants defined by

\bigskip 
\begin{equation}
S(V)^{G}=\{f(\mathbf{z})\in A[\mathbf{z}]:\text{ }M\circ f(\mathbf{z})=f(M%
\mathbf{z})=f(\mathbf{z})\text{ }\forall M\in G\}.  \tag{9.1}
\end{equation}

\bigskip\ If $X(G)$ is the set of all irreducible (complex, in general)
characters of $G,$ then $A[\mathbf{z}]$ can be decomposed into direct sum as
follows: $A[\mathbf{z}]=\coprod\limits_{\chi }S(V)_{\chi }^{G}$, where the
condition $f(\mathbf{z})\in S(V)_{\chi }^{G}$ means that 
\begin{equation}
S(V)_{\chi }^{G}=\{f\in A[\mathbf{z}]:\text{ }M\circ f(\mathbf{z})=\chi (M)f(%
\mathbf{z})\text{ }\forall M\in G\text{ and }\chi (M)\in X(G)\}.  \tag{9.2}
\end{equation}%
From this it follows, that earlier defined $S(V)^{G}=S(V)_{\varepsilon }^{G}$
where $\varepsilon $ denotes the trivial character.

Let $n=\dim V$ $\ $be the \textit{degree} of $G$ while $\left\vert
G\right\vert $ be its \textit{cardinality.} Emmy Noether [70] proved the
following theorem

\ 

\textbf{Theorem 9.3.(}Noether\textbf{) }\textit{Let }$G$\textit{\ be of
cardinality }$\left\vert G\right\vert $\textit{\ and }$n$\textit{\ is its
degree, then }$S(V)^{G}$\textit{\ is generated as an algebra over \textbf{C}
by no more than }$\mathit{(}%
\begin{array}{c}
\left\vert G\right\vert +n \\ 
n%
\end{array}%
)$\textit{\ homogenous invariants of degree not exceeding }$\left\vert
G\right\vert .$

\ 

The results of previous sections are in complete accord with this theorem.
However, now we are in the position to develop some refinements. This can be
accomplished in several steps. For instance, to put the results of Section
7.6.3. into proper perspective we need to introduce the following

\textbf{Definition 9.4}. The set $T:=(\mathbf{C}\backslash 0)^{n}=:(\mathbf{C%
}^{\ast })^{n}$ is called a \textit{complex algebraic torus}.

Since each $z\in \mathbf{C}^{\ast }$ can be written as $z=r\exp (i\theta )$
so that for $r>0$ the fiber: \{$z\in \mathbf{C}^{\ast }\mid \left\vert
z\right\vert =r\}$ is a circle of radius $r$, we can represent $T$ as the
product ($R_{>0})^{n}\times \left( S^{1}\right) ^{n}$. The product of $n$
circles $\left( S^{1}\right) ^{n}$ is the deformation retract of $T$. It is
indeed a topological torus. Following Fulton [24], we are going to call it a 
\textit{compact} \textit{torus} $S_{n}$. Hence, the algebraic torus is a
product of a compact torus and a vector space. This circumstance is helpful
since whatever we can prove for the deformation retract can be extended to
the whole torus $T$. \ As an illustration relevant to our calculations of
the Veneziano amplitude made in Part I, and to discussions we had \ in
Section 7, we \ would like following Fulton, Ref.[24], to consider a
deformation retract of the complex projective space \textbf{CP}$^{n}$. Such
a retraction is achieved by using the map 
\begin{equation*}
\text{ }\tau :\text{\ }\mathbf{CP}^{n}\rightarrow \mathbf{P}_{\geq }^{n}=%
\mathbf{R}_{\geq }^{n+1}\setminus \{0\}/\mathbf{R}^{+}
\end{equation*}%
or, explicitly, 
\begin{equation}
\tau :\text{ }(z_{0},...,z_{n})\mapsto \frac{1}{\sum\nolimits_{i}\left\vert
z_{i}\right\vert }(\left\vert z_{0}\right\vert ,...,\left\vert
z_{n}\right\vert )=(t_{0},...,t_{n})\text{ , }t_{i}\geq 0.  \tag{9.3}
\end{equation}%
The mapping $\tau $ is onto the standard $n$-simplex : $t_{i}\geq 0$, $%
t_{0}+...+t_{n}=1.$\ 

Since we are interested in the torus action on the algebraic variety the
above constructed deformation retract simplifies matters considerably. We
had a chance to see these simplifications in Part I when we performed our
calculations of the Veneziano amplitudes. Now, however, we would like to
consider more general cases. To this purpose we provide the following

\bigskip \textbf{Definition 9.5}\ An\textit{\ affine} algebraic variety $%
V\in \mathbf{C}^{n}$ is the set of zeros of the\ collection of polynomials
from the ring $A[\mathbf{z}].$ \ \ \ \ \ \ \ \ \ \ \ \ \ \ \ \ \ \ \ \ \ \ 

According to the famous Hilbert's Nullstellensatz a collection of such
polynomials is \textit{finite }and forms the set \ $I(\mathbf{z}):=\{f\in $
A[$\mathbf{z}$]$,$ $f(\mathbf{z})=0\}$ of maximal ideals usually denoted as 
\emph{Spec}A[\textbf{z}]. Using this fact, we provide the following formal
definition.

\textbf{Definition 9.6.} The zero set of a \textit{single} function which
belongs to $I(\mathbf{z})$ is called \textit{algebraic hypersurface. }%
Accordingly, the set\ $I(\mathbf{z})$ corresponds to \textit{intersection}
of a finite number of hypersurfaces\footnote{%
In \ Section 3 (and in Part I) the pole $Q=0$ of the period integral,
Eq.(3.3), defines the algebraic hypersurface. For the Veneziano amplitude it
is the Fermat hypersurface. More generally, earlier studies of the
scattering amplitudes using Feynman's diagrammatic rules [71] produced
similar types of period integrals. Typically, the denominator $Q$ for such
integrals is a product of several algebraic functions. Therefore, it should
be clear that methods developed for calculation of the Veneziano amplitudes
are fully consistent with earlier studies of scattering amplitudes.}.

\bigskip

Being armed with such results, we would like to construct the affine toric
variety and consider the torus action on such a variety. This is
accomplished in several steps. First, instead of considering the set of
Laurent monomials of the type $\lambda \mathbf{z}^{\mathbf{\alpha }}\equiv
\lambda z_{1}^{\alpha _{1}}\cdot \cdot \cdot z_{n}^{\alpha _{n}}\in $ A[$%
\mathbf{z}$], we would like to consider a subset made of \ monic monomials,
i.e. those with $\lambda =1.$ Such a subset forms a subring with respect to
the usual multiplication and addition. The crucial step forward is to assume
that for such monomials the exponent $\mathbf{\alpha \in }S_{\sigma }.$ The
monoid $S_{\sigma }$ will be defined momentarily. This fact allows us to
define the mapping 
\begin{equation}
u_{i}:=z^{a_{i}}  \tag{9.4}
\end{equation}%
with $a_{i}$ being \ one of the generators of the monoid $S_{\sigma }$ and$\
z\in \mathbf{C}.$ The monoid $S_{\sigma }$ \ can be defined now as follows.

\textbf{Definition 9.7.} \textit{A semi-group }$\mathit{S}$\textit{\ }that
is a non-empty set with associative\textit{\ }operation\textit{\ }is called 
\textit{monoid\ }if it is commutative, satisfies cancellation law\textit{\
(i.e. s+x=t+x }implies\textit{\ s=t for all s,t,x}$\in S)$ and has zero
element $(i.e.$ $s+0=s,s\in S)$. \textit{A monoid }$\mathit{S}_{\sigma }$%
\textit{\ is} \textit{finitely generated }if there exist some set of\textit{%
\ }$a_{1},...,a_{k}\in S$\textit{, }called $generators$, such that 
\begin{equation}
S_{\sigma }=\mathbf{Z}_{\geq 0}a_{1}+\cdot \cdot \cdot +\mathbf{Z}_{\geq
0}a_{k}.  \tag{9.5}
\end{equation}%
\ Based on this, we can make a \textit{crucial observation}: the mapping
given by Eq.(9.4) provides an isomorphism between the \textit{additive}
group of exponents $a_{i}$ and the \textit{multiplicative} group of monic
Laurent polynomials. Next, we recall that the function $\phi $ is considered
to be \textit{quasi homogenous} of degree $d$ with exponents \textit{l}$%
_{1},...,l_{n}$ if 
\begin{equation}
\phi (\lambda ^{\mathit{l}_{1}}x_{1},...,\lambda ^{\mathit{l}%
_{n}}x_{n})=\lambda ^{d}\phi (x_{1},...,x_{n}),  \tag{9.6}
\end{equation}%
provided that $\lambda \in \mathbf{C}^{\ast }.$ Applying this result to $z^{%
\mathbf{a}}\equiv z_{1}^{a_{1}}\cdot \cdot \cdot z_{n}^{a_{n}}$ we obtain
the Veneziano-like equation 
\begin{equation}
\sum\limits_{j}\left( l_{j}\right) _{i}a_{j}=d_{i}.  \tag{9.7}
\end{equation}%
Clearly, if the index $i$ is numbering different monomials, then \ the sum
in Eq.(9.7) (equal to $d_{i})$ belongs to the monoid $S_{\sigma }.$ The same
result can be achieved if instead we would consider the products of the type 
$u_{1}^{l_{1}}\cdot \cdot \cdot u_{n}^{l_{n}}$ and rescale all $%
z_{i}^{\prime }s$ by \textit{the same} factor $\lambda .$ Actually, Eq.(9.7)
should be understood as a scalar product between $\left( l_{j}\right)
_{i}^{\prime }s$ (living in the space \textit{dual} to $a_{j}^{\prime }s)$ \
and the $a_{j}^{\prime }s$. It is convenient at this point to define a cone.

\textbf{Definition 9.8.} A \textit{convex polyhedral} cone $\sigma $ is a
set 
\begin{equation}
\sigma =\{\sum\limits_{i=1}^{k}r_{i}a_{i},\text{ }r_{i}\geq 0\}.  \tag{9.8}
\end{equation}

\textbf{\ \ Remark 9.9. }In the case when earlier introduced generators $%
a_{1},...,a_{k}$ are considered as the basis of a vector space $V$, the
definitions of $S_{\sigma }$ and $\sigma $ \textit{describe the same object}%
. \ We shall always refer to it as a cone. In this case the dual cone $%
\sigma ^{\vee }$ is defined by

\textbf{Definition 9.10}. 
\begin{equation}
\sigma ^{\vee }=\{\mathbf{l\in }\text{ }V^{\ast }:\text{ }<\mathbf{l},%
\mathbf{y}>\geq 0\text{ }\forall \mathbf{y}\in \sigma \}.  \tag{9.9}
\end{equation}%
It explains why the set of $\left( l_{j}\right) _{i}^{\prime }s$
\textquotedblright lives" in the space dual to that for $a_{j}^{\prime }s$.
Next, in view of the results just described, we can rewrite Eq.(8.8) as 
\begin{equation}
f(\mathbf{z})=\sum\limits_{\mathbf{a\in S}_{\sigma }}\lambda _{\mathbf{a}}%
\mathbf{z}^{\mathbf{a}}=\sum\limits_{\mathbf{l}}\lambda _{\mathbf{l}}\mathbf{%
u}^{\mathbf{l}}.  \tag{9.10}
\end{equation}%
As before, these polynomials form a polynomial ring. The ideal for this ring
can be constructed based on observation that for the fixed\ $d_{i}$ \ and
the assigned set of \ cone generators $a_{i}^{\prime }$ there is more than
one set of generators for the dual cone. This redundancy produces relations
of the type 
\begin{equation}
u_{1}^{l_{1}}\cdot \cdot \cdot u_{k}^{l_{k}}=u_{1}^{\tilde{l}_{1}}\cdot
\cdot \cdot u_{k}^{\tilde{l}_{k}}.  \tag{9.11}
\end{equation}%
If now we require $u_{i}\in \mathbf{C}_{i},$ then the above equation belongs
to the ideal $I(\mathbf{z})$ of the above polynomial ring. In accord with
the Definition 9.6., Eq.(9.11) represents a hypersurface. Naturally, the
ideal $I(\mathbf{z})$ represents the intersection of these hypersurfaces.
The \textit{affine} \textit{toric} variety $X_{\sigma ^{\vee }}$ is made out
of the hypersurfaces which belong to $I(\mathbf{z}).$ The generators \{$%
u_{1},...,u_{k}\}\in \mathbf{C}^{k}$ are coordinates for $X_{\sigma ^{\vee
}} $. They represent the same point in $X_{\sigma ^{\vee }}$ if and only if 
\textbf{u}$^{\mathbf{l}}=$\textbf{u}$^{\mathbf{\tilde{l}}}.$ \ Thus formed
toric variety corresponds to just one (dual) cone. The set of cones having a
common origin can be assembled into a fan [24,25]. The fan is \textit{%
complete} if \ it spans the $k-$dimensional vector space. In fact, using
results of Appendix, part a), we notice that there is one-to-one
correspondence between the cones and the chambers defined in the Appendix.
From chambers one can construct a gallery and, hence, a building. So that a
complete fan is essentially a building. The information leading to the
design of a particular building can be thus used for construction of a toric
variety from the set $\Sigma $ (a complete fan) of affine toric varieties.
To do so, one needs the set of gluing maps \{$\Psi _{\sigma ^{\vee }\sigma
^{\vee }}\}.$Thus, we obtain the following

\textbf{Definition 9.11. }Let $\Sigma $ be a complete fan and $%
\coprod\nolimits_{\sigma ^{\vee }\in \Sigma }X_{\sigma ^{\vee }}$ be the
disjoint union of affine toric varieties. Then, using the set of gluing maps 
$\mathit{\{}$ $\Psi _{\sigma ^{\vee }\breve{\sigma}^{\vee }}\}$ such that
each of them identifies two points $x\in X_{\sigma ^{\vee }}$ and $\breve{x}%
\in X_{\breve{\sigma}^{\vee }}$ on respective affine varieties, one obtains
the toric variety $X_{\Sigma }$ determined by the fan $\Sigma .$

Thus constructed variety $X_{\Sigma }$ may contain singularities. This
should be obvious just by looking at Eq.(9.11). There is a procedure of
desingularization described, for example, in Ref.[24\textbf{] }which we are
going to by- pass. This is permissible in view of the results we have
discussed thus far. Clearly, if physically needed, such more complicated
varieties can be studied as well. Obtained results allow us to introduce the
following

\textbf{Definition} \textbf{9.12}. The \textit{torus action} is a continuous
map : $T\times X_{\Sigma }\rightarrow X_{\Sigma }$ such that for each affine
variety corresponding to the dual cone it is given by 
\begin{equation}
T\times X_{\sigma ^{\vee }}\rightarrow X_{\sigma ^{\vee }}\text{ , }%
(t,x)\mapsto tx:=(t^{a_{1}}x_{1},...,t^{a_{k}}x_{k}).  \tag{9.12}
\end{equation}%
Naturally, such an action should be compatible with the gluing maps thus
extending it from one cone (chamber) to the entire variety $X_{\Sigma }($%
building). The compatibility is easy to enforce since for each of Eq.s
(9.11) multiplication by $t-$factors will not affect the solutions set. This
can be formally stated as follows. Let $\Psi :$ $X_{\Sigma }\rightarrow X_{%
\tilde{\Sigma}}$ be a map and $\alpha :T\rightarrow T^{\prime }$ a
homomorphism, then the map $\Psi $ is called \textit{equivariant} if it
obeys the following rule compatible with earlier defined Eq.(9.2): 
\begin{equation}
\Psi (cx)=\chi (c)\Psi (x)\text{ for all }c\in T.  \tag{9.13}
\end{equation}%
As before, the factor $\chi (c)$ is a character of the algebraic (in our
case, torus) group. This fact is known as the Borel-Weil theorem [72]. As
such it belongs to the theory of the induced group representations [73].

\textbf{Remark 9.13}. In spite of its apparent simplicity, this theorem is
exceptionally deep.. It can be used for reconstruction of physical model
whose observables obey Eq.(9.13). Examples are given in Refs.[14,74]. Very
important contributions to the Borel-Weil-Bott theorem were made recently by
Teleman, Ref.s [75,76]. These are mainly applicable to the CFT since in his
case the loop groups need to be used as explained below. Evidently, models
reproducing the Veneziano and Veneziano-like amplitudes \ as well as all CFT
can be reconstructed \ exclusively with help of the Borel-Weil-Bott theorem
as the point of departure.

For physical applications we are also interested in maps $\Psi $ compatible
with Eq.(9.2) producing ring of symmetric invariants. Evidently, they are
given by 
\begin{equation}
\Psi (cx)=\Psi (x)  \tag{9.14}
\end{equation}%
To actually obtain \ these invariants we need to study the orbits of the
torus action. To this purpose, in view of Eq.(9.12), we need to consider the
following fixed point equation 
\begin{equation}
t^{a}x=x.  \tag{9.15}
\end{equation}%
Apart from trivial solutions$:x=0$ and $x=\infty $, there is a nontrivial
solution $t^{a}=1$ for any $x$. For integer $a^{\prime }s$ this is a
cyclotomic equation whose nontrivial $a-1$ solutions all lie on the circle $%
S^{^{1}}.$ In view of this circumstance, it is possible to construct the
invariants for this case as we would like to explain now. First, such an
invariant can be built as a \textit{ratio} of two equivariant mappings of
the type given by Eq.(9.13)\footnote{%
Clearly, the integrand in the period integral, Eq.(3.3), when written in the
projective form ( as explained in Section 3.1. of Part I) fits this
requirement.}.By construction, such a ratio is the\textbf{\ }\textit{%
projective} \textit{toric} variety. Unlike the affine case, such varieties
are \textit{not} represented by functions of homogenous coordinates in 
\textbf{CP}$^{n}.$Instead, they are just constants associated with points in 
\textbf{CP}$^{n}$ which they represent\footnote{%
This is nontrivial fact. It is used in Ref.[6] and Part III for development
of symplectic models reproducing the Veneziano amplitudes.}.Second option,
is to restrict the algebraic torus to a compact torus. These two options are
interrelated in an important way as we would like to explain now.

To this purpose we notice that Eq.(9.12) still holds if some of $t-$factors
are replaced by $1$'s. This means that one should take into account all
situations when one, two, etc. $t$-factors in Eq.(9.12) are replaced by 1's
and account for all permutations involving such cases. This observation
leads to the torus actions on toric subvarieties. It is very important that
different orbits which belong to different subvarieties do not overlap.
Thus, by design, $X_{\Sigma }$ is the \textit{disjoint union} of \textit{%
finite} number of orbits identified with the subvarieties of $X_{\Sigma }.$
This leads to the flag decomposition, etc. to be discussed further in Parts
III and IV. For the sake of space, in this part we consider only the main
ideas. For instance, let us consider a specific example of an action of the
map $\Psi $ on a monomial $\mathbf{u}^{\mathbf{l}}=u_{1}^{l_{1}}\cdot \cdot
\cdot u_{n}^{l_{n}}\equiv z_{1}^{l_{1}a_{1}}\cdot \cdot \cdot
z_{n}^{l_{n1}a_{n}}$. For such a map the character $\chi (c)$ is given by 
\begin{equation}
\chi (c)=t^{<\mathbf{l}\cdot \mathbf{a>}},  \tag{9.16}
\end{equation}%
where $<\mathbf{l}\cdot \mathbf{a}>=\sum\nolimits_{i}l_{i}a_{i}$ and both $%
l_{i}$ and $a_{i}$ are integers. Following Ref.[77], let us consider the
limit $t\rightarrow 0$ in the above expression. We obtain, 
\begin{equation}
c(t)=\left\{ 
\begin{array}{c}
1\text{ if }<\mathbf{l}\cdot \mathbf{a}>=0 \\ 
0\text{ if }<\mathbf{l}\cdot \mathbf{a}>\neq 0%
\end{array}%
\right. .  \tag{9.17}
\end{equation}%
The equation $<\mathbf{l}\cdot \mathbf{a}>=0$ describes a hyperplane or,
better, a set of hyperplanes for a given vector $\mathbf{a}$. Based on the
results of Appendix, such set forms at least one chamber. To be more
accurate, we would like to complicate matters a little bit by introducing a
subset $I$ $\subset $ $\{1,...,n\}$ such that, say, only those $%
l_{i}^{\prime }s$ which belong to this subset satisfy $\ <\mathbf{l}\cdot 
\mathbf{a}>=0.$ Naturally, one obtains the one- to- one correspondence
between such subsets and earlier mentioned flags. The set of such
constructed monomials forms the set of invariants of torus group action. In
view of Eq.(9.15), we would like to replace the limiting $t\rightarrow 0$
procedure by the limiting procedure requiring $t\rightarrow \xi ,$ where $%
\xi $ is the nontrivial $n$-th root of unity. After such a replacement we
are entering formally the domain of pseudo-reflection groups \ as explained
in Appendix, part d). Such groups are acting on hyperplanes $<\mathbf{l}%
\cdot \mathbf{a}>.$Replacing $t$ by $\xi $ causes us to change the rule,
Eq.(9.17), as follows 
\begin{equation}
c(\xi )=\left\{ \text{ \ \ }%
\begin{array}{c}
1\text{ if }<\mathbf{l}\cdot \mathbf{a}>=0\text{ }\func{mod}\text{\textit{n}}
\\ 
0\text{ if \ \ \ \ \ \ \ \ \ \ }<\mathbf{l}\cdot \mathbf{a}>\neq 0%
\end{array}%
\right. .  \tag{9.18}
\end{equation}%
At this point it is appropriate to recall Eq.(3.11a) of Part I. In view of
this equation, we shall call the equation $<\mathbf{l}\cdot \mathbf{a}>=n$ \
as the \textit{Veneziano condition} while the $\mathit{Kac-Moody-Bloch-Bragg}
$ $\mathit{(K-M-B-B)}$ \textit{condition},$\mathit{\ }$Eq.(3.22) of Part I,
can be written now as $\ \ \mathit{<}\mathbf{l}\mathit{\cdot }\mathbf{a}%
\mathit{>=0}$ $\func{mod}\mathit{n}$.

The results of the Appendix A, parts c)and d) indicate that the first option
(the Veneziano condition) is characteristic for the standard Weyl-Coxeter
(pseudo) reflection groups while the second is characteristic for the 
\textit{affine} Weyl-Coxeter groups thus leading to the Kac-Moody affine Lie
algebras.

\textbf{Remark 9.14. }In Chapter 4, Section 4.4, of Ref.[21].Ginzburg shows
how to recover finite dimensional representations, e.g.$sl_{k}(\mathbf{C}),$%
discussed in Section 8, even in the case when K-M-B-B condition is used
instead of the Veneziano. In this sense, in accord with Remark 9.13. , one
can design both CFT and high energy physics observables using the same
formalism. Still another option is discussed immediately below.

The arguments just presented provide ''physical'' proof of the theorem by
Serre.

\subsection{Additional uses of the theorem by Serre}

\subsubsection{Connections with the theory of hyperplane arrangements}

According to Appendix, part d), both real and complex reflection groups are
isometries of respectively \textbf{R}$^{n}$ and \textbf{C}$^{n}$ spaces
leaving some known quadratic forms invariant. The quadratic forms are
essentially the hyperplanes. Although the Corollary 2.3. to the theorem by
Solomon provides a very interesting connection between the hyperplanes and
pseudo-reflections, it should be clear that Eq.s(9.18) and (2.6) represent
different (but related) sets of hyperplanes (in view of the isomorphism,
Eq.(9.4)). The Weyl group $W$ introduced in Appendix permutes these
hyperplanes

\textbf{Definition 9.15.} A collection $\mathcal{A}$ of hyperplanes on which
the group $W$ acts transitively is called the \textit{reflection arrangement}%
. \ \ \ \ \ \ \ \ \ \ \ \ \ \ \ \ \ \ \ \ \ \ \ \ \ \ \ 

\textbf{Remark 9.16.} \ In mathematics literature one can find a large group
of researchers who take the above definition as a starting point of the
whole mathematical development presented in this paper, e.g. see
Refs.[33,34]. Such an approach is helpful in the following sense. The
hypergeometric integrals of the type given by Eq.(3.5) (and those discussed
in Part I) can be obtained as solutions of some differential equations (of
Picard-Fucs type) whose origin is naturally explained with help of the
theory of arrangements. The same equations can be obtained from the point of
view of singularity theory, e.g see Ref.[78]. In our earlier work, Ref.[10],
we have indeed obtained such type of equations for the Veneziano-type
integral, Eq.(3.5), using ideas from singularity theory. Since the theory of
arrangements was already applied successfully [79] to reproduce the results
of 2 dimensional CFT [80], it can be used, in principle, as unifying
formalism for both \textquotedblright new\textquotedblright\ string and
\textquotedblright old\textquotedblright\ CFT.

In Appendix, parts c) and d), we have explained in simple terms the
difference between the affine and standard (pseudo) reflection groups. The
associated with these groups polyhedra can be thought of as fundamental
domains for respective groups of isometries of \textbf{R}$^{n}$ and \textbf{C%
}$^{n}$. The action of such isometry groups causes tessellation of these
spaces. The situation here is analogous to that encountered in solid state
physics [81] (as we mentioned already in Part I) where it is well known that
the scattering processes in solids should be treated with account of
translational symmetry of the crystal so that the concepts of energy and
momentum \ loose their original meaning and should be modified to account
for periodicity\footnote{%
As a typical illustration we suggest to our readers to think about, is the
process of heat or electrical conduction in perfect crystals and its
explanation in the solid state physics literature [81].}.Such situation is
characteristic for \textit{all} CFT where one should use the affine
Weyl-Coxeter reflection groups and the associated with them Kac-Moody
algebras in accord with the Proposition A.1 of Appendix. At the same time,
for processes taking place in high energy physics, it is sufficient to
consider only the point group symmetries (in solid physics terminology),
i.e. the usual Weyl-Coxeter reflection groups and the complete fans
associated with them. Such a picture can be extended, if necessary, to
include the spherical, hyperbolic and complex hyperbolic spaces so that the
polyhedra associated with isometry groups of these spaces will represent the
respective fundamental domains. In the case of Kac-Moody algebras such a
program was actually implemented (e.g. for the hyperbolic spaces) as
described in the book by Kac, Ref.[57]. In the case of high energy physics
the next subsection can be used as a point of departure for analogous
development.

\subsubsection{From complex hyperbolic space to the Heisenberg group}

\bigskip

Earlier in Remark 7.4. we've noticed that the Hermitian quadratic form can
be extended so that the isometries of the resulting complex space are
complex hyperbolic. At that time such a possibility appeared no more than a
curiosity. However, upon discussing the exact solution of the Veneziano
model in Section 8 we made an observation that the Dirac--type equation
associated with the Hamiltonian, Eq.(8.11), is invariant with respect to the
Lorentz transformations and it is this invariance which eventually causes
connections with zeros of the Riemann zeta function. The connected part of
the Lorentz group describes isometries of the hyperbolic space, e.g. see [82%
\textbf{, }pages 64-66]. The complex hyperbolic space includes real
hyperbolic space as a subspace. According to Goldman, Ref.[47], the
pseudo-reflection groups are isometries of the complex hyperbolic space.

In our earlier work, Ref.[83], we discussed various properties of the real
hyperbolic space in connection with widely publicized AdS-CFT
correspondence. As can be seen either from the book by Thurston, Ref.[82],
or from our \ earlier work, the hyperbolic ball model of the real hyperbolic
space is quite adequate for description of many meaningful physical models.
In this case, the boundary of the hyperbolic space plays an important role.
For instance, the infinitesimal variations at the boundary of the Poincare$%
^{\prime }$ disc model- the simplest model of H$^{2}-$ produce naturally the
Virasoro algebra. Extension of the method producing this algebra to, say, H$%
^{3}$ is complicated by the Mostow rigidity theorem (as discussed in
Ref.[83]). This theorem tells us that the Teichm\"{u}ller space for the
hyperbolic 3-manifolds without boundaries is a point. Simply speaking, all
hyperbolic surfaces without boundaries in hyperbolic space are rigid
(nonbendable). This restriction can be lifted in certain cases.

\ Since the real hyperbolic space is a part of complex hyperbolic and since
the real hyperbolic space can be modeled by the hyperbolic ball model, it is
not too surprising that the complex hyperbolic space also can be modelled
with help of the complex hyperbolic ball model as it is demonstrated in
[47]. What is surprising however is that the isometry \ group at the
boundary of this ball model is the Heisenberg group. We would like to argue
that to make an extension as suggested in the Remark 7.4. is not an artifact
but, actually, a necessity. To demonstrate this we need to go to Part I and
to take into account the discussion on pages 14 and 15 related to phase
factors. This discussion is just an adaptation of results described in
Chapter 5 of the monograph by Lang, Ref.[84]. Specifically, on page 77 in
connection with calculation of the periods of the Fermat curve, he mentions
about a complex plane \textbf{C} with two points (0 and 1) deleted. Clearly,
the third point is $\infty $ and, therefore, such trice punctured plane has
the hyperbolic disk model as its universal cover. Due to factorization
property of the Veneziano amplitude (e.g. see Eq.(3.28) of Part I), the
model of complex projective space discussed earlier in Subsection 7.6.3.
will inherit the hyperbolicity coming from each complex plane \textbf{C}. In
fact, much more can be said following Ref.[46]. In particular, by analogy
with complex sphere (equivalent to a projective space \textbf{CP}$^{1})$
with $n$ points removed thus making it hyperbolic one can think about a
complex projective space \textbf{CP}$^{n}$ with 2n+1 hyperplanes removed. In
our case, using terminology of hyperplane arrangements, Ref[33,34], Eq.(2.6)
is called a \textit{defining polynomial} for such an arrangement. The
entries in such polynomial represent hyperplanes. The polynomial is zero
when at least one of its entries becomes zero.\textit{\ }In the simplest
case of 4 particle amplitude we have a polynomial $Q$ of the type: $Q=x(1-x)$
The third hyperplane is at infinity. Clearly, zeros of Q are the same as we
just mentioned. The general pattern can be deduced from Eq.(2.8) of Part I.
Hence, removal of certain number of hyperplanes from projective space can
indeed make it hyperbolic. This is the content of the theorem by Kiernan,
Ref [85], who proved that the manifold $M=$\textbf{CP}$^{n}\backslash \cup
_{i=1}^{2n+1}H_{i\text{ }}$ is hyperbolic if the hyperplanes $H_{i}$ are in
general position. We would like to prove this result using some physical
arguments relevant to calculations of the Veneziano amplitudes. For this
purpose it is sufficient only to take another look at our Eq.(1.22) of Part I%
\begin{equation}
P(k,t)\equiv \left( \frac{1}{1-t}\right) ^{k+1}=\dsum\limits_{n=0}^{\infty
}p(k,n)t^{n}  \tag{1.22, Part I}
\end{equation}%
As in Eq.(2.8) of Part I, we replace the $t-$variable in this equation by
the variable $t=u_{1}+u_{1}+\cdot \cdot \cdot +u_{n}$. Next, choosing the
parametrization of the projective space as described in Subsection 7.6.3.,
we would like to consider the generating function $K(\mathbf{z},\mathbf{\bar{%
z}})$ of the following type%
\begin{equation}
K(\mathbf{z},\mathbf{\bar{z}})=\frac{n!}{\pi ^{n}}\left( \frac{1}{1-t}%
\right) ^{n+1}  \tag{9.19}
\end{equation}%
where now $t=\dsum\limits_{i=1}^{n}z_{i}\bar{z}_{i}\equiv
\dsum\limits_{i=1}^{n}u_{i}$ (and the last sum is written as a deformation
retract ). The constants $n!$ and $\pi ^{n}$ are chosen in accord with Ref.s
[46, 47, page 79]. Such a function is known in geometric function theory,
Ref.[46], as the Bergman kernel. It is used as a potential for constriction
of (the Bergman) metric by the following rule:%
\begin{equation}
ds^{2}=\sum_{i,j=0}^{n}g_{i,j}(\mathbf{z},\mathbf{\bar{z}})dz_{i}d\bar{z}%
_{j}=\sum_{i,j=0}^{n}\left( \frac{\partial ^{2}}{\partial z_{i}\partial \bar{%
z}_{j}}\ln K(\mathbf{z},\mathbf{\bar{z}})\right) dz_{i}d\bar{z}_{j}. 
\tag{9.20}
\end{equation}%
Also, the same potential is used for construction of the fundamental (1,1)
form which (up to a factor $i/\pi )$ coincides with the first Chern class
[48, pages 219,220]. Using this observation along with standard facts from
the theory of characteristic classes it is rather straightforward now to
reproduce the Solomon algebra of invariants. Moreover, since the symplectic
manifolds all are of K\"{a}hler type and since in the present case the K\"{a}%
hler manifold is of Hodge type, Ref.[48, page 219], this fact can be used
for development of symplectic model reproducing the Veneziano amplitudes.
This is done in Ref.[6] and will be further discussed in Part III using more
rigorous mathematical arguments. The metric in Eq.(9.20)\ by design is the
metric of the complex hyperbolic ball $B^{n}$ model. Biholomorphic mappings
of $B^{n}$ are isometries of the Bergman metric, Ref.[47, page 79] and
Ref.[48, page 219]. \textit{It plays the same role for the} \textit{complex
hyperbolic space as the Lobachevsky metric for the real hyperbolic space}.
But, as we know already from our experience with \ Eq.s(1.19) and (2.8) of
Part I, Eq.(9.19) can be identified with the partition (generating) function
for the multiparticle Veneziano amplitudes! Thus, we just have demonstrated
that inseparable connections between the complex hyperbolic geometry and the
Veneziano amplitudes imply the existence of the Heisenberg group at the
boundary of $B^{n}$. The connections between the hyperbolic geometry inside $%
B^{n}$ and the Heisenberg group at the boundary of $B^{n}$ is explained in
detail in Goldman's monograph, Ref.[47]. Our earlier experience with the
AdS-CFT correspondence in real hyperbolic space, Ref.[83], suggests that
analogous constructions can be made in the complex hyperbolic space. The
intrinsic role of the Heisenberg group at the boundary of $B^{n}$ makes such
a project especially attractive.

Thus, the theory of polynomial invariants of finite (pseudo) reflection
groups and, especially, the Theorem 9.1.by Serre, not only allow us to
restore the scattering amplitudes and the generating function associated
with them but also impose \ very rigid constraints on analytical form of
such amplitudes thus making them to reflect the symmetries of space-time in
which they act. This puts the Veneziano amplitudes into very unique
position. Only future might tell if such a position should be replaced by
something even more fundamental.

\bigskip

\textbf{Notes added in proof}.

1. The results of Sections 8 and 9 can be put in a broader mathematical
context with help of the monograph on "Noncommutative Harmonic Analysis" by
M.E.Taylor, AMS Publishers, Providence , RI, 1986.

2.Relevance of spin chains to QCD had been discussed by Faddeev and
Korchemsky in the paper "High energy QCD\ as a completely integrable model",
arxiv: hep-th/9404173. The arxiv contains numerous follow up papers based on
that just cited.

3. In Part IV we shall discuss results by Faddeev and Korchemsky using
formalism developed in this work.

\ 

\textbf{Acknowledgement}.The author would like to thank the Editor of JGP,
Dr.Ugo Bruzzo, for careful selection of referee for this paper and the
anonymous referee for his thoughtful detailed remarks which helped to bring
this manuscript to its present form. The author also would like to thank
Prof. Boris Apanasov (U. of \ Oklahoma, Norman) who attracted his attention
to (the unpublished yet, at that time) notes by W. Goldman which
subsequently were converted into the book\ on complex hyperbolic geometry.
Also, the author would like to thank Professors Viktor Ginzburg (U. of \
Chicago) for some clarifications regarding the content of his book (with
Neil) and Michael Berry (U. of Bristol) for sharing his thoughts about
science and scientists and for sending CD ROM containing all of his works to
date (including that on $H=xp$).

\bigskip

\pagebreak

\bigskip

\textbf{Appendix: Some results from the theory of Weyl-Coxeter reflection
and pseudo-reflection groups}

\bigskip

a) \textit{The Weyl group. } Let $V$ be a finite dimensional vector space
endowed with a scalar product $<,>$ which is positive-definite symmetric
bilinear form. For each nonzero $\alpha \in V$ \ let $r_{\alpha }$ denote
the orthogonal reflection in the hyperplane H$_{\alpha }$ through the origin
perpendicular to $\alpha .$Clearly, the set of hyperplanes H$_{\alpha }$ is
in one-to- one correspondence with the set of $\alpha ^{\prime }s.$ For $%
v\in V$ we obtain, 
\begin{equation}
r_{\alpha }(v)=v-<v,\alpha ^{\vee }>\alpha ,  \tag{A.1.}
\end{equation}%
where $\alpha ^{\vee }=2\alpha /<\alpha ,\alpha >$ is the vector \textit{dual%
} to $\alpha .$ Thus defined reflection is an orthogonal transformation in a
sense that \ $<r_{\alpha }(v),r_{\alpha }(\mu )>=$\ $<\nu ,\mu >.$ In
addition,\ $\left[ r_{\alpha }(v)\right] ^{2}=1$ $\forall \alpha ,\nu .$
Conversely, these two properties imply the transformation law, Eq.(A.1).
From these results it follows that for $v=\alpha $ we get $r_{\alpha
}(\alpha )=-\alpha ,$ that is reflection in the hyperplane with change of
vector orientation. If the set of vectors $\alpha \in V$ is mutually
orthogonal, then $r_{\alpha }(v)=v$ for $v\neq \alpha $ but, in general, the
orthogonality is not required. Because of this, one introduces the \textit{%
root system }$\Delta $ of vectors which span $V$. Such a system is \textit{%
crystallographic} if for each pair $\alpha ,\beta \in \Delta $ one has 
\begin{equation}
<\alpha ^{\vee },\beta >\in \mathbf{Z}\text{ and }r_{\alpha }(\beta )\in
\Delta .  \tag{A.2.}
\end{equation}%
Thus, each reflection $r_{\alpha }$ ($\alpha \in \Delta )$ permutes $\Delta
. $ A finite collection of such reflections forms a group $W$ known as the 
\textit{Weyl group of }$\mathit{\Delta }$\textit{\ }. The vectors $\alpha
^{\vee }$ (for $\alpha \in \Delta )$ form a root system $\Delta ^{\vee }$ 
\textit{dual} to $\Delta $. Let $v\in \Delta $ be such that $<v,\alpha >\neq
0$ for each $\alpha \in \Delta $. Then, the set $\Delta ^{+}$ of roots $%
\alpha \in \Delta $ such that $<v$,$\alpha >>0$ is called a system of 
\textit{positive} roots of $\Delta $. A root $\alpha \in \Delta ^{+}$\ is 
\textit{simple} if it is not a sum of two elements from $\Delta ^{+}.$The
number of simple roots coincides with the dimension of the vector space $V$
and the root set $\Delta $ is made of the disjoint union $\Delta =\Delta
^{+}\amalg \Delta ^{-}$. The integral linear combinations of roots, i.e. $%
\sum\limits_{i}m_{i}\alpha _{i}$ \ with $m_{i}^{\prime }s$ being integers,
forms a root \textit{lattice} $Q$($\mathcal{\Delta )}$ in $V$ (that is free
abelian group of rank $n=\dim V).$ Clearly, the simple roots form a basis $%
\Sigma $ of $Q$($\mathcal{\Delta )}$. Accordingly, $Q$($\mathcal{\Delta }^{+}%
\mathcal{)}$ is made of combinations $\sum\limits_{i}m_{i}\alpha _{i}$ with $%
m_{i}^{\prime }s$ being nonnegative integers.

In view of one-to -one correspondence between the set of hyperplanes $\cup
_{\alpha }$H$_{\alpha }$ and the set of roots $\Delta ,$ it is convenient
sometimes to introduce the \textit{chambers} as connected components of the
complement of $\cup _{\alpha }$H$_{\alpha }$ in $V$. In the literature,
Ref.[17, page 70], this complement is known also as the \textit{Tits cone}.
Accordingly, for a given chamber $C_{i}$ its \textit{walls} are made of
hyperplanes H$_{\alpha }.$ The roots in $\Delta $ can therefore be
characterized as those roots which are orthogonal to some wall of $C_{i}$
and directed towards the interior of this chamber. A \textit{gallery}\textbf{%
\ }is a sequence \ ( $C_{0}$, $C_{1}$,..., $C_{\mathit{l}}$\ )\ of chambers
each of which is adjacent to and distinct from the next. Let $%
w=r_{i_{1}}...r_{i_{l}}$ then, treating the Weyl group $W$ as a chamber
system, a gallery from $1$ to $w$ can be formally written as ($1$, $%
r_{i_{1}},r_{i_{1}}r_{i_{2}},...,r_{i_{1}}...r_{i_{l}})$. If this gallery is
of the shortest possible $lenght$ $\ \mathit{l}$($w),$ then one is saying
that $r_{i_{1}}...r_{i_{l}}$ is \textit{reduced decomposition} for the word $%
w$ made of \textquotedblright letters\textquotedblright\ $r_{i_{j}}$. Let $%
C_{x}$ and $\ C_{y}$ be some distinct chambers which we shall call $x$ and $%
y $ for brevity. One can introduce the distance function $d(x,y)$ so that,
for example, if $w=r_{i_{1}}...r_{i_{l}}$ is the reduced decomposition, then 
$d(x,y)=w$ if and only if there is a gallery of the type $%
r_{i_{1}}...r_{i_{l}}$ from $x$ to $y$. If, for instance, $d(x,y)=r_{i}$ ,
this means simply that $x$ and $y$ are distinct and $i-$adjacent. A \textit{%
building} $\mathcal{B}$ is a chamber system having a distance function $%
d(x,y)$ taking values in the Weyl-Coxeter group $W$. Finally, an \textit{%
apartment} in a building $\mathcal{B}$ is a subcomplex $\mathcal{\hat{B}}$
of $\mathcal{B}$ which is isomorphic to $W$. There is a bijection $\varphi :$
$W\rightarrow \mathcal{\hat{B}}$ such that $\varphi (w)$ and $\varphi
(w^{\prime })$ are $i-$adjacent in $\mathcal{\hat{B}}$ if and only if $w$
and $w^{\prime }$ are adjacent in $W$, e.g. see Ref.[86].

\bigskip

b) \textit{The} \textit{Coxeter group. }The Coxeter group is related to the
Weyl group through the obviously looking type of relation between
reflections, 
\begin{equation}
\left( r_{\alpha }r_{\beta }\right) ^{m(\alpha ,\beta )}=1,  \tag{A.3}
\end{equation}%
where $m(\alpha ,\alpha )=1$ and $m(\alpha ,\beta )\geq 2$ for $\alpha \neq
\beta $. In particular, for \textit{finite} Weyl groups\ \ $m(\alpha ,\beta
)\in \{2,3,4,6\},$ Ref.[30\textbf{,} page 39], while for the \textit{affine}
Weyl groups (to be discussed below) $m(\alpha ,\beta )\in \{2,3,4,6,\infty
\} $, e.g. read Ref.[30, page 136], and the Proposition A.1$.$below.
Clearly, different refection groups will have different matrix $m(\alpha
,\beta )$ and, clearly, the matrix $m(\alpha ,\beta )$ is connected with the
bilinear form (the Cartan matrix, see below) for the Weyl's group $W$
[66,87]. As an example of \ use of the concept of building in the Weil
group, consider the set of \textit{fundamental} \textit{weights} defined as
follows.\ For the root basis $\Sigma $ (or $\Sigma ^{\vee })$ the set of
fundamental weights $\mathcal{D}=$\{$\omega _{1},...,\omega _{n}\}$\textit{\
with respect to} $\Sigma $ is defined by the rule: 
\begin{equation}
<\alpha _{i}^{\vee },\omega _{j}>=\delta _{ij}.  \tag{A.4}
\end{equation}%
The usefulness of such defined fundamental weights lies in the fact that
they allow to introduce the concept of the \textit{highest weight }$\lambda $
\textit{(}sometimes also known as the dominant weight, [88, page 203]. Thus
defined $\lambda $ can be presented as $\lambda
=\sum\nolimits_{i=1}^{d}a_{i}\omega _{j}$ with all $a_{i}\geq 0.$ Sometimes
it is convenient to relax the definition of fundamental weights to just
weights by comparing Eq.s(A.2) and (A.4). That is $\beta ^{\prime }s$ in
Eq.(A.2) are just weights. Thus, for instance, we have $\Delta $ as building
and a subcomplex $\mathcal{D}$ of fundamental weights as an apartment
complex.

To illustrate some of these concepts let us consider examples which are
intuitively appealing and immediately relevant to the discussion in the main
text. These are the root system $B_{d}$ and $C_{d}.$ They are made of vector
set $\{u_{1},...,u_{d}\}$ constituting an orthonormal basis of the $d-$%
dimensional cube. The vectors $u_{i}$ should not be necessarily of unit
length, Ref. [29, page 27]. It is important only that they all have the same
length. For B$_{d}$ system one normally chooses, Ref. [29, page 30], 
\begin{equation}
\Delta =\{\pm u_{i}\pm u_{j}\mid i\neq j\}\amalg \{\pm u_{i}\}.  \tag{A.5}
\end{equation}%
In this case, the reflections corresponding to elements of $\Delta $ can be
described by their effect on the set $\{u_{1},...,u_{d}\}.$ Specifically, $%
r_{u_{i}-u_{j}}=$permutation which interchanges $u_{i}$ and u$%
_{j};r_{u_{i}}= $sign change of $u_{i};r_{u_{i}+u_{j}}=$permutation which
interchanges $u_{i} $ and u$_{j}$ and changes their sign. The action of the
Weyl group on $\Delta $ can be summarized by the following formula 
\begin{equation}
W(\Delta )=\left( \mathbf{Z}/2\mathbf{Z}\right) ^{d}\unlhd \Sigma _{d} 
\tag{A.6}
\end{equation}%
with \ $\unlhd $ representing the semidirect product between the permutation
group $\Sigma _{d}$ and the dihedral group $\left( \mathbf{Z}/2\mathbf{Z}%
\right) ^{d}$ of sign changes both acting on $\{u_{1},...,u_{d}\}.$Thus
defined product constitutes the full symmetry group of the $d-$cube,
Ref.[29, page 31]. The same symmetry information is contained in $C_{d}$
root system defined by 
\begin{equation}
\Delta =\{\pm u_{i}\pm u_{j}\mid i\neq j\}\amalg \{\pm 2u_{i}\}  \tag{A.7}
\end{equation}%
Both systems possess the same root decomposition: $\Delta =\Delta ^{+}\amalg
\Delta ^{-}$ , Ref.\ [29, page 37]. In particular, considering a square as
an example we obtain the basis $\Sigma _{B_{2}}$ of $Q$($\mathcal{\Delta )}$
as 
\begin{equation}
\Sigma _{B_{2}}=\{u_{1}-u_{2},u_{2}\}.  \tag{A.8a}
\end{equation}%
From here the dual basis is given by 
\begin{equation}
\Sigma _{B_{2}}^{\vee }=\{u_{1}-u_{2},2u_{2}\}.  \tag{A.8b}
\end{equation}%
Using Eq.(A.4) we obtain the fundamental weights as $\omega _{1}=u_{1}$ \
and $\omega _{2}=\frac{1}{2}(u_{1}+u_{2})$ respectively. By design, they
obey the orthogonality condition , Eq.(A.4). The Dynkin diagram, Ref.[29,
page 122], for $B_{2}$ provides us with coefficients \ $a_{1}=1$ and $a_{2}$=%
$2$ obtained for the case when the expansion $\lambda
=\sum\nolimits_{i=1}^{d}a_{i}\omega _{j}$ is relaxed to $\lambda
=\sum\nolimits_{i=1}^{d}a_{i}\beta _{j}$ as discussed above. In view of
Eq.(A.8a) this produces at once: $\lambda _{B_{2}}=u_{1}+u_{2}.$
Analogously, for C$_{2}$ we obtain, 
\begin{equation}
\Sigma _{C_{2}}=\{u_{1}-u_{2},2u_{2}\},  \tag{A.9}
\end{equation}%
with coefficients $a_{1}=2$ and $a_{2}$=$2$ thus leading to $\lambda
_{C_{2}}=2\left( u_{1}+u_{2}\right) .$

For the square, these results are intuitively obvious. Evidently, the $d-$
dimensional case can be treated accordingly. The physical significance of
the highest weight should become obvious if one compares the Weyl-Coxeter
reflection\ group algebra with that for the angular momentum familiar to
physicists. In the last case, the highest weight means simply the largest
value of the projection of the angular momentum onto z-axis. The raising
operator will annihilate the wave vector for such a quantum state while the
lowering operator will produce all eigenvalues lesser than the maximal value
(up to the largest negative) and, naturally, all eigenfunctions. The
significance of the fundamental weights goes beyond this analogy, however.
Indeed, suppose we can expand some root $\alpha _{i}$ according to the rule 
\begin{equation}
\alpha _{i}=\sum\limits_{j}m_{ij}\omega _{j}.  \tag{A.10}
\end{equation}%
Then, substitution of such an expansion into Eq.(A.2) and use of Eq.(A.4)
produces: 
\begin{equation}
<\alpha _{k}^{\vee },\alpha _{i}>=\sum\limits_{j}m_{ij}<\alpha _{k}^{\vee
},\omega _{j}>=m_{ik}  \tag{A.11}
\end{equation}%
The expression $<\alpha _{k}^{\vee },\alpha _{i}>$ is known in the
literature as the Cartan matrix. It plays the central role in defining 
\textit{both} finite and infinite dimensional semisimple Lie algebras [57].
According to Eq.s(A.4),(A.10),(A.11), the transpose of the Cartan matrix
transforms the fundamental weights into the fundamental roots.

\bigskip\ 

c) \textit{The} \textit{affine Weyl-Coxeter groups. }Physical significance
of the affine Weyl-Coxeter reflection groups comes from the following
proposition

\textbf{Proposition A.1}. \textit{Let W be the Weyl group of any Kac-Moody
algebra. Then W is a Coxeter group for which }$m(\alpha ,\beta )\in
\{2,3,4,6,\infty \}.$\textit{\ Any Coxeter group with such }$m(\alpha ,\beta
)$\textit{\ is crystallographic (e.g. see Eq.(A.2) )}

\ \ 

The proof can be found in Ref.[87, pages 25-26]. To understand better the
affine Weyl-Coxeter groups, following Coxeter, Ref.[89], we would like to
explain in simple terms the origin and the physical meaning of these groups.
It is being hoped, that such a discussion might \ significantly facilitate
understanding of the results presented in the main text. We begin with the
quadratic form 
\begin{equation}
\Theta =\sum\limits_{i,j}a_{ij}x_{i}x_{j}  \tag{A.12}
\end{equation}%
having symmetric matrix $\left\Vert a_{ij}\right\Vert $ whose rank is $\rho
. $ Such a form is said to be \textit{positive definite }if it is positive
for all values of $\mathbf{x=\{}x_{1},...,x_{n}\}$ ( $n\geq \rho $ in
general !) except zero. \ It is \textit{positive semidefinite} if it is
never negative but vanishes for some $x_{i}^{\prime }s$ not all zero. The
form $\Theta $ is indefinite if it can be both positive for some $%
x_{i}^{\prime }s$ and negative for others.\footnote{%
For the purposes of comparison with existing mathematical physics literature
[26] it is suficient to consider only positive and positive semidefinite
forms.} If the positive semidefinite form vanishes for some $x_{i}=z_{i}$ $%
(i=1-n)$, then 
\begin{equation}
\sum\limits_{i}z_{i}a_{ij}=0,\text{ }j=1-n.  \tag{A.13}
\end{equation}%
For a given matrix $\left\Vert a_{ij}\right\Vert $ Eq.(A.13) can be
considered as the system of linear algebraic equations for $z_{i}^{\prime
}s. $ Let $\mathcal{N}=n-\rho $ be the \textit{nullity} of the form $\Theta $%
. Then, it is a simple matter to show that \textit{every positive
semidefinite connected }$\Theta $\textit{\ form is of nullity 1}. The form
is connected if it cannot be presented as a sum of two forms involving
separate sets of variables. The following two propositions play the key role
in causing differences between the infinite affine Weyl-Coxeter (Kac-Moody)
algebras and their finite counterparts

\textbf{Proposition A.2}. \textit{For any positive semidefinite connected }$%
\Theta $\textit{\ form there exist \ unique (up to multiplication by the
common constant)\textbf{\ positive} numbers z}$_{i}$\textit{\ satisfying
Eq.(A.13).}

\textbf{Proposition A.3. }\textit{If we modify a positive semidefinite
connected \ }$\Theta $ \textit{form by making one of the variables vanish,
the obtained form becomes positive definite.}

\ 

\bigskip Next, we consider the quadratic form $\Theta $ as the norm and the
matrix $a_{ij}$ as the metric tensor. Then, as usual, we have $\mathbf{x}%
\cdot \mathbf{x=}$ $\Theta =\left\vert \mathbf{x}\right\vert ^{2}$ and, in
addition, $\mathbf{x}\cdot \mathbf{y=}\sum\limits_{i,j}a_{ij}x_{i}y_{j}$ $%
\equiv \sum\limits_{i}x^{i}y_{i}=\sum\limits_{i}x_{i}y^{i}$ so that if
vectors $\mathbf{x}$ and $\mathbf{y}$ are orthogonal we get $%
\sum\limits_{i,j}a_{ij}x_{i}y_{j}=0$ as required. Each vector \textbf{x}
determines a point (\textbf{x}) and a hyperplane [\textbf{x}] with respect
to some reference point \textbf{0} chosen as an origin. The distance $\emph{l%
}$ between a point (\textbf{x}) and a hyperplane [\textbf{y}] measured along
the perpendicular is the projection of \textbf{x} along the direction of 
\textbf{y}, i.e. 
\begin{equation}
\emph{l}=\frac{\mathbf{x}\cdot \mathbf{y}}{\left\vert \mathbf{y}\right\vert }%
.  \tag{A.14}
\end{equation}%
Let now (\textbf{x}$^{\prime })$ be the image of (\textbf{x}) by reflection
in the hyperplane [\textbf{y}]. Then, $\mathbf{x}-\mathbf{x}^{\prime }$ is a
vector parallel to \textbf{y} of magnitude 2\emph{l.} Thus, 
\begin{equation}
\mathbf{x}^{\prime }=\mathbf{x}-2\frac{\mathbf{x}\cdot \mathbf{y}}{%
\left\vert \mathbf{y}\right\vert }\mathbf{y,}  \tag{A.15}
\end{equation}%
in accord with Eq.(A.1). From here, the equation for the reflecting
hyperplane is just $\mathbf{x}\cdot \mathbf{y=}0\mathbf{.}$ Let the vector 
\textbf{y} be pre assigned, then taking into account Propositions A.2 and
A.3 we conclude that for the nullity \ $\mathcal{N}=0$ the only solution
possible is \textbf{x}=\textbf{0. }That is to say, in such a case $n$
reflecting hyperplanes have the point \textbf{0 }as the only\textbf{\ }%
common intersection point. A complement of these hyperplanes in \textbf{R}$%
^{n}$ forms a chamber system\textbf{\ }discussed already in a). In the main
text it is called a complete fan in accordance with existing terminology
[24,25]. For $\mathcal{N}$=1 the equation $\mathbf{x}\cdot \mathbf{y=}0$ may
have many \textit{nonnegative} solutions for \textbf{x. }Actually\textbf{, }%
such reflecting hyperplanes occur in a finite number of different
directions. More accurately, such hyperplanes belong to a \textit{finite}
number of families, each consisting of hyperplanes \textit{parallel \ to
each other}. If we choose a single representative from each family in a such
a way that it passes through \textbf{0}, then the complement of such
representatives is going to form a polyhedral cone as before. But now, in
addition, we have a group of translations $\mathit{T}$ \ for each
representative of the hyperplane family so that the total affine Weyl group $%
W_{aff\text{ }}$ is the semidirect product : $W_{aff\text{ }}$ =$T\unlhd W.$
The fundamental region for $W_{aff\text{ }}$ is a simplex (to be precise, an
open simplex, Bourbaki, Ref.[17], Chr.5, Proposition 10) \ called \textit{%
alcove}\textbf{\ }bounded by $n+1$ hyperplanes (walls) $n$ of which are
reflecting hyperplanes passing through \textbf{0} while the remaining one
serves to reflect \textbf{0} into another point $\mathbf{0}^{\prime }.$ If
one connects \textbf{0} with $\mathbf{0}^{\prime }$ and reflects this line
in other hyperplanes one obtains a lattice. By analogy with solid state
physics [81\textbf{]} one can construct a dual lattice (just like in a) and
b) \ above) \ the fundamental cell of which is known in physics as the
Brilluin zone. For the alcove the fundamental region of the dual lattice
(the Brilluin zone) is the polytope having \textbf{0} for its centre of
symmetry, i.e. zonotope [89].

\ 

d) \textit{The} \textit{pseudo-reflection groups. }Although the
pseudo-reflection groups are also formally described in the monograph by
Bourbaki, Ref. [17], their geometrical (and potentially physical) meaning is
beautifully explained in the book by McMullen [32]. In particular, all \
earlier presented reflection groups are isometries of the Euclidean space.
Their action preserves some quadratic form which is real. More generally,
one can think of \ reflections in spherical and hyperbolic spaces. From this
point of view earlier described polytopes (polyhedra) represent the
fundamental regions for respective isometry groups. Action of these groups
on fundamental regions causes tesselation of these spaces (without gaps).
The collection of spaces can be enlarged by considering reflections in the
complex $n$-dimensional space \textbf{C}$^{n}.$ In this case the Euclidean
quadratic form is replaced by the positive definite Hermitian form. Since
locally \textbf{CP}$^{n}$ is the same as $\mathbf{C}^{n+1}$ and since 
\textbf{CP}$^{n}$ is at the same time a symplectic manifold with well known
symplectic two- form $\Omega $ [27], this makes the pseudo-reflection groups
(which leave $\Omega $ invariant) especially attractive for physical
applications (e.g. see Section 9 and Part III) The pseudo-reflections are
easily described. By analogy with Eq.(A.1) (or (A.15)) one writes 
\begin{equation}
r_{\alpha }(v)=v+(\xi -1)<v,\alpha ^{\vee }>\alpha  \tag{A.16}
\end{equation}%
where $\xi $ is the nontrivial solution of the cyclotomic equation $x^{h}=x$
and $\alpha ^{\vee }=\alpha /<\alpha ,\alpha >$ with $<x,y>$ being a
positive definite Hermitian form satisfying as before $<r_{\alpha
}(v),r_{\alpha }(\mu )>=$\ $<\nu ,\mu >$ with $\alpha $ being an eigenvector
such that $r_{\alpha }(\alpha )=\xi \alpha \footnote{%
According to Bourbaki, Ref. [17\textbf{]}, Chr5, paragraph 6, if $\xi $ is
an eigenvalue of the psedo-reflection operator, then $\xi ^{-1}$ is also an
eigenvalue with the same multiplicity.}.$ In addition, $\left[ r_{\alpha
}(\alpha )\right] ^{k}=\xi ^{k}\alpha $ for $1\leq k\leq h-1.$ This follows
from the fact that 
\begin{equation}
\left[ r_{\alpha }(\nu )\right] ^{k}=\nu +(1+\xi +\cdot \cdot \cdot +\xi
^{k-1})(\xi -1)<v,\alpha ^{\vee }>\alpha  \tag{A.17}
\end{equation}%
and taking into account that $(1+\xi +\cdot \cdot \cdot +\xi ^{k-1})(\xi
-1)=\xi ^{k}-1.$

Finally, the Weyl-Coxeter reflection groups considered earlier in this
Appendix can be treated as pseudo-reflection groups if one replaces a single
Euclidean reflection by the so called Coxeter element [19,29] \ $\omega $
which is a product of individual reflections belonging to the distinct roots
of $\Delta .$ Hence, the Euclidean Weyl-Coxeter reflection groups can be
considered as a subset of the pseudo-reflection groups so that all useful
information about these groups can be obtained from considering the same
problems for the pseudo-reflection groups. It can be shown [19,29] that the
Coxeter element $\omega $ has eigenvalues $\xi ^{m_{1}},...,\xi ^{m_{l}}$
with \emph{l} being dimension of the vector space $\Delta $ while the
exponents $m_{1},...,m_{l}$ being positive integers less than $h$ and such
that $\sum\nolimits_{i=1}^{l}(h-m_{i})=\sum\nolimits_{i=1}^{l}m_{i}$ . This
result implies that the number $\sum\nolimits_{i=1}^{l}m_{i}=N$ - the number
of positive roots in the Weyl-Coxeter group is connected with the Coxeter
number $h$ via relation : $N=\frac{1}{2}lh,$ Ref$.[$30$,$ page79].

\bigskip \pagebreak

\bigskip

\bigskip

\textbf{References}

\bigskip

[1] \ \ A. Kholodenko, New strings for old Veneziano amplitudes.

\ \ \ \ \ \ I. Analytical treatment, J.Geom.Physics 55 (2005) 50-74.

[2] \ \ R.Stanley, Combinatorics and Commutative Algebra,

\ \ \ \ \ \ Birkh\"{a}user, Boston,1996.

[3] \ \ S.Ghorpade, G. Lachaud, Hyperplane sections of Grassmannians

\ \ \ \ \ \ and the number of MDS linear codes,

\ \ \ \ \ \ Finite Fields and Their Applications 7 (2001) 468-506.

[4] \ \ R.Stanley, Enumerative Combinatorics, Vol.1.,

\ \ \ \ \ \ \ Cambridge University Press, Cambridge, UK, 1999

[5] \ \ M.Aigner, Combinatorial Theory,

\ \ \ \ \ \ Springer-Verlag, Berlin, 1979.

[6] \ \ A.Kholodenko, New models for Veneziano amplitudes:combinatorial,

\ \ \ \ \ \ \ symplectic and supersymmetric aspects, Int. J.of Geom.Methods
in

\ \ \ \ \ \ \ Mod.Phys.2 (4) (2005) 1-22.

[7] \ \ \ A.Kholodenko, Kontsevich-Witten model from 2+1 gravity:

\ \ \ \ \ \ \ new exact combinatorial solution, J.Geom.Phys. 43 (2002) 45-91.

[8] \ \ \ G.Andrews, K.Eriksson, Integer Partitions,

\ \ \ \ \ \ \ Cambridge University Press, Cambridge, UK, 2004.

[9] \ \ \ R.Bott, L.Tu, Differential Forms in Algebraic Topology,

\ \ \ \ \ \ \ Springer-Verlag, Berlin, 1982.

[10] \ \ A.Kholodenko, New Veneziano amplitudes from old

\ \ \ \ \ \ \ \ Fermat (hyper)surfaces, in: C.Benton Editor, Trends in

\ \ \ \ \ \ \ \ Mathematical Physics Research, pp 1-94,

\ \ \ \ \ \ \ \ Nova Science Publishers, Inc., New York, 2004.

[11] \ \ A.Weil, Numbers of solutions of equations in finite fields,

\ \ \ \ \ \ \ \ AMS Bulletin, 55(6) (1949) 497-508.

[12] \ \ J.Schwartz, Differential Geometry and Topology,

\ \ \ \ \ \ \ \ Gordon and Breach Inc., New York,1968.

[13]. \ M.Stone, Supersymmetry and the quantum mechanics of spin,

\ \ \ \ \ \ \ \ Nucl.Phys.B314 (1989) 557-586.

[14] \ \ O.Alvarez, I.Singer, p.Windey, Quantum mechanics and the

\ \ \ \ \ \ \ \ geometry of the Weyl character formula,

\ \ \ \ \ \ \ \ Nucl.Phys.B337(1990) 467-486.

[15] \ \ A.Polyakov, Gauge Fields and Strings,

\ \ \ \ \ \ \ \ Harwood Academic Publishers, New York, 1987.

[16] \ \ A.Kholodenko, V.Nesterenko, Classical dynamics of rigid string

\ \ \ \ \ \ \ \ from Willmore functional, J.Geom.Phys.16 (1995) 15-26.

[17] \ \ N.Bourbaki, Groupes et Algebres de Lie (Chapitre 4-6).

\ \ \ \ \ \ \ \ Hermann, Paris, 1968.

[18] \ \ L. Solomon, Invariants of finite reflection groups,

\ \ \ \ \ \ \ \ Nagoya Math.Journ.22(1963) 57-64.

[19] \ \ R. Carter, Simple Groups of Lie Type,

\ \ \ \ \ \ \ \ John Wiley \& Sons Inc., New York, 1972.

[20] \ \ R. Stanley, Relative invariants of finite groups generated by

\ \ \ \ \ \ \ \ pseudoreflections, J.of Algebra 49 (1977) 134-148.

[21] \ \ V. Ginzburg, Representation Theory and Complex Geometry\textit{,}

\ \ \ \ \ \ \ \ Birkh\"{a}user Verlag, Inc., Boston, 1997.

[22] \ \ M.Atiyah, R. Bott, The moment map and equivariant cohomology,

\ \ \ \ \ \ \ \ Topology 23(1984) 1-28.

[23] \ \ V.Guillemin, S.Sternberg, Supersymmetry and Equivariant

\ \ \ \textit{\ \ \ \ \ }de Rham Theory\textit{, }Springer-Verlag Inc.,
Berlin, 1999.

[24] \ \ W.Fulton, Introduction to Toric Varieties,

\ \ \ \ \ \ \ \ Ann.Math.Studies 131,

\ \ \ \ \ \ \ \ \ Princeton University Press, Princeton,1993.

[25] \ \ \ G.Ewald, Combinatorial convexity and Algebraic Geometry,

\ \ \ \ \ \ \ \ \ Springer-Verlag, Inc., Berlin, 1996.

[26] \ \ \ V. Guillemin,V. Ginzburg,Y. Karshon, Moment Maps,

\textit{\ \ \ \ \ \ \ \ }Cobordisms and Hamiltonian Group Actions\textit{, }

\ \ \ \ \ \ \ \ AMS Publishers, Providence, RI, 2002.

[27] \ \ M.Atiyah, Angular momentum, convex polyhedra and algebraic

\ \ \ \ \ \ \ \ geometry, Proceedings of the Edinburg

\ \ \ \ \ \ \ \ Math.Society 26 (1983) 121-138.

[28] \ \ A. Koushnirenko, The Newton polygon and the number of solutions

\ \ \ \ \ \ \ \ of a system of $k$ equations in $k$ unknowns,

\ \ \ \ \ \ \ \ Uspekhi.Math.Nauk 30 (1975) 302-303.

[29] \ \ R. Kane, Reflection Groups and Invariant Theory\textit{,}

\ \ \ \ \ \ \ \ Springer-Verag, Inc., Berlin, 2001.

[30] \ \ J.Humphreys, Reflection Groups and Coxeter Groups,

\ \ \ \ \ \ \ \ Cambridge University Press, Cambridge,UK,1997.

[31] \ \ G. Shepard, J. Todd, Finite unitary reflection groups,

\ \ \ \ \ \ \ \ \ Canadian J.of Math. 6 (1974) 274-304.

[32] \ \ P. McMullen, Absract Regular Polytopes,

\ \ \ \ \ \ \ \ Cambridge University Press, Cambridge,UK, 2002.

[33]\ \ \ P. Orlik, H. Terrao, Arrangements and Hypergeometric Integrals%
\textit{,}

\ \ \ \ \ \ \ \ Math.Soc. Japan Memoirs, Vol.9.,

\ \ \ \ \ \ \ \ \ Japan Publ.Trading Co., Tokyo, 2001.

[34] \ \ P.Orlik, H.Terao, Arrangements of Hyperplanes\textit{,}

\ \ \ \ \ \ \ \ Springer-Verlag, Inc., Berlin, 1992.

[35] \ \ W. Lerche, C.Vafa, N.Warner, Chiral rings in N=2

\ \ \ \ \ \ \ \ superconformal theories,

\ \ \ \ \ \ \ \ Nucl.Phys.B324 (1989) 427-474.

[36] \ H.Hiller, Geometry of Coxeter Groups,

\ \ \ \ \ \ \ Pitman Advanced Publishing Program, Boston, 1982.

[37] \ I.Berenstein, I.Gelfand and S.Gelfand, Schubert cells and

\ \ \ \ \ \ \ cohomology of the space G/P, Russian Math.Surveys 28 (1973)
17-32.

[38] \ A.Mishchenko, Vector Bundles and Their Applications,

\ \ \ \ \ \ \ Kl\"{u}wer Academic Publishers, Dordrecht, 1998.

[39] \ D.Husem\"{o}ller, Fibre Bundles,

\ \ \ \ \ \ \ Springer-Verlag, Inc., Berlin, 1994.

[40]\ \ W.Fulton, Young Tableaux,

\ \ \ \ \ \ \ Cambridge U.Press, Cambridge,UK, 1997.

[41] \ J.Harris, Algebraic Geometry. A First Course,

\ \ \ \ \ \ \ Springer-Verlag, Inc., Berlin, 1992.

[42] \ R. Parthasarathy, K. Viswanathan, Geometric properties of QCD

\ \ \ \ \ \ \ string from Willmore functional,

\ \ \ \ \ \ \ J. Geom. Phys. 38 (2001) 207--216.

[43] \ R.Stanley, Enumerative Combinatorics, Vol.2.,

\ \ \ \ \ \ \ Cambridge U.Press, Cambridge,UK, 1999.

[44] \ I.Macdonald, Symmetric Functions and Orthogonal Polynomials,

\ \ \ \ \ \ \ AMS Publishers, Providence, RI, 1998.

[45] \ T.Miwa, M.Jimbo, E.Date, Solitons: Differential Equations,

\ \ \ \ \ \ \ Symmetries and Infinite Dimensional Algebras,

\ \ \ \ \ \ \ Cambridge U.Press, Cambridge, UK, 2000.

[46] \ B.Shabat, Introduction to Complex Analysis, Vol.2,

\ \ \ \ \ \ \ AMS Publishers, Providence, RI, 1992.

[47] \ W.Goldman, Complex Hyperbolic Geometry,

\ \ \ \ \ \ \ Clarendon Press, Oxford, 1999.

[48] \ R.Wells, Differential Analysis on Complex Manifolds,

\ \ \ \ \ \ \ Springer-Verlag, Inc., Berlin, 1980.

[49] \ E.Witten, Supersymmetry and Morse theory,

\ \ \ \ \ \ \ \ J.Diff.Geom. 17(1982) 661-692.

[50] \ T.Frankel, Fixed points and torsion on K\"{a}hler manifolds,

\ \ \ \ \ \ \ Ann.Math.70 (1959) 1-8.

[51] \ V.Guillemin, Moment Maps and Combinatorial Invariants of

\ \ \ \ \ \ \ Hamiltonian T$^{n}-$spaces, Birkh\"{a}user, Boston, 1994.

[52] \ N.Berline, E.Getzler, M.Vergne, Heat Kernels and Dirac Operators,

\ \ \ \ \ \ \ Springer-Verlag, Inc., Berlin, 1992.

[53] \ J.Humphreys, Introduction to Lie Algebra and Representation Theory,

\ \ \ \ \ \ \ Springer-Verlag, Inc., Berlin, 1992.

[54]\ \ R.Stanley, Invariants of finite groups and their applications to

\ \ \ \ \ \ \ combinatorics, AMS Bulletin (New Series) 1(1979) 475-511.

[55] \ J.Dixmier, Enveloping Algebras,

\ \ \ \ \ \ \ Elsevier, Amsterdam, 1977.

[56] \ J-P. Serre, Algebres de Lie Semi-Simples Complexes,

\ \ \ \ \ \ \ Benjamin Inc., New York, 1966.

[57] \ V.Kac, Infinite Dimensional Lie Algebras,

\ \ \ \ \ \ \ Cambridge U.Press, UK, 1990.

[58] \ A.Huckelberry, T.Wurzbacher, Infinite Dimensional K\"{a}hler
Manifolds,

\ \ \ \ \ \ \ Birkh\"{a}user, Boston, 2001.

[59] \ M.Mimura, H.Toda, Topology of Lie Groups, I and II,

\ \ \ \ \ \ \ AMS Publishers, Providence, RI, 1991.

[60] \ M.Berry, J.Keating, The Riemann zeros and eigenvalue asymptotics,

\ \ \ \ \ \ \ SIAM Review 41 (1999) 236-266.

[61] \ F.Faure, S.Nonnenmacher, S.De Bievre, Scarred eigenstates for

\ \ \ \ \ \ quantum cat maps of mininal periods,

\ \ \ \ \ \ Comm.Math.Phys. 239 (2003) 449-492.

[62] \ S.Nonnenmacher, A.Voros, Eigenstate structures around a hyperbolic

\ \ \ \ \ \ \ point, J.Phys.A 30 (1997) 295-315.

[63] \ J.Keating, F.Mezzadri, M.Robbins, Quantum boundary conditions for

\ \ \ \ \ \ \ torus maps, Nonlinearity 12 (1999) 579-591.

[64] \ S.Okubo, Lorentz-invariant Hamiltonian and Riemann hypothesis,

\ \ \ \ \ \ \ J.Phys.A 31 (1998) 1049-1057.

[65] \ R.Feres, Dynamical systems and Semisimple Groups: An Introduction,

\ \ \ \ \ \ \ Cambridge U.Press, Cambridge, UK, 1998.

[66] \ I. Macdonald, Linear Algebraic Groups,

\ \ \ \ \ \ \ LMS Student Texts , Vol.32,

\ \ \ \ \ \ \ Cambridge U.Press, Cambridge, UK, 1990.

[67] \ J-P. Serre, Groupes finis d'automorphismes d'anneaux

\ \ \ \ \ \ \ locaux regulieres, Colloq, d'Alg. Ec.Norm. Sup.

\ \ \ \ \ \ \ de Jeunes Filles, Paris 8-01-8-11, 1967.

[68] \ D.Benson, Polynomial Invariants of Finite Groups,

\ \ \ \ \ \ \ LMS Lecture Notes, Vol.190,

\ \ \ \ \ \ \ Cambridge U.Press, Cambridge,UK, 1993.

[69]\ \ L.Smith, Polynomial Invariants of Finite Groups,

\ \ \ \ \ \ \ A.K.Peters, Ltd., Wellesley, Ma, 1995.

[70] \ E.Noether, Der Endlischkeitsatz der Invarianten endlicher

\ \ \ \ \ \ \ \ Gruppen, Math.Ann.77 (1916) 89-92.

[71] \ R.Hwa, V.Teplitz, Homology and Feynman Integrals,

\ \ \ \ \ \ \ \ W.A.Benjamin, Inc. New York, 1966.

[72] \ J.Taylor, Several Complex Variables with Connections to

\ \ \ \ \ \ \ \ Algebraic Geometry and Lie Groups,

\ \ \ \ \ \ \ AMS Publications, Providence, RI, 2002.

[73] \ A.Kirillov, Elements of the Theory of Representations\textit{,}

\ \ \ \ \ \ \ (in Russian), Nauka, Moscow,1972.

[74] \ O.Alvarez, I.Singer, P.Windey,The supersymmetric $\sigma -$model and

\ \ \ \ \ \ \ the geometry of the Weyl-Kac character formula,

\ \ \ \ \ \ \ Nucl.Phys.B373(1992) 647-687.

[75] \ C.Teleman, The quantization conjecture revisited,

\ \ \ \ \ \ \ Ann.Math.152 (2000) 1-43.

[76] \ S.Fishel, I.Grojnowski, C.Teleman, The strong Macdonald conjecture

\ \ \ \ \ \ \ and Hodge theory on the loop Grassmannian,

\ \ \ \ \ \ \ arxiv: math.AG/0411355.

[77] \ G.Kempf, F.Knudsen, D.Mumford, B.Saint-Donat,

\ \ \ \ \ \ \ Toroidal Embeddings I,

\ \ \ \ \ \ \ LNM 339, Springer-Verlag, Berlin, 1973.

[78] \ V. Arnol'd, S.Gussein-Zade, A.Varchenko,

\ \ \ \ \ \ \ Singularities of Differentiable Maps,Vol.2,

\ \ \ \ \ \ \ Birkh\"{a}user, Boston, 1988.

[79] \ A.Varchenko, Multidimensional Hypergeometric Functions

\ \ \ \ \ \ \ and Representation Theory of Lie Algebras and Quantum Groups,

\ \ \ \ \ \ \ World Scientific, Singapore, 1995.

[80] \ P. Di Francesco, P.Mathieu, D.Senechal, Conformal Field Theory,

\ \ \ \ \ \ \ Springer-Verlag, Berlin, 1997.

[81] \ N. Ashcroft, D. Mermin, Solid State Physics\textit{,}

\ \ \ \ \ \ \ Saunders College Press, Philadelphia, 1976.

[82] \ W.Thurston, Three Dimensional Geometry and Topology, Vol.1.,

\ \ \ \ \ \ \ Princeton University Press, Princeton, NJ, 1997.

[83]\ \ A.Kholodenko, Boundary conformal field theories,

\ \ \ \ \ \ \ limit sets of Kleinian groups and holography,

\ \ \ \ \ \ \ J.Geom.Phys. 35 (200) 193-238.

[84] \ \ S.Lang, Introduction to Algebraic and Abelian Functions,

\ \ \ \ \ \ \ Springer-Verlag, Berlin, 1982.

[85] \ P.Kiernan, Hyperbolically imbedded spaces and the big

\ \ \ \ \ \ \ Picard theorem, Math.Ann.204 (1973) 203-209.

[86] \ W.Kantor, R. Liebler, S. Payne, E.Schult, Finite Geometries,

\ \ \ \ \ \ \ Buildings and Related Topics, Clarendon Press, Oxford, 1990.

[87] \ S.Kumar, Kac-Moody Groups, Their Flag Varieties

\ \ \ \ \ \ \ and Representation Theory, Birkh\"{a}user Inc., Boston, 2002.

[88] \ W.Fulton, J. Harris, Representation Theory. A first Course\textit{,}

\ \ \ \ \ \ \ Springer-Verlag, Inc., Berlin, 1991.

[89] \ H.Coxeter, Regular Polytopes,

\ \ \ \ \ \ The Macmillan Co., New York, 1963.

\bigskip

\bigskip

\bigskip

\bigskip

\bigskip

\bigskip

\end{document}